\documentclass[superscriptaddress,aps,prd,amsmath,amssymb,amsfonts,nofootinbib,preprint]{revtex4-1}
\usepackage{amssymb,amsfonts}
\usepackage{mathrsfs}
\usepackage{tikz}
\usetikzlibrary{external}
\usepackage{amsthm}

\usepackage[colorlinks, allcolors=blue!70!black, linktocpage]{hyperref}
\usepackage{accents}
\usepackage{graphicx}
\usepackage[utf8x]{inputenc}
\usepackage{float}
\usepackage{amsmath}
\DeclareMathAlphabet{\mathpzc}{OT1}{pzc}{m}{it}
\usetikzlibrary{patterns}
\usepackage{titlesec}
\usepackage{cases}
\usepackage{mathtools}
\usepackage{color}
\usepackage{cleveref}
\usepackage{braket,mathtools,float,tikz,amsfonts}
\usepackage{extarrows}

\newcommand{\ex}[1]{\mathrm{e}^{#1}}

\newcommand{\pa}[1]{\left(#1 \right)}

\newcommand{\ca}[1]{\mathcal{#1}}

\newcommand{\abs}[1]{\left|#1\right|}

\newcommand{\ar}[1]{\xrightarrow[#1]{}}

\newcommand{\fr}{\frac}

\def\be{\begin{equation}}
\def\ee{\end{equation}}
\def\ba{\begin{eqnarray}}
\def\ea{\end{eqnarray}}

 \def\ba{{\bar{\alpha}}}

 \def\b{{\beta}}

\newcommand{\bea}{\begin{eqnarray}}
\newcommand{\eea}{\end{eqnarray}}

\definecolor{Gray}{gray}{0.95}
\definecolor{LightCyan}{rgb}{0.88,1,1}

\def\Tr{{\text{Tr}}}

\crefname{lem}{lemma}{lemmas}
\crefname{thm}{theorem}{theorems}
\crefname{cor}{corollary}{corollaries}
\crefname{rem}{remark}{remarks}
\crefname{prop}{proposition}{propositions}

\begin{document}
\count\footins = 1000

\preprint{YITP-20-05}

\title{Correlation measures and the entanglement wedge cross-section after quantum quenches in two-dimensional conformal field theories}

\author{Jonah Kudler-Flam}
\email{jkudlerflam@uchicago.edu}
\affiliation{Kadanoff Center for Theoretical Physics, University of Chicago, Chicago, IL~60637, USA}
% \affiliation{James Franck Institute, University of Chicago, Chicago, Illinois 60637, USA}

\author{Yuya Kusuki}
\email{yuya.kusuki@yukawa.kyoto-u.ac.jp}
\affiliation{Center for Gravitational Physics, 
Yukawa Institute for Theoretical Physics (YITP), Kyoto University,
Kitashirakawa Oiwakecho, Sakyo-ku, Kyoto 606-8502, Japan.}

\author{Shinsei Ryu}
\email{ryuu@uchicago.edu}
\affiliation{Kadanoff Center for Theoretical Physics, University of Chicago, Chicago, IL~60637, USA}
\affiliation{James Franck Institute, University of Chicago, Chicago, Illinois 60637, USA}

\begin{abstract}
We consider the time
evolution of mixed state 
correlation measures 
in two-dimensional 
conformal field theories,
such as 
logarithmic negativity, 
odd entropy,
and reflected entropy,
after quantum quenches of 
various kinds.
These correlation measures,
in the holographic context,
% AdS$_3$/CFT$_2$,
are all associated to the entanglement wedge cross section.
% To see the implication of 
% these correlation measures 
% on quantum information scrambling,
We contrast various classes of conformal field theories, 
both rational and irrational (pure) conformal field theories.
First, 
for rational conformal field theories,
whose dynamics can be well described 
by the quasi-particle picture,
we find 
all four quantities for disjoint intervals to be proportional, 
regardless of the specific quench protocol. 
Second, 
using the light cone bootstrap, we generalize our results to irrational conformal field theories where we find sharp distinctions from the quasi-particle results and striking differences between mutual information and the other measures.
The large surplus of logarithmic negativity 
relative to mutual information
forces us to reconsider what mutual information and logarithmic negativity really measure.
We interpret these results as a signature of information scrambling and chaos in irrational theories. These CFT results perfectly agree with our gravitational (holographic) calculations. 
Furthermore, using holography, we are able to generalize the results to outside of the light cone limit. 
Finally,
due to the breakdown of the quasi-particle picture for irrational theories, we appeal to the ``line-tension picture," motivated by random unitary circuits, as a phenomenological description. We observe that random unitary circuits, with local Hilbert space dimension determined by the Cardy formula, have precisely the same entanglement dynamics as irrational (including holographic) conformal field theories.
\end{abstract}

\maketitle
\tableofcontents

\maketitle
\flushbottom
\newpage

\section{Introduction}

%%%%%%%%%%%%%%%%%%%%%%%%%%%%%%%%%%%%%%%%%%%%%%%%%%%%%%%%%%%%%%%%%%%%%%%%%%%%%%%%%%%%%%%%%%%%%%
% \subsection{Introduction}
%%%%%%%%%%%%%%%%%%%%%%%%%%%%%%%%%%%%%%%%%%%%%%%%%%%%%%%%%%%%%%%%%%%%%%%%%%%%%%%%%%%%%%%%%%%%%%

A fundamental objective in the study of quantum systems is to understand how information and correlations flow in space and time. Significant progress has been made by studying the dynamics of von Neumann entropy in non-equilibrium situations \cite{2015NatPh..11..124E,2016AdPhy..65..239D,2016RPPh...79e6001G,2016JSMTE..06.4003C}. However, the von Neumann entropy is only a proper measure of correlations for pure quantum states. In order to have a deeper understanding of the behavior of correlations, both quantum and classical, in non-equilibrium settings, one needs to study mixed state correlation measures. 

Information theorists have introduced an abundance of such measures, though the vast majority are formulated in terms of intractable (NP-hard) optimization procedures \cite{2009RvMP...81..865H}. 
One main exception 
is the logarithmic negativity, which may be operationally defined in terms of the reduced density matrix,
and is expected to capture only quantum correlations
\cite{PhysRevLett.77.1413,1996PhLA..223....1H,1999JMOp...46..145E,2000PhRvL..84.2726S,2002PhRvA..65c2314V,2005PhRvL..95i0503P}. 
This can be contrasted with the mutual information,
which captures 
total (both quantum and classical) correlations,
though  
is also useful and tractable. 
In addition to the negativity and mutual information, in this paper, we also study two recently introduced information theoretic quantities, the odd entropy and the reflected entropy \cite{PhysRevLett.122.141601,2019arXiv190500577D}. Knowledge of the correlation properties of each quantity is quite limited, so our work serves to build intuition for what exactly they are measuring.

What makes these new quantities particularly intriguing is that, like negativity \cite{2019PhRvD..99j6014K,PhysRevLett.123.131603}, they are intimately related, in the holographic context,
to a specific geometric object known as the entanglement wedge cross-section \cite{2018NatPh..14..573U,2018JHEP...01..098N}. 
The entanglement wedge cross section 
%is 
can be thought of as
a natural generalization of the Ryu-Takayanagi surface \cite{2006PhRvL..96r1602R,2006JHEP...08..045R,2007JHEP...07..062H} in holographic conformal field theories and has recently received considerable attention.

The archetypal way to probe non-equilibrium dynamics in quantum systems is by inducing a quantum quench. Quantum quenches come in several flavors. 
In a \textit{global quantum quench}, 
one first prepares the system in a short-ranged entangled state e.g.~the ground state of a gapped Hamiltonian, $H'$. 
Afterwards, one instantaneously changes to the quench Hamiltonian, $H$, so that the system is highly excited with respect to $H$. One can then track how various quantities evolve in time. 
More specifically,
we start with the initial state 
\begin{align}
    \ket{\Psi_0} = e^{-\beta H /4} \ket{B}
\end{align}
where $\ket{B}$ is a short-range entangled state 
(a boundary state in the case of CFTs).
We have added a ``smearing" or
``regularization" factor $e^{-\beta H /4}$,
which makes
the state have correlations over a 
length scale of order $\beta$. 
We will often call the smearing parameter $\beta$
``temperature", although we should remember that
we are always dealing with pure states.
We then evolve in Lorentzian time with the gapless Hamiltonian to quench
\begin{align}
    \ket{\Psi(t)} = e^{-iHt}\ket{\Psi_0}.
\end{align}
This has the effect of an instantaneous homogeneous injection of energy with 
effective temperature $\beta$. 
Global quenches are experimentally relevant
\cite{2015arXiv150901160I,
2019Sci...364..256L, 
2019Sci...364..260B}
and have been intensely studied both numerically and analytically.

While homogeneous global quenches bring to the forefront many interesting phenomena, they miss a great deal of the richness of non-equilibrium dynamics due to the translational invariance. This may be remedied in several ways. One way is to provide inhomogeneity in the quench protocol by injecting more energy in certain areas than others. Another, is to excite the system locally. With local quenches, one can isolate the coherent propagation of quantum information. Using techniques from conformal field theory, we will study all of the aforementioned correlations measures in (in)homogeneous global and local quenches.

In discussing non-equilibrium dynamics,
we naturally distinguish 
different CFTs in terms of the complexity 
of their spectra.
One class of theories of our interest 
are rational CFTs --
for these, we expect that their dynamics
can be largely described by the quasi-particle picture.
The quasi-particle posits that for all integrable systems, when the system is excited above the ground state, the entanglement dynamics can be fully accounted for by tracking the trajectories of local ``quasi-particles" \cite{2005JSMTE..04..010C,2017PNAS..114.7947A,2018ScPP....4...17A,2018arXiv180909119A}. For a global quench, quasi-particle pairs are created homogeneously across the entire system. Each quasi-particle pair contains certain entanglement content dependent on its (free-streaming) velocity. At a given time after the quench, the entanglement between two arbitrary regions can be computed by the number of quasi-particle pairs the regions share, weighted by the entanglement content. The velocities of the quasi-particles are determined by the dispersion relation. Because we only consider massless quantum field theories in this paper, all quasi-particles will move at the speed of light in the following. The entanglement content of quasi-particles for the R\'enyi entropies may be extracted from generalized Gibbs ensemble (GGE) thermodynamic entropy. 

On the other hand, 
we expect the quasi-particle picture fails
for irrational conformal field theories.
In this paper, we discuss
pure CFTs --
these are defined as a specific family of irrational conformal field theories, in that, they are unitary compact $c>1$ CFTs with no conserved currents except for the stress-energy tensor. At large central charge, these CFTs may be described holographically.

%%%%%%%%%%%%%%%%%%%%%%%%%%%%%%%%%%%%%%%%%%%%%%%%%%%%%%%%%%%%%%%%%%%%%%%%%%%%%%%%%%%%%%%%%%%%%%
\subsection{Summary of main results}
%%%%%%%%%%%%%%%%%%%%%%%%%%%%%%%%%%%%%%%%%%%%%%%%%%%%%%%%%%%%%%%%%%%%%%%%%%%%%%%%%%%%%%%%%%%%%%

The purpose of the paper is to study 
the time-evolution of 
the mixed state correlation measures 
after quantum quench,
for rational and irrational (pure) conformal field theories. 
Here we briefly summarize our results.

\begin{itemize}

\item Universal contribution to correlation in generic quenches (Section \ref{univ_sec})

We study the theory independent part of correlation functions in boundary conformal field theory on the upper half plane. This allows us to determine the ``universal" contribution to all correlation measures of interest: logarithmic negativity, mutual information, odd entropy, and reflected entropy. By computing these correlation functions on the upper half plane, we are able to conformally map the results to determine how each quantity evolves following global (in)homogeneous and local quantum quenches. Intriguingly, we find that they are all proportional regardless of the quench protocol. 
Furthermore, all results are consistent with the quasi-particle picture, providing credence that the newly introduced odd and reflected entropies truly measure correlations and are well-behaved in complex dynamical settings.
 
\item Reflected entropy vs.~mutual information after global quenches (Section \ref{light cone_sec})

The reflected entropy 
measures entanglement and classical correlation between subsystems.
For this reason, it is natural to study the dynamics of correlations under a global quench by utilizing the reflected entropy and comparing it to the mutual information \cite{2015JHEP...09..110A}.
In this section, we compute these quantities
in the light-cone limit, in which
$\beta \ll t,d, L$.
We consider two subsystems with length $L$, which are separated by $d<L$ as shown in Fig.~\ref{fig:setup}. 
This parameter range, which has been previously overlooked, shows a striking difference between the mutual information and reflected entropy; this is
displayed in Fig.~\ref{fig:result12} where we denote the mutual information by $I$ and the reflected entropy by $S_R$. 
As a signature of chaos, we find {\it missing entanglement} (left of Fig.~\ref{fig:result12}) and {\it mysterious correlation} which can be detected by the reflected entropy, odd entropy, and negativity but not the mutual information (right of Fig.~\ref{fig:result12}).
This result raises a potential question about quantum entanglement and classical correlation.
Given this mysterious correlation, we try to reconcile 
% give a physical explanation for this mysterious in Section \ref{light cone_sec} from 
the following two facts: (i) the reflected entropy is more sensitive to classical correlations than the mutual information \cite{2019arXiv190712555U,2019arXiv190706646K,2019arXiv190906790K}, and (ii) the negativity is bounded from below by the reflected entropy for holographic CFTs \cite{PhysRevLett.123.131603}.

\begin{figure}[t]
\centering
  \includegraphics[width=10.0cm,clip]{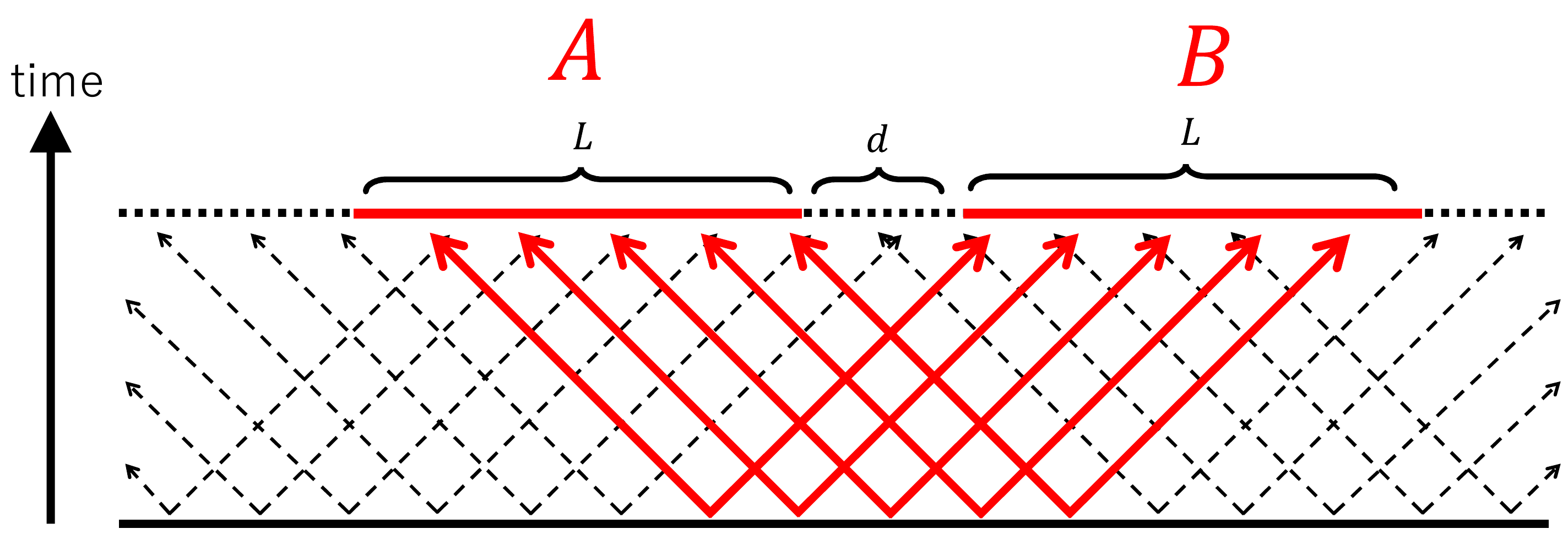}
 \caption{The setup we consider in the present paper. The arrows show the quasi-particle picture of the information spreading under the global quench.}
 \label{fig:setup}
\end{figure}

\begin{figure}[t]
\centering
  \includegraphics[width=8.0cm,clip]{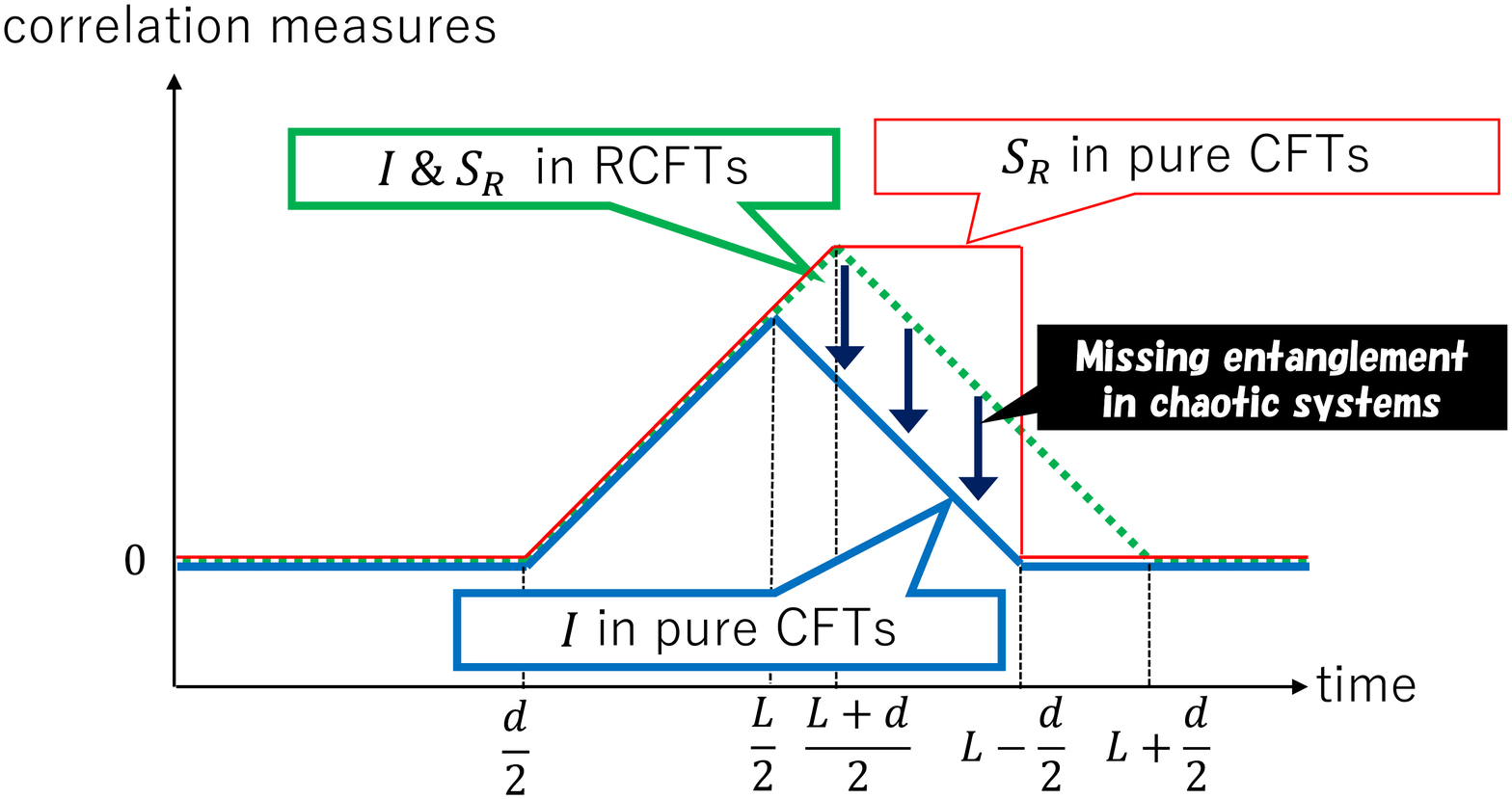}
  \includegraphics[width=8.0cm,clip]{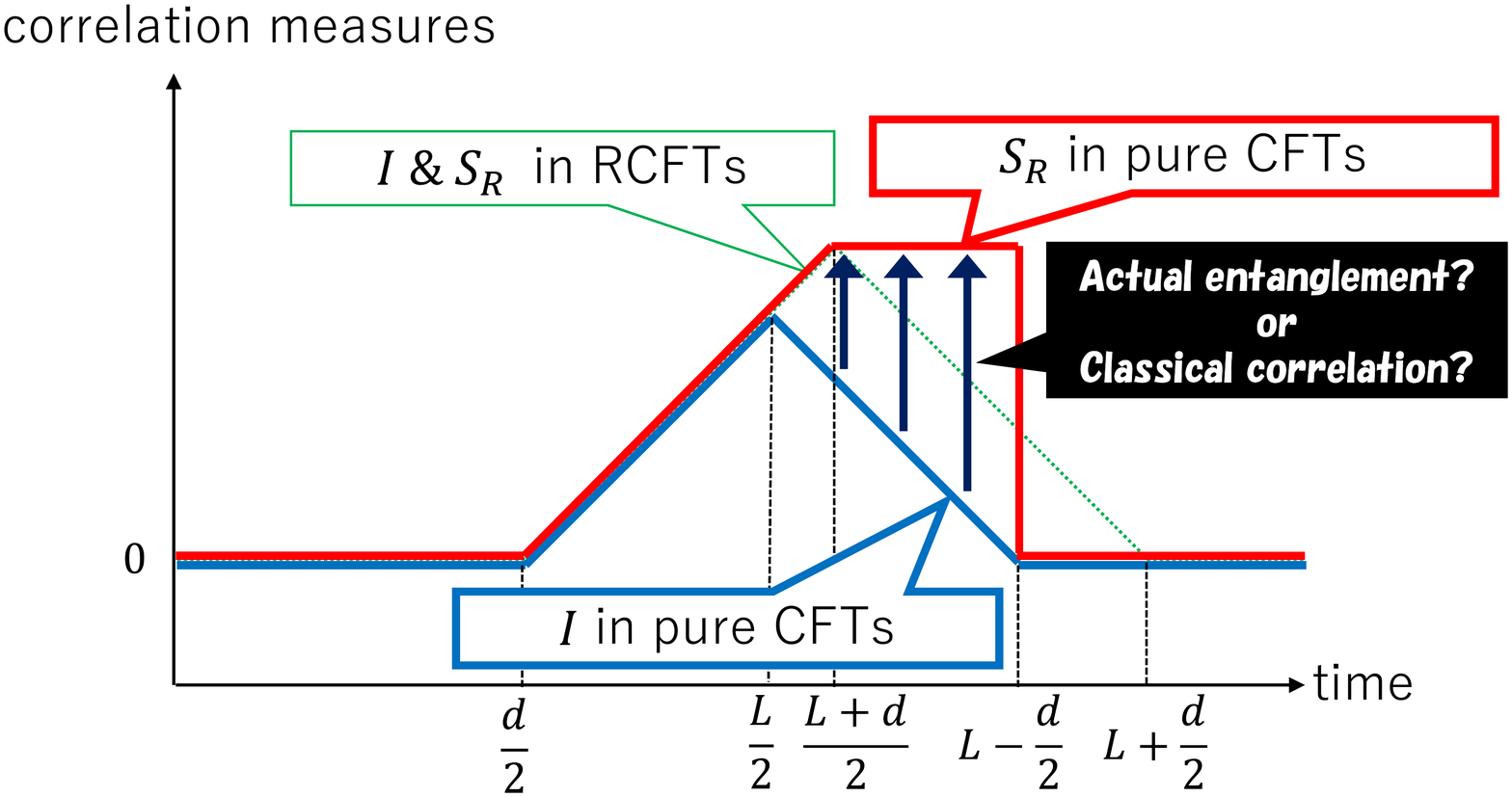}
 \caption{Left: The comparison of the mutual information in integrable systems (RCFTs) and chaotic systems (pure CFTs). The growth in RCFTs can be captured by the quasi-particle picture (see Figure \ref{fig:setup}). However, the growth in pure CFTs shows the breakdown of the quasi-particle picture. It seems that the entanglement is missing in chaotic systems. Right: The comparison between the mutual information ($I$) and the reflected entropy ($S_R$). We find a remarkable discrepancy between these two measures in chaotic systems.}
 \label{fig:result12}
\end{figure}

\item Holographic duals of quantum quenches (Section \ref{holo_sec})

We provide an independent derivation of mutual information and reflected entropy after global quantum quenches in irrational CFTs by studying the holographic dual to the quantum quench using AdS/BCFT. The area of the entanglement wedge cross-section precisely agrees with the results derived in Section \ref{light cone_sec} and further generalizes them beyond the light cone limit. 
We find that the mysterious correlation persists for any temperature. We are also able to generalize to inhomogeneous and local quenches by constructing their holographic dual geometries. These show interesting and distinct violations of the quasi-particle picture. We find that if the injected energy density is fixed, in all quench protocols, there exists a critical distance between disjoint intervals such that the shared correlations are trivial, pointing to maximal scrambling of information, a feature absent in rational CFTs. 

\item Random unitary circuits and the line-tension picture (Section \ref{randU_sec})

Random unitary circuits provide the minimal toy model of local and chaotic unitary dynamics. 
We use their solvability at large on-site Hilbert space dimension to interpret the results from irrational CFT. 
The coarse-grained dynamics of entanglement in chaotic systems is described by a phenomenological line-tension picture which complements the quasi-particle picture for integrable systems. We propose a line-tension picture for the reflected entropy. Using this formalism, we find remarkable agreement between irrational CFTs and random unitary circuits with large bond dimension. In fact, we find that the complete dynamics of mutual information and reflected entropy after global quantum quenches may be exactly predicted by the line-tension picture with local Hilbert space dimension determined by the Cardy entropy. This suggests that the dynamics of correlations in irrational CFTs are identical to local Haar random unitary circuits. Finally, we perform numerical simulations of random Clifford circuits. While we find the missing entanglement seen in irrational CFTs, we find that the plateau in the reflected entropy describing the mysterious correlation is missing. We interpret this as a consequence of the special nature of stabilizer states and small local Hilbert space dimension in Clifford circuits ($q=2$)  which suggests that the signatures of strong chaos in mixed state entanglement measures are large-$q$ phenomena, consistent with analysis from Ref.\ \cite{2019arXiv190709581W}.

\item
In Section \ref{discussion_sec}, we further discuss interpretations and future directions.

\end{itemize}

\subsection{Mixed state correlation measures}

Before closing this section,
we summarize
correlation measures of our interest in this paper.
\paragraph{Mutual information}
Mutual information is a canonical measure of correlations between disjoint subregions $A$ and $B$. It is defined as a linear combination of von Neumann entropies
\begin{align}
    I(A:B) \equiv S(A) + S(B) -S(A\cup B), \quad S(\Omega) \equiv - \Tr \rho_{\Omega} \log \rho_{\Omega}
\end{align}
where $\rho_{\Omega}$ is the reduced density matrix on $\Omega$. The mutual information sets a bound on the connected correlation functions between the two regions \cite{2008PhRvL.100g0502W}
\begin{align}
    I(A:B) \geq \frac{\left(\langle \mathcal{O}_A \mathcal{O}_B\rangle - \langle \mathcal{O}_A\rangle \langle \mathcal{O}_B\rangle \right)^2}{2\left| \mathcal{O}_A\right|^2\left| \mathcal{O}_B\right|^2}.
    \label{connected_correlator}
\end{align}
The mutual information is monotonic under inclusions
\begin{align}
    I(A:B \cup C) \geq I(A:B),
\end{align}
so it makes sense as a correlation measure. It is, however, important to note that it is not generally monotonic under local operations and classical communications (LOCC), so it is not a valid measure of quantum correlations. To compute the entanglement entropies, we use the replica trick by computing the partition function of an n-sheeted Riemann surface to find the R\'enyi entropies, then analytically continue to $n=1$
\begin{align}
    S(\Omega) = \lim_{n\rightarrow 1} \frac{1}{1-n} \log  \Tr \rho_{\Omega}^n.
\end{align}
Mutual information 
may be computed using the twist operator formalism \cite{DIXON198713,knizhnik1987,2001CMaPh.219..399L,2008JSP...130..129C}. 

\paragraph{Negativity}
%We progress to 
The logarithmic negativity which, unlike mutual information, is a suitable measure of quantum correlations for mixed states 
%because 
in that 
it is monotonic under LOCC \cite{2005PhRvL..95i0503P}. It is defined as
\begin{align}
    \mathcal{E}(A:B) = \log \left| \rho_{AB}^{T_B} \right|_1
\end{align}
where $\left| \cdot \right|_1 $ is the trace norm and $\cdot^{T_B}$ is the partial transpose operation. The moments of the partially transposed reduced density matrix may also be computed in conformal field theory using correlation functions of twist fields, though with a different ordering \cite{2012PhRvL.109m0502C}.

\paragraph{Odd entropy}
The odd entropy is defined in a similar manner to the negativity in that it involves the moments of the partial transpose. However, it is the odd moments that are needed \cite{PhysRevLett.122.141601}
\begin{align}
    S_o(A:B) = \lim_{n\rightarrow 1} \frac{1}{1-n} \log \left(\rho_{AB}^{T_B} \right)^n.
\end{align}
Little is known about the information theoretic properties of the odd entropy, though we believe the results in this paper are supportive evidence that it measures some sort of correlation, at least in conformal field theory. 
We should subtract off the von Neumann entropy to get the quantity which is captured by the entanglement wedge cross section in holographic theories \cite{PhysRevLett.122.141601}
\begin{align}
    \mathcal{E}_W(A:B) \equiv S_o(A:B) - S(A\cup B).
\end{align}

\paragraph{Reflected entropy}
The reflected entropy was introduced as the von Neumann entropy of a ``canonical purification" of the reduced density matrix \cite{2019arXiv190500577D}. The canonical purification maps the density operator to a state in a doubled Hilbert space $\mathcal{H}_A \otimes \mathcal{H}^*_{A^*} \otimes \mathcal{H}_B \otimes \mathcal{H}^*_{B^*}$ using the GNS construction
\begin{align}
    \rho_{AB} \rightarrow \ket{\rho_{AB}^{1/2}}.
\end{align}
The reflected entropy is then the von Neumann entropy of the state reduced to $A A^*$
\begin{align}
    S_R(A:B) = S_{vN}\left(\Tr_{B B^*} \ket{\rho_{AB}^{1/2}} \bra{\rho_{AB}^{1/2}} \right).
\end{align}
The reflected entropy satisfies multiple properties including a lower bound by the mutual information. Because of this bound, we do not expect the reflected entropy to be monotonic under LOCC. However, the integer R\'enyi reflected entropies can be shown to be monotonic under inclusions, so it is suspected that the reflected entropy is a sensible measure of total correlations \cite{2019arXiv190500577D}. 

A replica trick was formulated for the reflected entropy in Ref.~\cite{2019arXiv190500577D} that involves two replica indices, $n$ and $m$. $n$ represents the usual R\'enyi index while $m$ determines the resulting state $\ket{\rho_{AB}^{m/2}}$. For the canoncial purification, one must take the continuation of even $m$ to one. There are two correlation functions of twist fields needed, one for the entropy and one for the normalization of the purified state. In total, the reflected entropy is the following combination of $mn$-sheeted partition functions
\begin{equation}
S_R(A:B)= \lim_{n,m \to 1} \frac{1}{1-n} \log \frac{Z_{n,m}}{   \pa{Z_{1,m}}^n}.
\label{SR_partition_function_definition}
\end{equation}
See Ref.~\cite{2019arXiv190500577D} for more details.

\section{Universal contribution}
\label{univ_sec}

In this section, we consider the universal contributions to the correlation measures i.e.~the behavior in various operator product expansion (OPE) limits. 
In addition to these contributions present 
for any conformal field theories, 
the correlation measures receive contributions
from the conformal blocks of the twist operators,
which encode theory specific details,  
e.g.~, irrationality v.s.~rationality.

As we will argue, the universal contributions,
being determined only by the kinematics (conformal symmetry),
can be accounted for by the quasi-particle picture,
i.e., the free-propagation of EPR pairs which carry
a ``unit" of entanglement entropy
for the case of entanglement entropy, or
more generally a unit of entanglement content.

It was noted that the entanglement content for negativity can be found by relating it to the R\'enyi entropy for $n=1/2$ \cite{2018arXiv180909119A}. 
Then, by using the quasi-particle picture, 
Alba and Calabrese showed 
that the entanglement negativity and 
the R\'enyi mutual information 
are simply related to each other,
${\cal E} = I^{(1/2)}/2$
\cite{2018arXiv180909119A}.
Below, we will rederive this relation. 
In addition,
to date, quasi-particle pictures for the odd and reflected entropies have not been formulated. In this work, we provide evidence that they should exist. Their entanglement content may also be extracted from GGE thermodynamic entropy by considering appropriate pure state limits. We will have more to say on this issue in future work \cite{LN_op_quench_draft}. 
Similar to the Alba and Calabrese relation, 
we will find a simple relation
within the quasi-particle picture
among
the R\'enyi mutual information,
negativity, 
odd-entropy, and reflected entropy, 
\begin{align}
    {S_R^{(n)} = I^{(n)} = 2\mathcal{E}^{(n)}_W=\frac{2(n+1)}{3n}\mathcal{E}}.
\end{align}
This is the same proportionality seen in holographic theories when the contribution is universal, such as that for adjacent intervals \cite{2019PhRvD..99j6014K,PhysRevLett.122.141601,2019arXiv190500577D}.

\paragraph{Twist operator correlation function on the upper half plane}
We now proceed to the technical calculations.
The correlation measures after quantum quench 
can be computed by BCFT, where 
we treat
the gapped (short-range entangled) initial state
as a spacetime boundary.
By global conformal invariance, four-point correlation functions of primary fields on the upper half plane have the following structure
\begin{align}
    \langle \mathcal{O}(z_1) \mathcal{O}(z_2) \mathcal{O}(z_3) \mathcal{O}(z_4)\rangle_{UHP} = \frac{1}{\prod_{a=1}^4 \left|z_a-\bar{z}_a \right|^{\Delta_{\mathcal{O}}}}\frac{1}{\eta_{1,2}^{\Delta_{\mathcal{O}}}\eta_{3,4}^{\Delta_{\mathcal{O}}}}\left(\frac{\eta_{1,4}\eta_{2,3}}{\eta_{1,3}\eta_{2,4}} \right)^{\Delta_{\mathcal{O}_p}/2-\Delta_{\mathcal{O}}}\mathcal{G}\left( \left\{\eta_{j,k} \right\}\right),
\end{align}
where the conformally invariant cross-ratios are
\begin{align}
    \eta_{i,j} = \frac{(z_i - z_j)(\bar{z}_i-\bar{z}_j)}{(z_i-\bar{z}_j)(\bar{z}_i-z_j)}
\end{align}
and $\mathcal{G}$ is a theory-dependent function that is fixed by 
considering various OPE limits of the correlator (when the cross-ratios go to 0 and 1) such that all power-law behavior is encoded in the prefactor. We first consider the universal contribution to the four-point function i.e.~we drop $\mathcal{G}$. This is justified in the above limits of the cross-ratio and generally corresponds to the quasi-particle picture \cite{2014JSMTE..12..017C,2015PhRvB..92g5109W,2019arXiv190607639K}. We now apply this to the various correlation measures of interest.

\paragraph{Mutual information}

Mutual information may be computed using the twist operator formalism \cite{DIXON198713,knizhnik1987,2001CMaPh.219..399L,2008JSP...130..129C}. First, we compute two-point functions of twist fields on the upper half plane to find the individual entropies for $A$ and $B$. Dropping the theory dependent function $\mathcal{G}$,
\begin{align}
    \Tr \rho_A^n = \langle \sigma_n(z_1) \bar{\sigma}_n(z_2) \rangle_{UHP} =
    \left( \frac{1}{\eta_{1,2}|z_1 -\bar{z}_1||z_2 -\bar{z}_2|}\right)^{ \Delta_n}
\end{align}
where $\Delta_n$ is the conformal weight of the twist-field
\begin{align}
    \Delta_n = \frac{c}{12}\left(n - \frac{1}{n} \right) (\equiv 2h_n) .
\end{align}
An analogous expression holds for $\Tr \rho_B^n$. The R\'enyi entropies are
\begin{align}
    S^{(n)}(A) &= \frac{c(n+1)}{12n} \log \left( {\eta_{1,2}|z_1 -\bar{z}_1||z_2 -\bar{z}_2|}\right)
    \nonumber
    \\
    S^{(n)}(B) &= \frac{c(n+1)}{12n} \log \left( {\eta_{3,4}|z_3 -\bar{z}_3||z_4 -\bar{z}_4|}\right)
    \label{indiv_renyis}
\end{align}
To find the R\'enyi mutual information, we need the four-point function
\begin{align}
    \Tr \rho_{A\cup B}^n = \langle \sigma_{n}(z_1) \bar{\sigma}_n(z_2) {\sigma}_n(z_3) \bar{\sigma}_n(z_4)\rangle_{UHP} = \frac{1}{\prod_{a=1}^4 \left|z_a-\bar{z}_a \right|^{\Delta_n}}\frac{1}{\eta_{1,2}^{\Delta_n}\eta_{3,4}^{\Delta_n}}\left(\frac{\eta_{1,4}\eta_{2,3}}{\eta_{1,3}\eta_{2,4}} \right)^{-\Delta_n},
\end{align}
where we have used the fact that the leading operator in the $\sigma \times \bar{\sigma}$ OPE is the identity. 
The R\'enyi entropy is
\begin{align}
    S^{(n)}(A\cup B) &= \frac{1}{1-n} \log \frac{1}{\prod_{a=1}^4 \left|z_a-\bar{z}_a \right|^{\Delta_n}}\frac{1}{\eta_{1,2}^{\Delta_n}\eta_{3,4}^{\Delta_n}}\left(\frac{\eta_{1,4}\eta_{2,3}}{\eta_{1,3}\eta_{2,4}} \right)^{-\Delta_n}
    \nonumber 
    \\
    &= -\frac{c(1+n)}{12n}\log {\prod_{a=1}^4 \frac{1}{\left|z_a-\bar{z}_a \right|}}\frac{1}{\eta_{1,2}\eta_{3,4}}\left(\frac{\eta_{1,4}\eta_{2,3}}{\eta_{1,3}\eta_{2,4}} \right)^{-1} 
    % + \frac{1}{1-n} \log \tilde{c}
    \label{renyi_univ}
\end{align}
We compute the R\'enyi mutual informations by subtracting (\ref{renyi_univ}) from the individual R\'enyi entropies \eqref{indiv_renyis} to find
\begin{align}
    I^{(n)} = -\frac{c(n+1)}{12n } \log \left(\frac{\eta_{1,4}\eta_{2,3}}{\eta_{1,3}\eta_{2,4}} \right).
\end{align}
The replica limit is given by
\begin{align}
    I = -\frac{c}{6}\log \left(\frac{\eta_{1,4}\eta_{2,3}}{\eta_{1,3}\eta_{2,4}} \right).
\end{align}

\paragraph{Negativity}
The universal contribution to the correlator for negativity is
\begin{align}
    \Tr \left(\rho_{AB}^{T_B} \right)^{n_e} &= \langle \sigma_{n}(z_1) \bar{\sigma}_{n_e}(z_2) \bar{\sigma}_{n_e}(z_3) \sigma_n(z_4)\rangle_{UHP} \nonumber \\&= \frac{1}{\prod_{a=1}^4 \left|z_a-\bar{z}_a \right|^{\Delta_{n_e}}}\frac{1}{\eta_{1,2}^{\Delta_{n_e}}\eta_{3,4}^{\Delta_{n_e}}}\left(\frac{\eta_{1,4}\eta_{2,3}}{\eta_{1,3}\eta_{2,4}} \right)^{\Delta^{(2)}_{n_e}/2-\Delta_{n_e}}
\end{align}
where $n_e$ means that we only consider even integers. The double-twist operator is the leading primary in the $\sigma \times \sigma$ OPE and has conformal weight
\begin{align}
     \Delta_{n_e}^{(2)} = \frac{c}{6}\left(\frac{n_e}{2} - \frac{2}{n_e} \right).
\end{align}
We take the replica limit to find the negativity
\begin{align}
    \mathcal{E}(A:B) = \lim_{n_e \rightarrow 1} \Tr \left(\rho_{AB}^{T_B} \right)^{n_e} = -\frac{c}{8} \log  \left(\frac{\eta_{1,4}\eta_{2,3}}{\eta_{1,3}\eta_{2,4}} \right)
\end{align}
Note that this is consistent with the statement from Ref.~\cite{2018arXiv180909119A} that in integrable theories in the scaling limit
\begin{align}
    \Delta \mathcal{E}(A:B) = \frac{ \Delta I^{(1/2)}(A:B)}{2}
\end{align}
where $\Delta$ means the change from the ground state value\footnote{We note that ``$\Delta$" was not explicitly written in Ref.~\cite{2018arXiv180909119A} though it was implied because they were discussing only the excitations above the ground state.}.

\paragraph{Odd entropy}
For odd replica index, the double twist field has the same dimension as the single twist field, so 
\begin{align}
    \left(\rho_{AB}^{T_B} \right)^n = \langle \sigma_{n}(z_1) \bar{\sigma}_n(z_2) \bar{\sigma}_n(z_3) \sigma_n(z_4)\rangle_{UHP} = \frac{1}{\prod_{a=1}^4 \left|z_a-\bar{z}_a \right|^{\Delta_n}}\frac{1}{\eta_{1,2}^{\Delta_n}\eta_{3,4}^{\Delta_n}}\left(\frac{\eta_{1,4}\eta_{2,3}}{\eta_{1,3}\eta_{2,4}} \right)^{-\Delta_n/2}.
\end{align}
The universal contribution to the R\'enyi odd entropy is
\begin{align}
    S_o^{(n)}(A:B) &= \frac{1}{1-n} \log \frac{1}{\prod_{a=1}^4 \left|z_a-\bar{z}_a \right|^{\Delta_n}}\frac{1}{\eta_{1,2}^{\Delta_n}\eta_{3,4}^{\Delta_n}}\left(\frac{\eta_{1,4}\eta_{2,3}}{\eta_{1,3}\eta_{2,4}} \right)^{-\Delta_n/2}
    \nonumber 
    \\
    &= \frac{c(n-\frac{1}{n})}{12(1-n)}\log {\prod_{a=1}^4 \frac{1}{\left|z_a-\bar{z}_a \right|}}\frac{1}{\eta_{1,2}\eta_{3,4}}\left(\frac{\eta_{1,4}\eta_{2,3}}{\eta_{1,3}\eta_{2,4}} \right)^{-1/2} 
\end{align}
In the replica limit, this is
\begin{align}
    S_o(A:B) &= -\frac{c}{6}\log\left[ {\prod_{a=1}^4 \frac{1}{\left|z_a-\bar{z}_a \right|}}\frac{1}{\eta_{1,2}\eta_{3,4}}\left(\frac{\eta_{1,4}\eta_{2,3}}{\eta_{1,3}\eta_{2,4}} \right)^{-1/2}\right] 
    % + \lim_{n\rightarrow 1} \frac{1}{1-n}\log \tilde{c}.
\end{align}
Subtracting off the von Neumann entropy,
$\mathcal{E}_W(A:B) \equiv S_o(A:B) - S(A\cup B)$,
we find 
\begin{align}
    \mathcal{E}_W(A:B) = -\frac{c}{12}  \log \left(\frac{\eta_{1,4}\eta_{2,3}}{\eta_{1,3}\eta_{2,4}} \right) ,
    % = \frac{2}{3} \mathcal{E}.
\end{align}
with the R\'enyi version
\begin{align}
    \mathcal{E}_W^{(n)}(A:B) = -\frac{c(n+1)}{24n}  \log \left(\frac{\eta_{1,4}\eta_{2,3}}{\eta_{1,3}\eta_{2,4}} \right) .
\end{align}

\paragraph{Reflected entropy}

For disjoint intervals $A$ and $B$, the universal contribution to the numerator 
of \eqref{SR_partition_function_definition}
is computed by
\begin{align}
    Z_{n,m} = \langle \sigma_{g_A}(z_1) {\sigma}_{g_A^{-1}}(z_2) {\sigma}_{g_B}(z_3) \sigma_{g_B^{-1}}(z_4)\rangle_{UHP} = \frac{1}{\prod_{a=1}^4 \left|z_a-\bar{z}_a \right|^{n\Delta_{m}}}\frac{1}{\eta_{1,2}^{n\Delta_{m}}\eta_{3,4}^{n\Delta_{m}}}\left(\frac{\eta_{1,4}\eta_{2,3}}{\eta_{1,3}\eta_{2,4}} \right)^{\Delta_n-n\Delta_{m}}
    \label{Znm_correlator}
\end{align}
where the $\sigma$'s are generalized twist fields with conformal dimensions
\begin{align}
    h_{g^{\ }_B} = h_{g^{-1}_B}= h_{g_A^{\ }} = h_{g_A^{-1}} = \frac{cn(m^2-1)}{24m}.
\end{align}
The leading operator in the $\sigma_{g_A^{-1} }\times \sigma_{g_B}$ OPE, $\sigma_{g_A^{-1} g_B}$, has conformal dimension
\begin{align}
    h_{g^{\ }_B g_A^{-1}} = \frac{2c(n^2-1)}{24n}.
\end{align}
The normalization is
\begin{align}
    \pa{Z_{1,m}}^n = \langle \sigma_{m}(z_1) \bar{\sigma}_m(z_2) {\sigma}_m(z_3) \bar{\sigma}_m(z_4)\rangle_{UHP}^{n} = \frac{1}{\prod_{a=1}^4 \left|z_a-\bar{z}_a \right|^{n\Delta_m}}\frac{1}{\eta_{1,2}^{n\Delta_m}\eta_{3,4}^{n\Delta_m}}\left(\frac{\eta_{1,4}\eta_{2,3}}{\eta_{1,3}\eta_{2,4}} \right)^{-n\Delta_m}.
\end{align}
Thus, the R\'enyi reflected entropy is
\begin{align}
    S_R^{(n)} &= \frac{1}{1-n}\log \left(\frac{\eta_{1,4}\eta_{2,3}}{\eta_{1,3}\eta_{2,4}} \right)^{\Delta_n}
    = -\frac{c(n+1)}{12n}\log \left(\frac{\eta_{1,4}\eta_{2,3}}{\eta_{1,3}\eta_{2,4}} \right).
\end{align}
Interestingly, the universal term does not depend on the purification labeled by $m$, rather all of this dependence arises from the non-universal term $\mathcal{G}$. $m$ is not meaningful in the context of purifications because $\ket{\rho^{m/2}}$ is only a purification of $\rho$ when $m=1$. However, $m \neq 1$ has been given meaning in the context of operator entanglement in Ref.~\cite{2019arXiv190709581W}. The replica limit gives
\begin{align}
    S_R = -\frac{c}{6}\log \left(\frac{\eta_{1,4}\eta_{2,3}}{\eta_{1,3}\eta_{2,4}} \right).
\end{align}

To recap, we find that all correlation measures are proportional 
\begin{align}
    {S_R^{(n)} = I^{(n)} = 2\mathcal{E}^{(n)}_W=\frac{2(n+1)}{3n}\mathcal{E}}.
\end{align}
This is the same proportionality seen in holographic theories when the contribution is universal, such as that for adjacent intervals \cite{2019PhRvD..99j6014K,PhysRevLett.122.141601,2019arXiv190500577D}. With this motivation, we limit our studies to
\begin{align}
    E_W \equiv -\frac{c}{12}\log \left(\frac{\eta_{1,4}\eta_{2,3}}{\eta_{1,3}\eta_{2,4}} \right)
    \label{EW_quasi}
\end{align}
which may then be related to each quantity by an overall factor. 

\subsection{Global homogeneous quench}
\begin{figure}
    \centering
    \includegraphics[width = \textwidth]{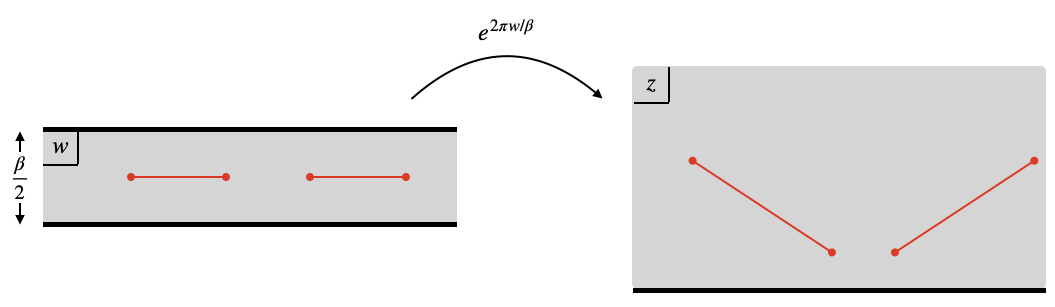}
    \caption{The conformal map from the infinite strip to the upper half plane used in the evaluation of the global quench. The red dots represent the twist operator insertions and the thick black lines represent boundaries of the manifold. }
    \label{glob_quench_map_fig}
\end{figure}

Now that we have understood the general form of the correlation measures on the upper half plane, we apply our results to specific dynamical situations. We begin with a global quantum quench. The initial state is prepared by smearing a boundary state in the imaginary time direction
\begin{align}
    \ket{\Psi_0} = e^{-\beta H /4} \ket{B}
    \label{psi0_glob}
\end{align}
where $\ket{B}$ is the boundary state. This state has only short-ranged (order $\beta$) entanglement. We then evolve in Lorentzian time with the gapless Hamiltonian to quench
\begin{align}
    \ket{\Psi(t)} = e^{-iHt}\ket{\Psi_0}.
\end{align}
This has the effect of an instantaneous homogeneous injection of energy with effective temperature $\beta$. 

To compute correlation measures, we insert twist fields into the strip Euclidean path integral that prepared \eqref{psi0_glob}. We can conformally map the strip to the upper half plane with (Fig.~\ref{glob_quench_map_fig})
\begin{align}
    z = e^{2 \pi w /\beta}.
    \label{glob_quench_map}
\end{align}

\begin{figure}
    \centering
    \includegraphics[width = 7cm]{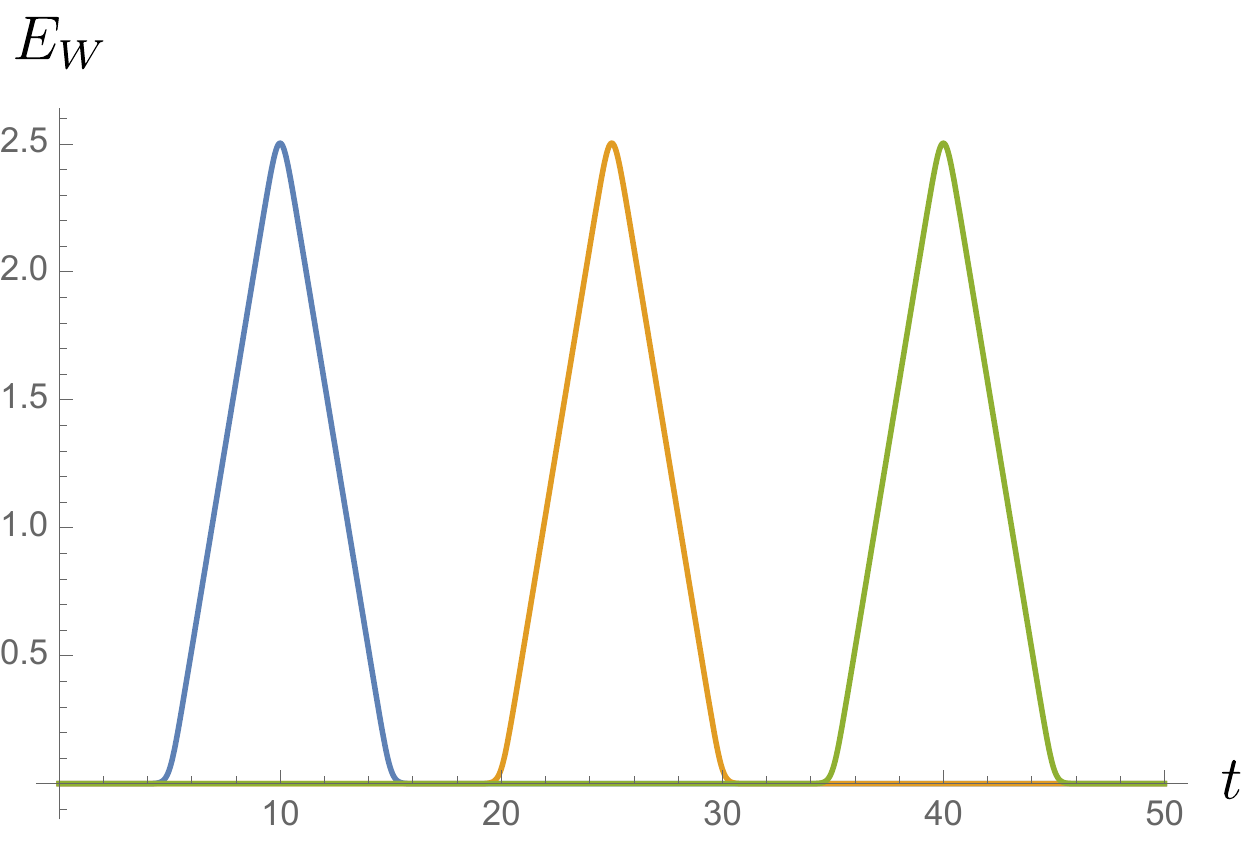}
    \includegraphics[width = 7cm]{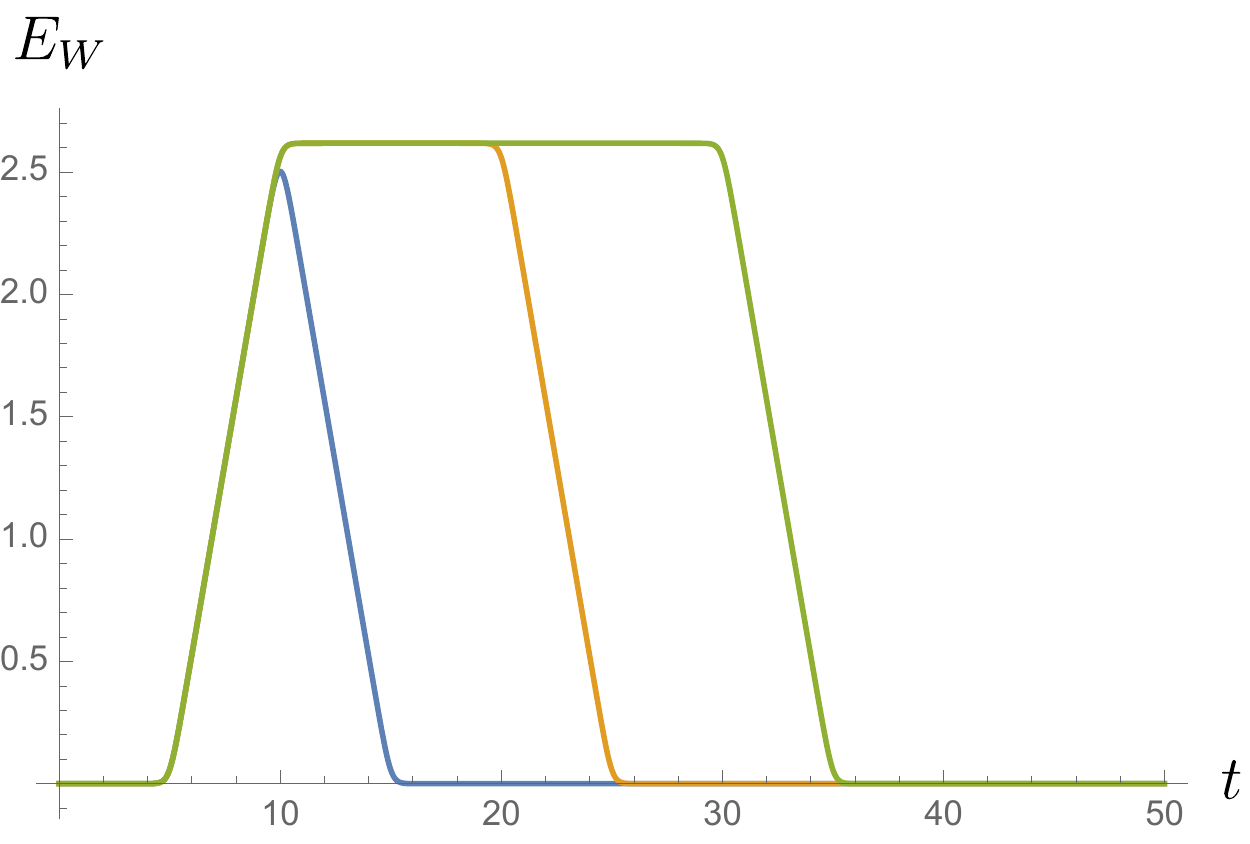}
    \caption{Global homogeneous quench: $\beta = 2$, $c = 1$ (left) $l_1 = l_2 = 10$, $d = \{10,25,40\}$ (right) $l_1 = 10$, $l_2 = \{10,30,50\}$, $d = 10$. The maximum value of $E_W$ reached is dictated by half of the thermal entropy of the smaller interval. This is because right-moving quasi-particles in the right interval are maximally entangled with the left-moving quasi-particles in the left interval. However, the left (right)-moving particles in the right (left) interval are entangled with the environment. At late times, $E_W$ is exponentially small, given by the thermal value \eqref{EW_univ_glob_thermal}.}
    \label{global_univ}
\end{figure}

In order to evaluate $E_W$ using \eqref{EW_quasi}, we only need to find the cross-ratios analytically continued to Lorentzian signature
\begin{align}
    \eta_{i,j} = \frac{\sinh^2\left( \frac{\pi}{\beta} (x_i - x_j)\right)}{\cosh\left(\frac{\pi}{\beta}(x_i -x_j -2t)\right)\cosh\left(\frac{\pi}{\beta}(x_i -x_j +2t)\right)}.
\end{align}
For disjoint intervals $A = (x_1, x_1+l_1)$, $B = (x_1 + l_1 + d,x_1 + l_1+l_2 + d )$, this leads to
\begin{align}
    E_W &= -\frac{c}{6}\log \left(\frac{{\sinh\left( \frac{\pi}{\beta} (l_1+l_2+d)\right)}
    {\sinh\left( \frac{\pi d}{\beta} \right)}}{{\sinh\left( \frac{\pi}{\beta} (l_1+d)\right)}{\sinh\left( \frac{\pi}{\beta} (l_2 + d)\right)}}\right)
    \nonumber
    \\
    &-\frac{c}{12}
    \left(\frac{
    \cosh\left(\frac{\pi}{\beta}(l_1+d -2t)\right)\cosh\left(\frac{\pi}{\beta}(l_1+d +2t)\right)\cosh\left(\frac{\pi}{\beta}(l_2 + d -2t)\right)\cosh\left(\frac{\pi}{\beta}(l_2 + d +2t)\right)}
    {{\cosh\left(\frac{\pi}{\beta}(d -2t)\right)\cosh\left(\frac{\pi}{\beta}(d +2t)\right)}{\cosh\left(\frac{\pi}{\beta}(l_1+l_2+d -2t)\right)\cosh\left(\frac{\pi}{\beta}(l_1+l_2+d +2t)\right)}{}}\right).
\end{align}
Given that the quench is translationally invariant, the expression is naturally independent of $x_1$. In the high temperature limit, $l_1, l_2,d \gg \beta$, 
\begin{align}
    E_W 
    &= \frac{c \pi}{12 \beta }\left(| d -2t| + |l_1+l_2+d -2t| - |l_1 + d -2t|-  |l_2 + d -2t|\right).
\end{align}
This perfectly describes the quasi-particle picture. Representative cases are plotted in Fig.~\ref{global_univ}.

We find the late-time value of $E_W$ to be equal to a universal thermal value at inverse temperature $\beta$
\begin{align}
       E_W \xrightarrow[t\rightarrow \infty]{} \frac{c}{6} \log \left(\frac{\sinh\left(\frac{\pi(d+l_1)}{\beta}\right)\sinh\left(\frac{\pi(d+l_2)}{\beta}\right)}{\sinh\left(\frac{\pi d}{\beta}\right)\sinh\left(\frac{\pi(d+l_1+l_2)}{\beta}\right)} \right).
       \label{EW_univ_glob_thermal}
\end{align}
This means that the subsystems have thermalized in the sense that, from the perspective of $E_W$, the subsystems are indistinguishable from a global thermal state at inverse temperature $\beta$, even though the global state is pure. The rest of the system is acting as a thermal bath for the subsystems.

\subsection{Global inhomogeneous quench}

\begin{figure}
    \centering
    \includegraphics[width = \textwidth]{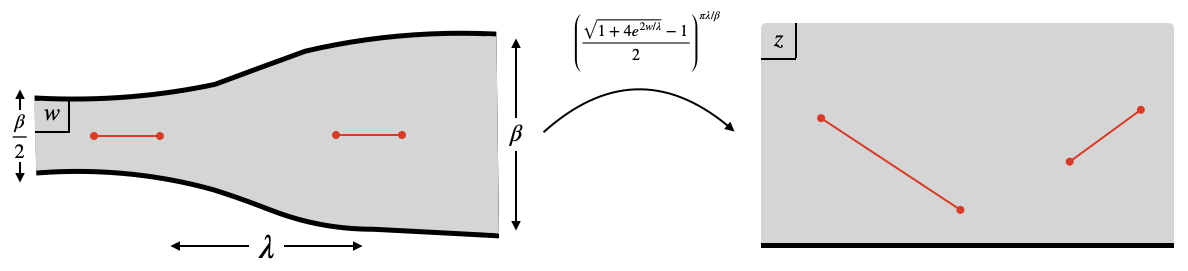}
    \caption{The inhomogeneous strip is shown on the left which smoothly interpolates between width $\beta/2$ and $\beta$ with interpolation region of length $\lambda$. The red dots represent operator insertions and the thick black lines represent conformal boundaries. In our setup, we are generally concerned with the configuration in which the intervals lie on either side of the interpolation region such that they initially observe different effective temperatures.}
    \label{glob_inhom_quench_map_fig}
\end{figure}
\begin{figure}
    \centering
    \includegraphics[width = 7cm]{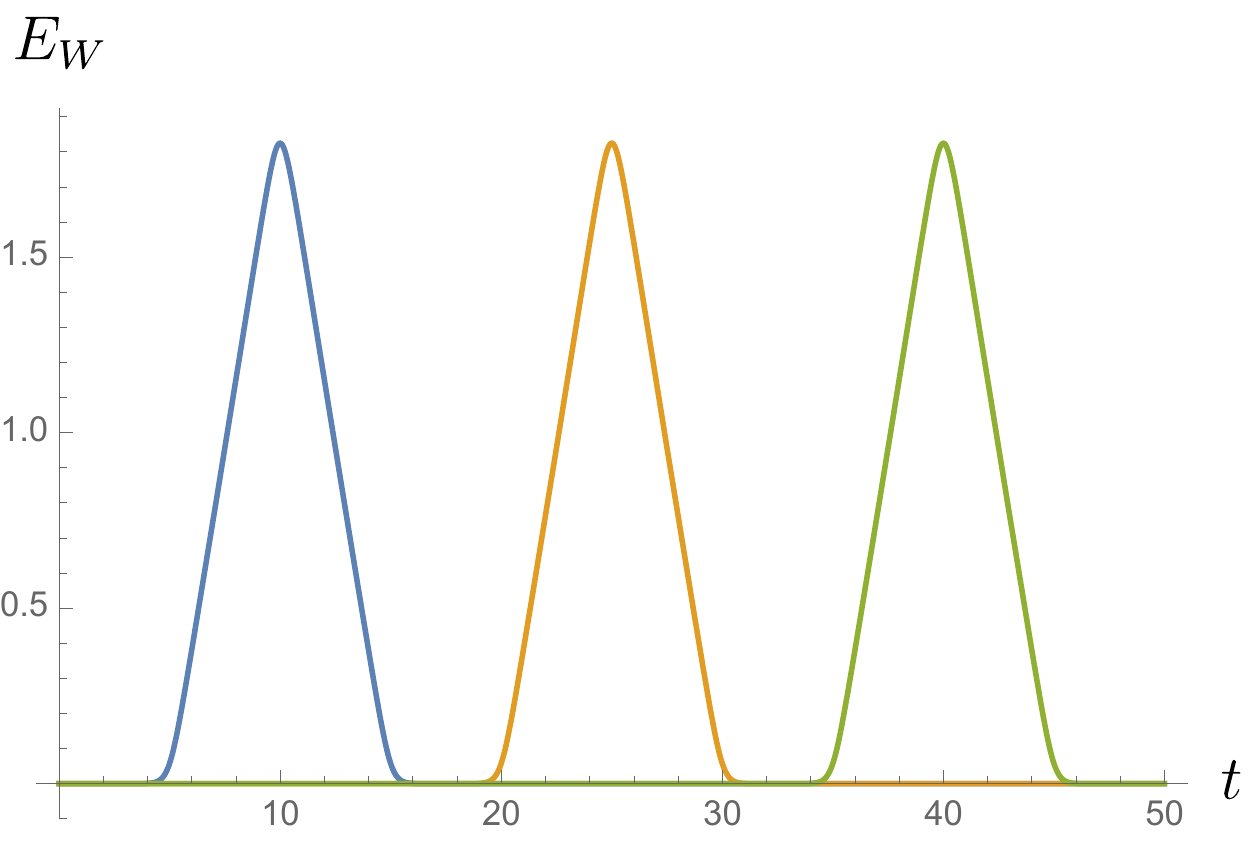}
    \includegraphics[width = 7cm]{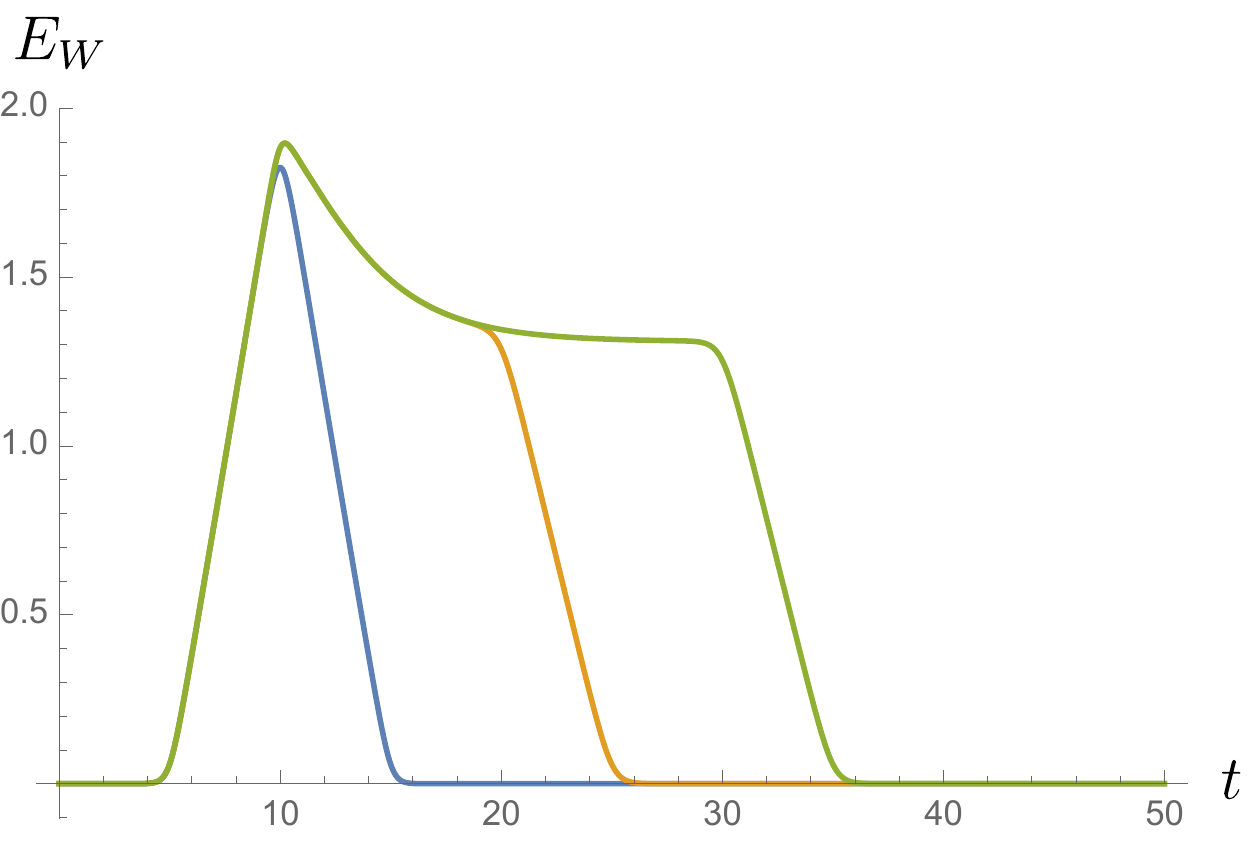}
    \caption{Global inhomogeneous quench: $\beta = 2, c = 1, \lambda = 3$ (left) $l_1 = l_2 = 10, d = \{10,25,40\}$ (right) $l_1 = 10, l_2 = \{10,30,50\}, d = 10$. The late-time value is, again, exponentially small, given by the thermal value \eqref{thermal_EW_inhom_univ}.}
    \label{inhom_univ}
\end{figure}

Of course, we would like to probe more complex dynamics than the homogeneous global quench. A tractable way to proceed is by preparing the initial state using an infinite strip Euclidean path integral that has spatially-dependent width. The smaller the width, the more energy is injected once the quench to the gapless Hamiltonian ensues. We take the strip to be asymptotically width $\beta/2$ on one side and $\beta$ on the other with a characteristic distance between these regimes $\lambda$. This is shown in Fig.~\ref{glob_inhom_quench_map_fig}. The conformal map that takes this inhomogeneous strip to the upper half plane is
\begin{align}
    z = \left(\frac{\sqrt{1 + 4e^{2 w /\lambda}}-1}{2} \right)^{\pi \lambda/\beta}.
\end{align}
Following analytic continuation, the cross-ratios for this quench are
\begin{align}
    \eta_{ij} &= \frac{\left(\frac{\sqrt{1 + 4e^{2 (x_i+t) /\lambda}}-1}{2} \right)^{\pi \lambda/\beta} -\left(\frac{\sqrt{1 + 4e^{2 (x_j+t) /\lambda}}-1}{2} \right)^{\pi \lambda/\beta} }{\left(\frac{\sqrt{1 + 4e^{2 (x_i-t) /\lambda}}-1}{2} \right)^{\pi \lambda/\beta} +\left(\frac{\sqrt{1 + 4e^{2 (x_j+t) /\lambda}}-1}{2} \right)^{\pi \lambda/\beta} }
    \nonumber
    \\
    &\qquad\qquad\qquad\qquad\times \frac{\left(\frac{\sqrt{1 + 4e^{2 (x_i-t) /\lambda}}-1}{2} \right)^{\pi \lambda/\beta} -\left(\frac{\sqrt{1 + 4e^{2 (x_j-t) /\lambda}}-1}{2} \right)^{\pi \lambda/\beta} }{\left(\frac{\sqrt{1 + 4e^{2 (x_i+t) /\lambda}}-1}{2} \right)^{\pi \lambda/\beta} +\left(\frac{\sqrt{1 + 4e^{2 (x_j-t) /\lambda}}-1}{2} \right)^{\pi \lambda/\beta} }
\end{align}
This cross-ratio is quite complicated, so we plot illuminating cases in Fig.~\ref{inhom_univ} where we find behavior similar to, though distinct from, the homogeneous quench. This may be interpreted through the quasi-particle picture with the quasi-particles injected on the left side having twice the entanglement content as those from the right due to how we prepared the initial state.
At late times, we find $E_W$ to thermalize
\begin{align}
    E_W \xrightarrow[t\rightarrow \infty]{} \frac{c}{12} \log \left(\frac{\sinh\left(\frac{\pi(d+l_1)}{\beta}\right)\sinh\left(\frac{\pi(d+l_2)}{\beta}\right)}{\sinh\left(\frac{\pi d}{\beta}\right)\sinh\left(\frac{\pi(d+l_1+l_2)}{\beta}\right)} \right) + \frac{c}{12} \log \left(\frac{\sinh\left(\frac{\pi(d+l_1)}{2\beta}\right)\sinh\left(\frac{\pi(d+l_2)}{2\beta}\right)}{\sinh\left(\frac{\pi d}{2\beta}\right)\sinh\left(\frac{\pi(d+l_1+l_2)}{2\beta}\right)} \right).
    \label{thermal_EW_inhom_univ}
\end{align}
Because this quenches at two different effective temperatures $\beta$ and $2\beta$, the system equilibrates somewhere inbetween. Interestingly, $E_W$ is an equal sum of the values at each temperature rather than the value of $E_W$ at $3\beta/2$ that was discussed in Ref.~\cite{2013arXiv1311.2562U}. However, in the limit that $\beta\rightarrow 0$, these are equivalent. This result makes sense from the quasi-particle perspective because the entanglement content of quasi-particles from the left side is twice that of quasi-particles generated from the right side. See Ref.~\cite{2008JSMTE..11..003S} for a thorough explanation of the quasi-particle picture for inhomogeneous quenches in the context of entanglement entropy.

\subsection{Local joining quench}
\begin{figure}
    \centering
    \includegraphics[width = \textwidth]{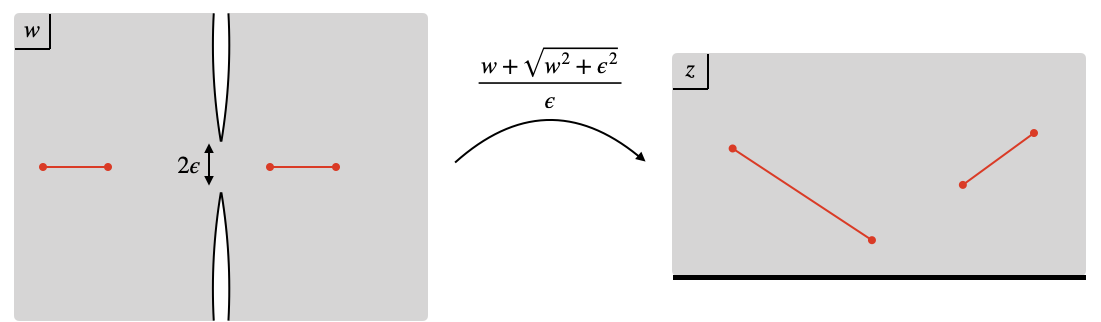}
    \caption{The path integral to prepare the initial state of the local joining quench is shown on the left. We exaggerate the size of the cuts on the imaginary axis. The boundaries on these cuts get mapped to the real axis after applying \eqref{loc_quench_map}.}
    \label{loc_quench_map_fig}
\end{figure}

\begin{figure}
    \centering
    \includegraphics[width = 7cm]{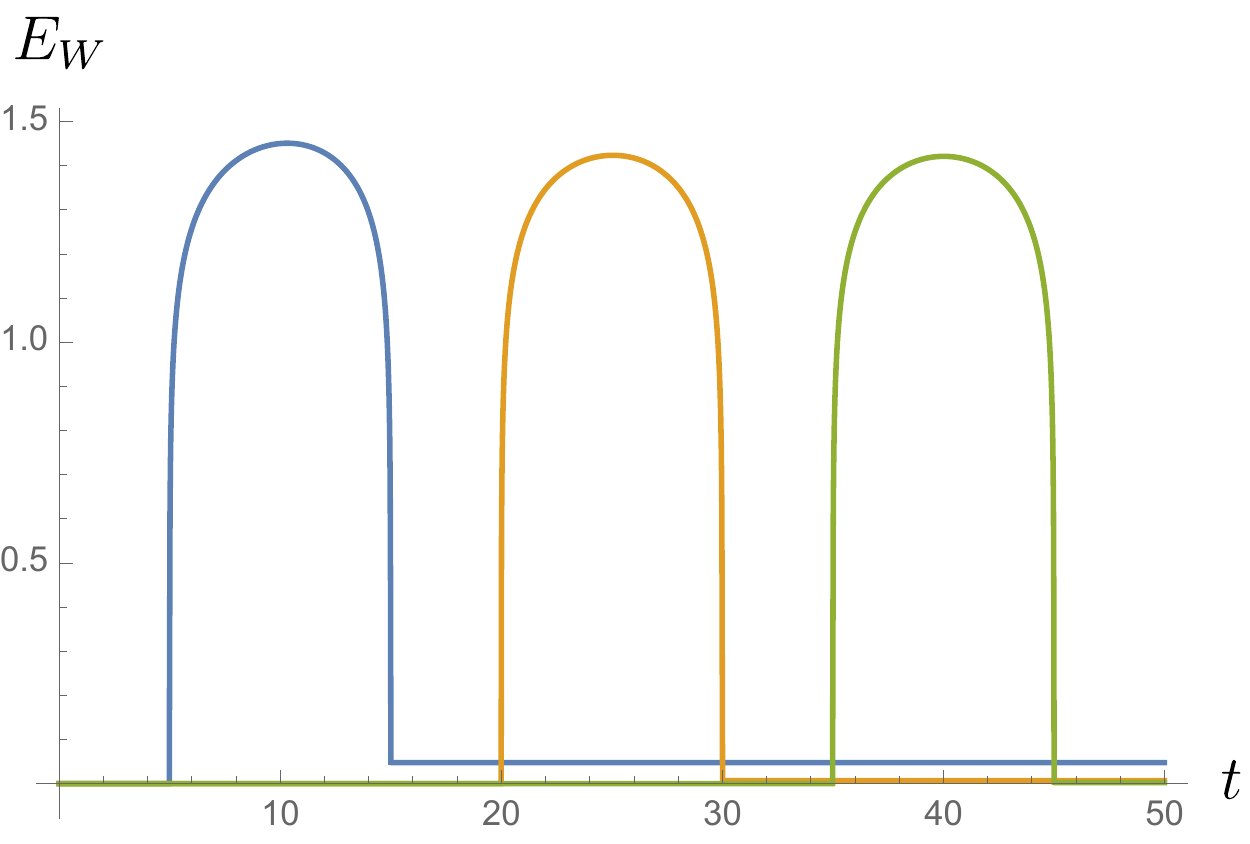}
    \includegraphics[width = 7cm]{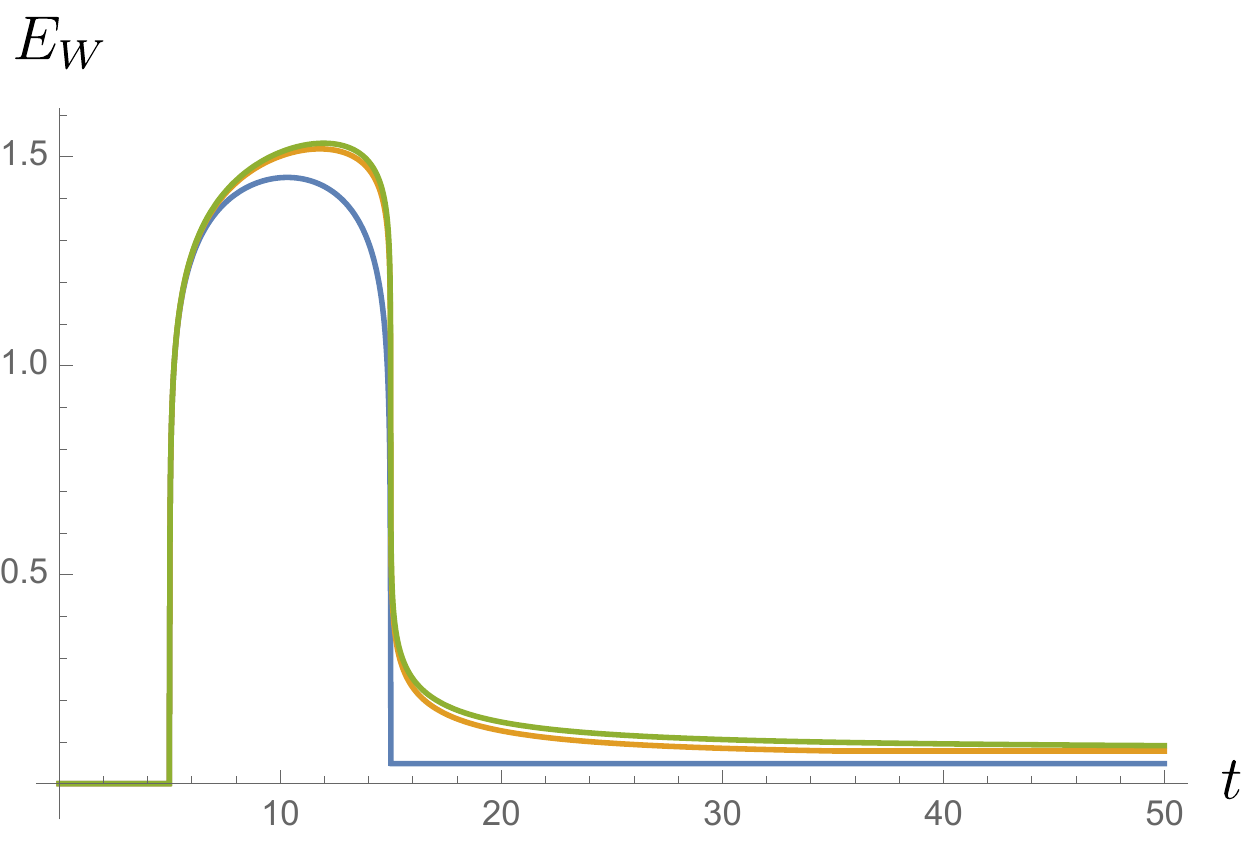}
    \caption{Local joining quench: $\epsilon = 10^{-3}$, $c = 1$ (left) $l_1 = l_2 = 10,$ $d = \{10,25,40\}$ (right) $l_1 = 10$, $l_2 = \{10,30,50\}$, $d = 10$.}
    \label{univ_loc_joining}
\end{figure}

An alternative way to probe inhomogeneous dynamics is from a local quantum quench. We are able to model the coupling of two boundary CFTs defined on half lines by preparing the initial state as the Euclidean complex plane with semi-infinite cuts along the positive and negative imaginary axes (Fig.~\ref{loc_quench_map_fig}). This effectively models an injection of energy at the origin. It is an extreme limit of the inhomogeneity from the previous section because the ``strip" width is infinite everywhere except at the origin where it is $2\epsilon$. The conformal map from this Riemann surface to the upper half plane is 
\begin{align}
    z = \frac{w + \sqrt{w^2+\epsilon^2}}{\epsilon}.
    \label{loc_quench_map}
\end{align}
After analytic continuation, the cross-ratios are
\begin{align}
    \eta_{ij} = \frac{\left( x_i - x_j + \sqrt{\left(x_i + t\right)^2+\epsilon^2 }- \sqrt{\left(x_j + t\right)^2+\epsilon^2 }\right)\left( x_i - x_j + \sqrt{\left(x_i - t\right)^2+\epsilon^2 }-  \sqrt{\left(x_j - t\right)^2+\epsilon^2 }\right)}{\left( x_i + x_j + \sqrt{\left(x_i + t\right)^2+\epsilon^2 }- \sqrt{\left(x_j - t\right)^2+\epsilon^2 }\right)\left( x_i + x_j + \sqrt{\left(x_i - t\right)^2+\epsilon^2 }-  \sqrt{\left(x_j + t\right)^2+\epsilon^2 }\right)}.
\end{align}
This leads to a quasi-particle-like picture with an entangled pair propagating from the origin (see Fig.~\ref{univ_loc_joining}). This quasi-particle picture is not quite as sharp as those discussed above for global quenches or for local quenches created by primary operator insertions \cite{2014PhRvL.112k1602N}. At late times, after the quasi-particle has moved through the intervals, $E_W$ relaxes to its universal ground state value
\begin{align}
    E_W \xrightarrow[t\rightarrow \infty]{} \frac{c}{6} \log \left(\frac{(l_1 + d)(l_2+d)}{d(l_1 + l_2 + d)} \right).
\end{align}
The relaxation to the ground state value, as opposed to a thermal value, is expected because the quasi-particle coherently moves through the interval in CFT. The energy injection density across the total system is trivial, in contrast to the global quenches.

%%%%%%%%%%%%%%%%%%%%%%%%%%%%%%%%%%%%%%%%%%%%%%%%%%%%%%%%%%%%%%%%%%%%%%%%%%%%%%%%%%%%%%%%%%%%%%
%%%%%%%%%%%%%%%%%%%%%%%%%%%%%%%%%%%%%%%%%%%%%%%%%%%%%%%%%%%%%%%%%%%%%%%%%%%%%%%%%%%%%%%%%%%%%%
\section{Global quench from the light cone limit}\label{light cone_sec}
%%%%%%%%%%%%%%%%%%%%%%%%%%%%%%%%%%%%%%%%%%%%%%%%%%%%%%%%%%%%%%%%%%%%%%%%%%%%%%%%%%%%%%%%%%%%%%
%%%%%%%%%%%%%%%%%%%%%%%%%%%%%%%%%%%%%%%%%%%%%%%%%%%%%%%%%%%%%%%%%%%%%%%%%%%%%%%%%%%%%%%%%%%%%%

We can make further progress analytically for the global quantum quench by making use of the light cone limit of conformal blocks. This allows us to make universal statements about irrational conformal field theories with no conserved currents beyond the stress tensor which we call denote as {\it pure CFTs} for abbreviation. These theories have been studied in the context of quantum chaos and information scrambling in e.g.~Refs.~\cite{2015JHEP...09..110A,2019JHEP...08..063K, 2019arXiv191014575K}.

%%%%%%%%%%%%%%%%%%%%%%%%%%%%%%%%%%%%%%%%%%%%%%%%%%%%%%%%%%%%%%%%%%%%%%%%%%%%%%%%%%%%%%%%%%%%%%
\subsection{Irrational CFT}
%%%%%%%%%%%%%%%%%%%%%%%%%%%%%%%%%%%%%%%%%%%%%%%%%%%%%%%%%%%%%%%%%%%%%%%%%%%%%%%%%%%%%%%%%%%%%%

In the following, let us restrict ourselves to the high temperature limit $t,L,d \gg\beta$, which simplifies our CFT calculation.
In this limit, the behavior of the eight-point correlator, \eqref{Znm_correlator}, is controlled by some OPE singularities.
We expect that the singular asymptotics are universal for any boundary state (which was first assumed in Ref.~\cite{2015JHEP...09..110A}). Therefore, we consider, instead of a conformal boundary state, \eqref{psi0_glob}, the thermofield double (TFD) state \cite{2013JHEP...05..014H},
\begin{equation}\label{eq:TFD}
\ket{\Psi(t)}=  \sum_n   \ex{-H \tau -H\frac{\b}{2}  } \ket{n}_1 \ket{n}_2 ,
\end{equation}
where the subscript 1,2 means CFT${}_1$ and its copy CFT${}_2$.
Here we perform the standard Wick rotation of $t \to -i\tau$. 

\begin{figure}[t]
\centering
  \includegraphics[width=15.0cm,clip]{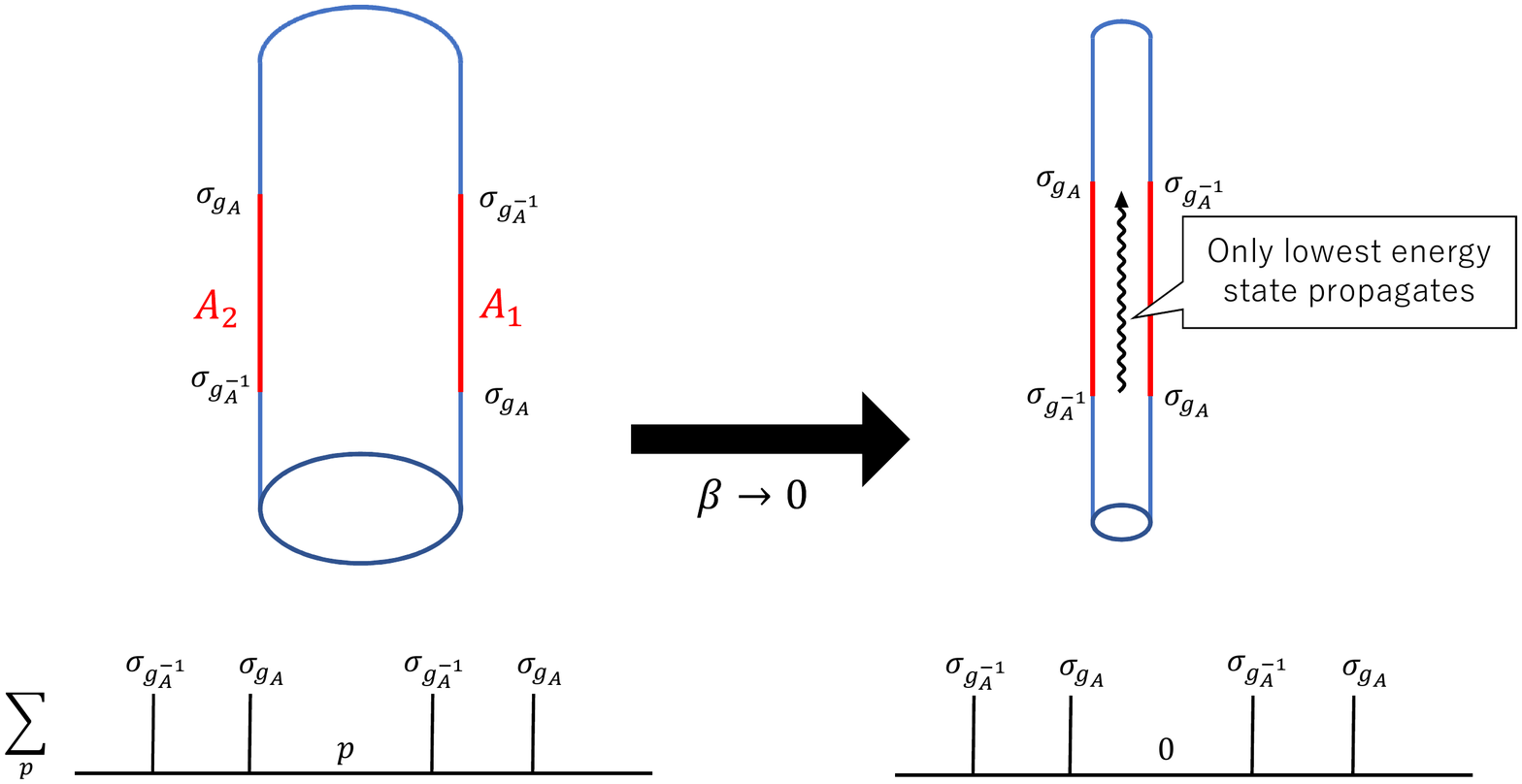}

 \caption{This figure shows how the high temperature limit simplifies the evaluation of the correlator.
In this limit, the distance between two pairs $\{   \sigma_{g_A}  ,  \sigma_{g_A^{-1}} \}$, $\{  \sigma_{g_A^{-1}},   \sigma_{g_A}    \}$ becomes much larger than other scales, therefore, only the vacuum can propagate between these two OPEs. This means that the correlator can be approximated by the vacuum block.
}
 \label{fig:high}
\end{figure}

For the TFD state, the canonical purification leads to the actual eight-point correlation function,
\begin{equation}
S_R(A:B)= \lim_{n,m \to 1} \frac{1}{1-n} \log \frac{Z_{n,m}}{   \pa{Z_{1,m}}^n}, 
\end{equation}
where the replica partition function $Z_{n,m} $ is defined by
\begin{equation}\label{eq:dRenyi}
\Braket{\sigma_{g_A}(w_1,\bar{w}_1)\sigma_{g_A^{-1}}(w_2,\bar{w}_2)   \sigma_{g_B}(w_3,\bar{w}_3) \sigma_{g_B^{-1}}(w_4,\bar{w}_4) \sigma_{g_A^{-1}}(w_5,\bar{w}_5)\sigma_{g_A}(w_6,\bar{w}_6)   \sigma_{g_B^{-1}}(w_7,\bar{w}_7) \sigma_{g_B}(w_8,\bar{w}_8) }_{\text{CFT}^{\otimes mn} }.
\end{equation}
If we set $A_1=A_2=[0,L]$ and $B_1=B_2=[L+d, 2L+d]$, their coordinates become
\begin{equation}
\begin{aligned}
&w_1= \ex{\frac{2\pi}{\b}\pa{-t+i\frac{\beta}{4}}}, \ \ \ \  \ \ \ \ \ \bar{w}_1= \ex{\frac{2\pi}{\b}\pa{t-i\frac{\beta}{4}}}, \\
&w_2= \ex{\frac{2\pi}{\b}\pa{L-t+i\frac{\beta}{4}}}, \ \ \ \ \ \ \ \ \bar{w}_2= \ex{\frac{2\pi}{\b}\pa{L+t-i\frac{\beta}{4}}}, \\
&w_3= \ex{\frac{2\pi}{\b}\pa{L+d-t+i\frac{\beta}{4}}}, \ \ \ \ \ \bar{w}_3= \ex{\frac{2\pi}{\b}\pa{L+d+t-i\frac{\beta}{4}}}, \\
&w_4= \ex{\frac{2\pi}{\b}\pa{2L+d-t+i\frac{\beta}{4}}}, \ \ \ \ \bar{w}_4= \ex{\frac{2\pi}{\b}\pa{2L+d+t-i\frac{\beta}{4}}}, \\
\end{aligned}
\end{equation}
and $w_{i+4}=\bar{w}_i$ $(i=1,2,3,4)$.
In the high temperature limit, the distances between two operators become very large; therefore, only the conformal families with the lowest energy can propagate.
In other words, the eight-point correlator can be approximated by the vacuum block (see Fig. \ref{fig:high})\footnote{
We can show this vacuum dominance explicitly in terms of the cross ratio.
As explained in Ref.~\cite{2015JHEP...09..110A}, the high temperature limit leads to the light cone limit $ z\ll1-\bar{z} \ll 1 $. Thus we can straightforwardly show the vacuum dominance from the high temperature limit.
Note that sometimes the limit $z,1-\bar{z} \ll 1$ is also called as the (double) light cone limit, but here we do not refer to this limit.
},

\newsavebox{\boxpa}
\sbox{\boxpa}{\includegraphics[width=180pt]{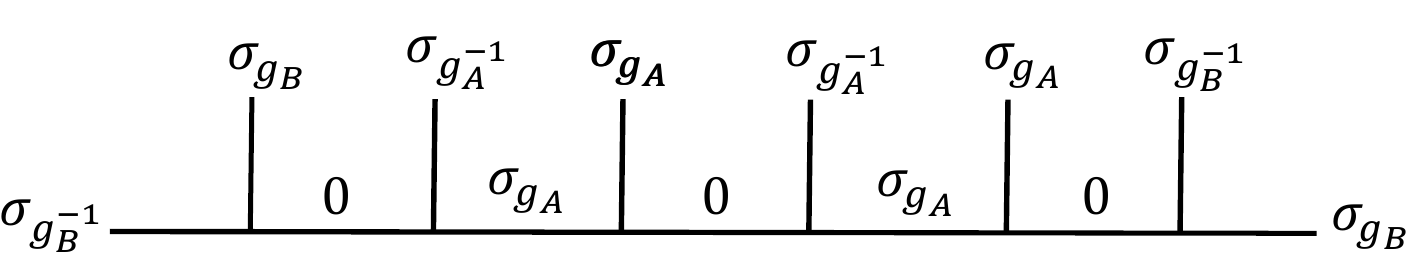}}
\newlength{\paw}
\settowidth{\paw}{\usebox{\boxpa}} 

\begin{equation}\label{eq:localchannel}
\parbox{\paw}{\usebox{\boxpa}} \times (\text{anti-holomorphic}).
\end{equation}

Let us consider the time evolution. 
We first comment on the case $d>L$. In that case, the dominant channel of the eight-point correlator is always given by the disconnected entanglement wedge, therefore, the reflected entropy is trivial, $S_R(A:B)=0$, at all times. In the context of mutual information, this is considered to be a signature of maximal scrambling \cite{2015JHEP...09..110A}. To find nontrivial results, we will instead focus on the case $d<L$ in the following.

If we set $d<L$, the dominant channel can be changed from (\ref{eq:localchannel}).
In fact, there are four time regions: (i) $0<t<\frac{d}{2}$, (ii) $\frac{d}{2}<t<\frac{L+d}{2}$, (iii) $\frac{L+d}{2}<t<L-\frac{d}{2}$ ,and (iv) $L-\frac{d}{2}<t$.
The dominant channel for each region can be expressed by
\newsavebox{\boxpba}
\sbox{\boxpba}{\includegraphics[width=160pt]{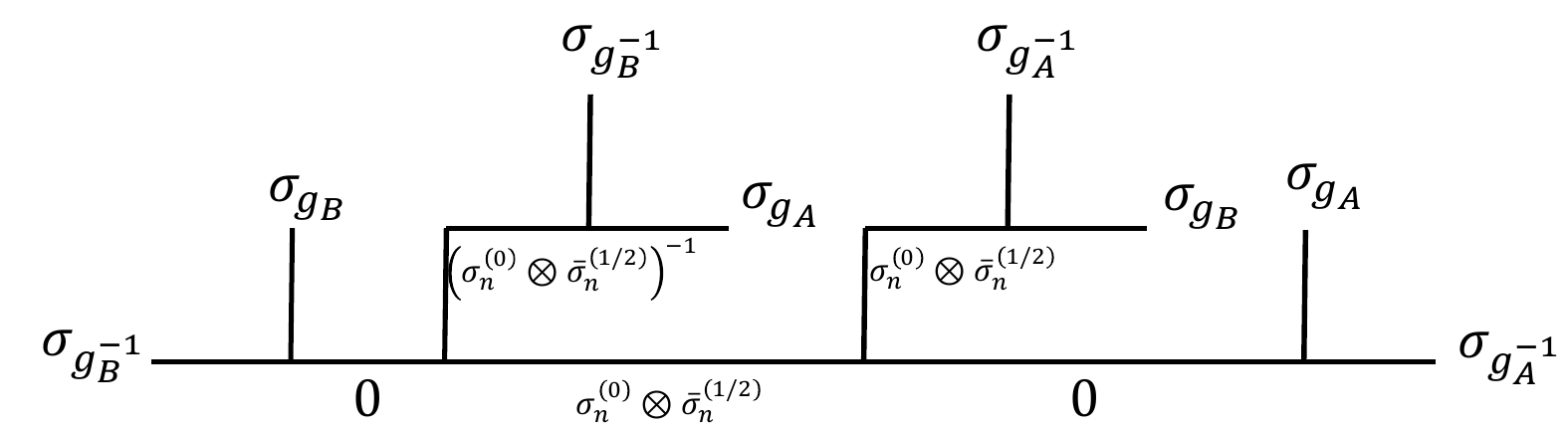}}
\newlength{\pbaw}
\settowidth{\pbaw}{\usebox{\boxpba}} 

\newsavebox{\boxpbb}
\sbox{\boxpbb}{\includegraphics[width=160pt]{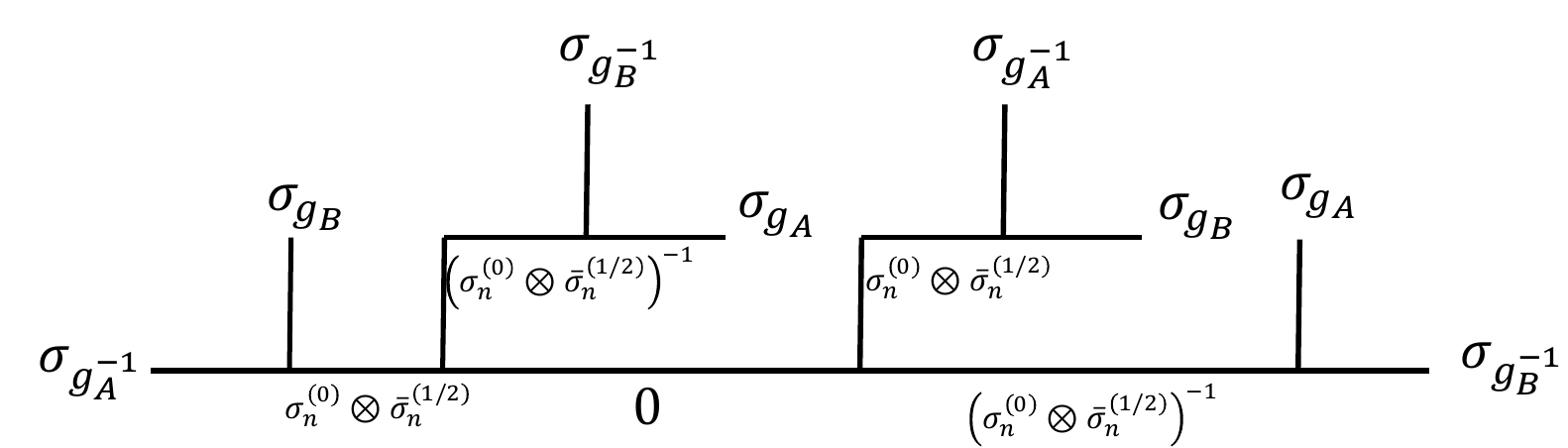}}
\newlength{\pbbw}
\settowidth{\pbbw}{\usebox{\boxpbb}}

\newsavebox{\boxpbc}
\sbox{\boxpbc}{\includegraphics[width=160pt]{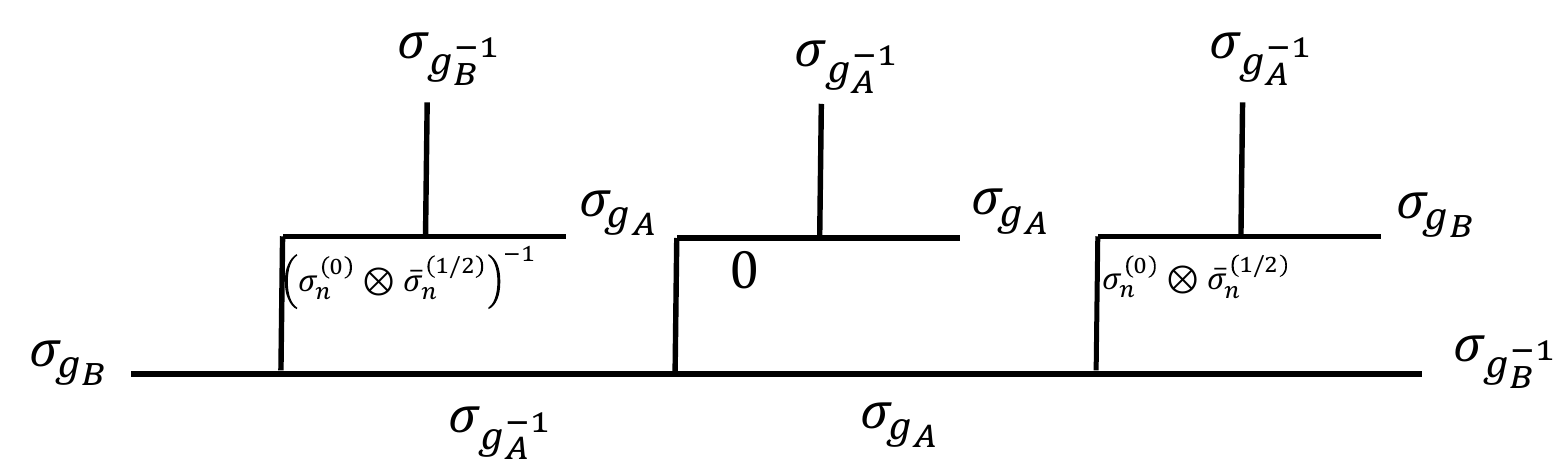}}
\newlength{\pbcw}
\settowidth{\pbcw}{\usebox{\boxpbc}}

\begin{description}

\item[(i)]

\begin{equation}\label{eq:blocki}
\parbox{\paw}{\usebox{\boxpa}} \times (\text{anti-holomorphic}).
\end{equation}

\item[(ii)]

\begin{equation}\label{eq:blockii}
 \parbox{\pbaw}{\usebox{\boxpba}} \times (\text{anti-holomorphic}).
\end{equation}

\item[(iii)]

\begin{equation}\label{eq:blockiii}
 \parbox{\pbcw}{\usebox{\boxpbc}} \times (\text{anti-holomorphic}).
\end{equation}

\item[(iv)]

\begin{equation}\label{eq:blockiv}
\parbox{\paw}{\usebox{\boxpa}} \times (\text{anti-holomorphic}).
\end{equation}

\end{description}
Here we have to mention that the dominant channel is chosen by the assumption $\Delta_n \ll \Delta_m$, which is necessary for reproducing the correct entanglement wedge.
Under this assumption, we need to evaluate the {\it light cone singularity} of the eight-point function, which is attributed to the high-temperature limit. An important point is that this light cone singularity does not contribute to the calculation of the mutual information (i.e., entanglement wedge) in pure CFTs (see \cite{Kusuki2019e, 2019JHEP...08..063K}). We emphasize that this is not the case in general, as we will show for rational CFTs later.
As a result, we obtain the dominant channels (\ref{eq:blocki}) $\sim$ (\ref{eq:blockiv}).
More detailed calculation can be refereed to in Ref.~\cite{Kusuki2019e}, which is the generalized version of the result in Ref.~\cite{2015JHEP...09..110A}.

After choosing the dominant channel, we can evaluate the approximated eight-point function by first taking the limit $m \to 1$ \cite{2019arXiv190706646K}.
This limit simplifies the evaluation. As a result, (i) and (iv) reduce to unity, whereas (ii) and (iii) reduce to non-trivial conformal blocks as follows\footnote{
We use the semi-classical $n$-point block. If the reader is interested in the details of its evaluation, please refer to the appendix of Ref.~\cite{2019arXiv190906790K}.
},

\newsavebox{\boxpca}
\sbox{\boxpca}{\includegraphics[width=160pt]{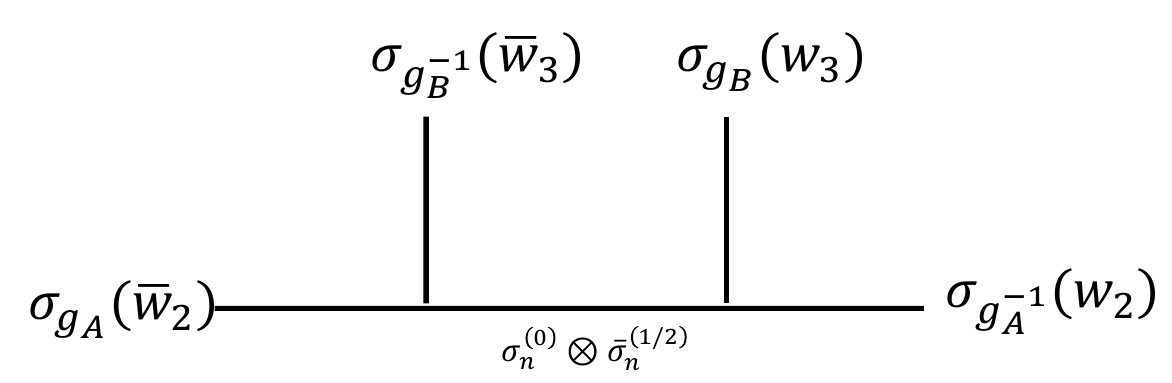}}
\newlength{\pcaw}
\settowidth{\pcaw}{\usebox{\boxpca}}

\newsavebox{\boxpcb}
\sbox{\boxpcb}{\includegraphics[width=200pt]{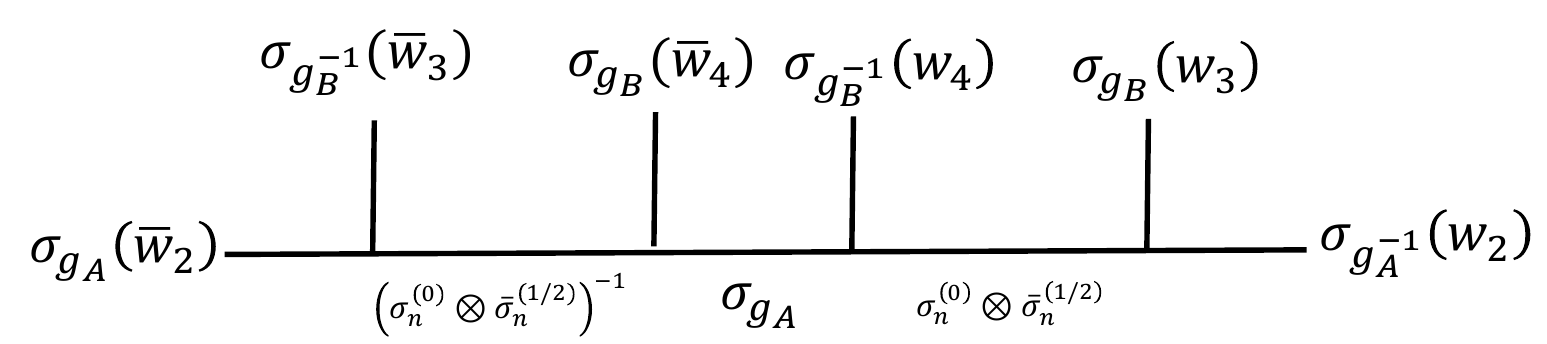}}

\newlength{\pcbw}
\settowidth{\pcbw}{\usebox{\boxpcb}}

\begin{description}

\item[(ii)]

\begin{equation}\label{eq:ablockii}
 \parbox{\pcaw}{\usebox{\boxpca}} \times (\text{anti-holomorphic}) \ar{\b \to 0} \abs{\ex{-2h_n \frac{2\pi}{\b}\pa{2t-d}}}^2,
\end{equation}

\item[(iii)]

\begin{equation}\label{eq:ablockiii}
 \parbox{\pcbw}{\usebox{\boxpcb}} \times (\text{anti-holomorphic})  \ar{\b \to 0} \abs{\ex{-2h_n \frac{2\pi}{\b}L}}^2.
\end{equation}

\end{description}
As a result, we obtain
\begin{equation}\label{eq:SRresult}
\begin{aligned}
S_R(A:B)&=\left\{
    \begin{array}{ll}
 0   ,& \text{if }  0<t<\frac{d}{2} ,\\
       2\frac{c}{3} \frac{2 \pi}{\b} \pa{t-\frac{d}{2}}  ,& \text{if } \frac{d}{2}<t<\frac{L+d}{2},  \\
	  2\frac{c}{3} \frac{2 \pi}{\b} \frac{L}{2}  ,& \text{if }  \frac{L+d}{2}<t<L-\frac{d}{2} ,\\
 	0	 ,& \text{if } L-\frac{d}{2}<t  .\\
 \end{array}
  \right.\\
\end{aligned}
\end{equation}
Note that the analysis in Ref.~\cite{2019arXiv190709581W} has related calculations from the gravitational side, though their results are obtained by a different order of limits than ours. Therefore, their result has an additional term. If we consider the adjacent intervals limit, $d \ll \b$, the singularity is changed from (\ref{eq:ablockii}) and (\ref{eq:ablockiii}) to
\begin{equation}
\begin{aligned}
\left\{
    \begin{array}{ll}
    C_{n,1}^2 \abs{\pa{ \ex{\frac{2\pi}{\b}d}-1 }^{4h_n} \ex{-2h_n \frac{2\pi}{\b}2t}}^2 = \abs{\pa{ \frac{ \pi d}{\b}}^{4h_n}  \ex{-2h_n \frac{2 \pi}{\b} 2t} }^2 ,& \text{if } 0<t<\frac{L}{2},  \\
  	C_{n,1}^4  \abs{\pa{ \ex{\frac{2\pi}{\b}d}-1 }^{4h_n}  \ex{-2h_n \frac{2\pi}{\b}L}}^2 =  \abs{ \pa{ \frac{ \pi d}{2\b}}^{4h_n} \ex{-2h_n \frac{2 \pi}{\b}L}}^2	,& \text{if }  \frac{L}{2}<t<L, \\
  	C_{n,1}^4 \abs{\pa{ \ex{\frac{2\pi}{\b}d}-1 }^{4h_n}}^2  = \abs{\pa{ \frac{ \pi d}{2\b}}^{4h_n} }^2	,& \text{if }  L<t,\\
    \end{array}
  \right.\\
\end{aligned}
\end{equation}
where $C_{n,m}$ is the OPE coefficient $\braket{\sigma_{g_A^{-1}}| \sigma_{g_B} (1) |\sigma_{g_B g_A^{-1}}}$. This coefficient can be calculated by the method developed in Ref.~\cite{2001CMaPh.219..399L}
\begin{equation}\label{eq:Mathur}
C_{n,m}=(2m)^{-4h_n}.
\end{equation}
The details of this derivation can be found in Appendix C of Ref.~\cite{2019arXiv190500577D}.
As a result, the reflected entropy is given by
\begin{equation}
\label{eq:SRresult_adjacent}
\begin{aligned}
S_R(A:B)&=\left\{
    \begin{array}{ll}
 2\pa{\frac{c}{3} \log \frac{\b}{ \pi d}+\frac{c}{3} \frac{2\pi}{\b}t } ,& \text{if } 0<t<\frac{L}{2},  \\
	  2\pa{\frac{c}{3} \log \frac{2\b}{ \pi d}   + \frac{c}{3} \frac{2 \pi}{\b} \frac{L}{2}   }   ,& \text{if }  \frac{L}{2}<t<L ,\\
  	 2\pa{\frac{c}{3} \log \frac{2\b}{ \pi d}}	,& \text{if }  L<t .\\
    \end{array}
  \right.\\
\end{aligned}
\end{equation}
The prefactor $2$ comes from the doubling to create the TFD state, so to compare with the boundary state quench, \eqref{psi0_glob}, we must divide everything by 2.

At the end of this subsection, we would like to comment that our analysis does not rely on the large-$c$ limit.
This means that our result is exact in $c$. According to Refs.~\cite{2015JHEP...09..110A, 2019arXiv190709581W}, the result from the gravity side is unique in the holographic CFT. 
% Tt is thought of as the signature of {\it maximal} scrambling because its behavior is quite different from that in integrable systems. 
However, from our result and Ref.~\cite{Kusuki2019e}, the same behavior can be also seen in pure CFTs (including the holographic CFTs). This implies that pure CFTs are also maximally scrambling.
% or the entanglement entropy and the reflected entropy do not allow us to diagnose the {\it maximality} of scrambling?
% We will address this question in the future.

%%%%%%%%%%%%%%%%%%%%%%%%%%%%%%%%%%%%%%%%%%%%%%%%%%%%%%%%%%%%%%%%%%%%%%%%%%%%%%%%%%%%%%%%%%%%%%
\subsection{RCFT}
%%%%%%%%%%%%%%%%%%%%%%%%%%%%%%%%%%%%%%%%%%%%%%%%%%%%%%%%%%%%%%%%%%%%%%%%%%%%%%%%%%%%%%%%%%%%%%

As we have already mentioned in the previous section, we have to take care of the {\it light cone singularity} in the calculation of the Lorentzian correlator in the high temperature limit.
In pure CFTs, the contribution to the entanglement entropy from the light cone singularity disappears, but it becomes important in RCFTs.

To explain this more simply, let us consider the four-point function,
\begin{equation}
G(z,\bar{z})=\braket{O(\infty) O (1) O (z,\bar{z}) O(0)  }.
\end{equation}
The bootstrap equation leads to
\begin{equation}
\braket{O(\infty) O (1) O (1-z,1-\bar{z}) O(0)  } = \sum_p C_{OOp}^2 \ca{F}(p|z) \overline{\ca{F}(p|\bar{z})},
\end{equation}
where $\ca{F}$ is the Virasoro conformal block and the summation is taken over all primary states in our theory.
From the left hand side, one can immediately find that the singularity of the correlator $G(1-z,1-\bar{z})$ in the limit $\bar{z} \to 1$ is given by
\begin{equation}\label{eq:sing}
G(1-z,1-\bar{z}) \ar{\bar{z} \to 1} (1-\bar{z})^{-2h_O}.
\end{equation}
This means that the summation of the blocks on the right hand side should reproduce this singularity.
Here is where the constraint that our theory is rational becomes important.
% important whether our theory is rational or irrational.
In RCFTs, there are only a finite number of primaries. Therefore, each conformal block in the summation should have a strong singularity. On the other hand, in irrational CFTs, there are an infinite number of primaries, which can reproduce the singularity in (\ref{eq:sing}) even if the individual blocks do not have this strong of a singularity.
This is the reason why the light cone singularity does not contribute to the entanglement entropy in pure CFTs; this disappearance cannot happen in RCFTs\footnote{
More precisely and technically, this difference of the light cone singularity between RCFTs and pure CFTs is directly seen from the pole structure of the {\it fusion matrix}. If one is interested in details, see Appendix A of \cite{Kusuki2019e}.
}.

The strong light cone singularity in RCFTs drastically changes the behavior of the entanglement entropy.
This has already studied for mutual information in Ref.~\cite{2015JHEP...09..110A}, hence we do not give the details of the calculation of the entanglement entropy here.
We can straightforwardly generalize this analysis to find the dominant conformal block of the eight-point correlator for reflected entropy, (\ref{eq:dRenyi}), in the high temperature limit.
Evaluating the dominant block in the light cone limit, we obtain

\begin{equation}
\begin{aligned}
S_R(A:B)&=\left\{
    \begin{array}{ll}
 0   ,& \text{if }  0<t<\frac{d}{2} ,\\
       2\frac{c}{3} \frac{2 \pi}{\b} \pa{t-\frac{d}{2}}  ,& \text{if } \frac{d}{2}<t<\frac{L+d}{2},  \\
	  2\frac{c}{3} \frac{2 \pi}{\b} \pa{L+\fr{d}{2}-t}  ,& \text{if }  \frac{L+d}{2}<t<L+\fr{d}{2} ,\\
 	0	 ,& \text{if } L+\frac{d}{2}<t  .\\
 \end{array}
  \right.\\
\end{aligned}
\end{equation}
This growth is drastically different from that in pure CFTs (\ref{eq:SRresult}).
This implies that the reflected entropy is sensitive to chaoticity of a given system.
Moreover, this is explained by the quasi-particle picture.
One important finding is that this perfectly matches the behavior of the mutual information \cite{2015JHEP...09..110A},
\begin{equation}
S_R(A:B)=I(A:B),
\end{equation}
at all times. This is precisely what was predicted by the universal contribution in Section \ref{univ_sec}. As we saw in the previous subsection (and explain in the following sections), this is not the case in general.

%%%%%%%%%%%%%%%%%%%%%%%%%%%%%%%%%%%%%%%%%%%%%%%%%%%%%%%%%%%%%%%%%%%%%%%%%%%%%%%%%%%%%%%%%%%%%%
%%%%%%%%%%%%%%%%%%%%%%%%%%%%%%%%%%%%%%%%%%%%%%%%%%%%%%%%%%%%%%%%%%%%%%%%%%%%%%%%%%%%%%%%%%%%%%
\subsection{Comparison to mutual information} %%%%%%%%%%%%%%%%%%%%%%%%%%%%%%%%%%%%%%%%%%%%%%%%%%%%%%%%%%%%%%%%%%%%%%%%%%%%%%%%%%%%%%%%%%%%%%
%%%%%%%%%%%%%%%%%%%%%%%%%%%%%%%%%%%%%%%%%%%%%%%%%%%%%%%%%%%%%%%%%%%%%%%%%%%%%%%%%%%%%%%%%%%%%%

\begin{figure}[t]
\centering
  \includegraphics[width=7.0cm,clip]{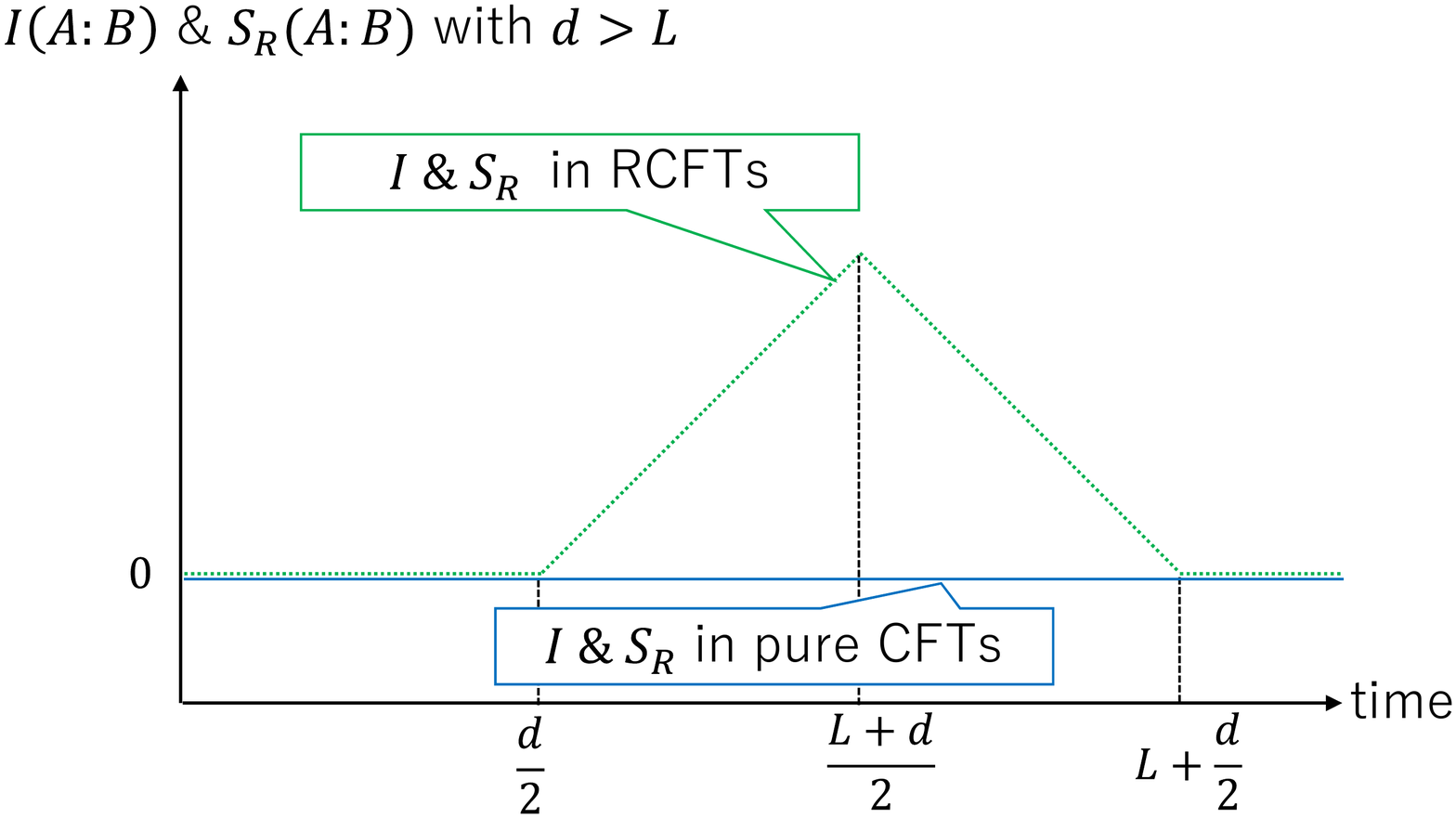}
  \includegraphics[width=7.0cm,clip]{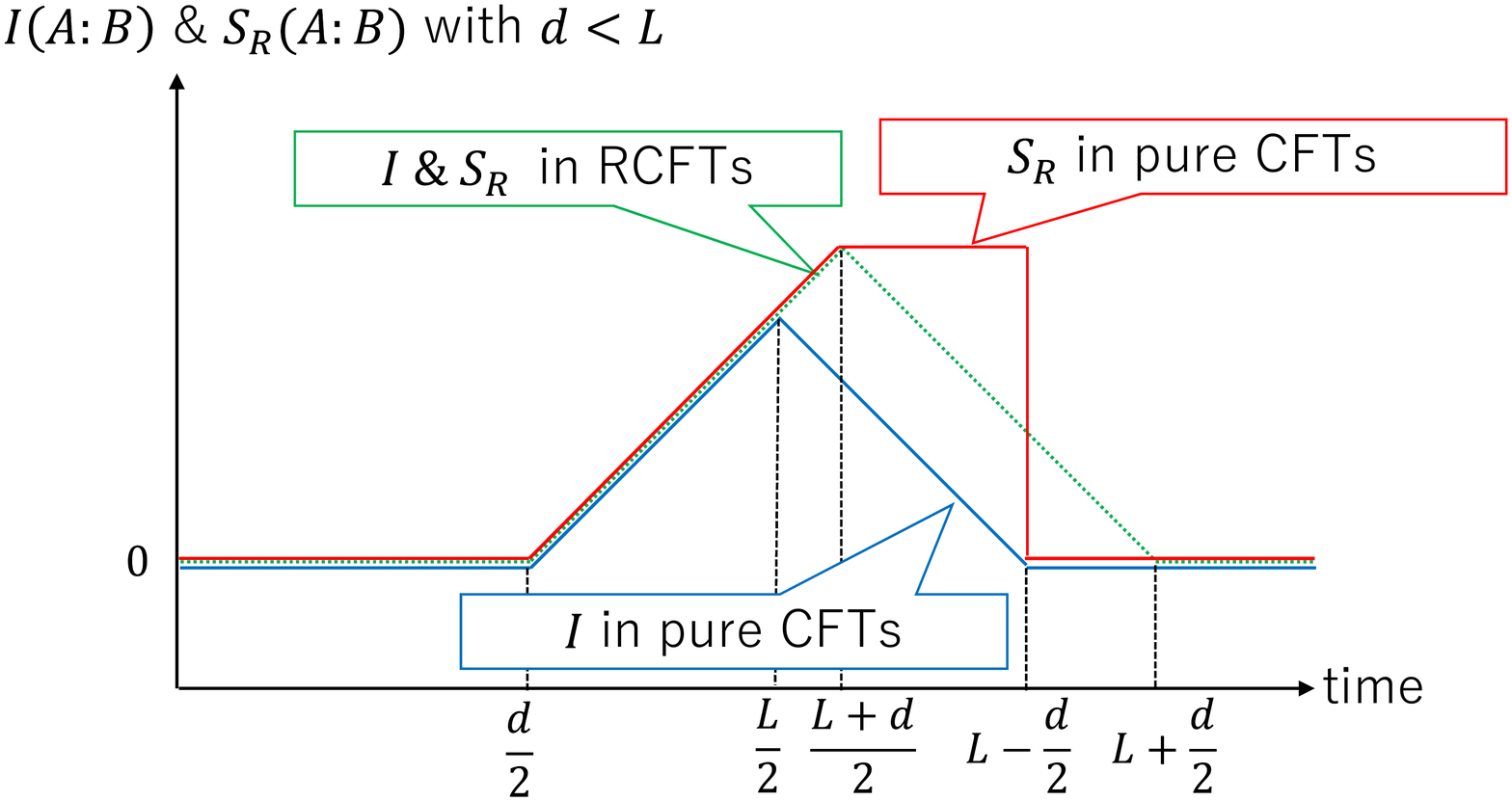}
 \caption{
Left: The time evolution of the mutual information and the reflected entropy under the global quench with $d>L$.
The holographic result (blue line) is always zero, which strongly violates the quasi-particle picture (green dashed line).
Right:
The time evolution when $d<L$. In this case, the reflected entropy behaves differently from the mutual information, which is a new signature of scrambling.
}
 \label{fig:plot}
\end{figure}

In this section, we compare the reflected entropy and the mutual information.
This comparison is motivated by the fact that the reflected entropy can be thought of as a measure of correlations between two intervals, like the mutual information.
Therefore, it is natural to utilize the reflected entropy to clarify how the information is spreading.
From this viewpoint, our results should be compared with the results from  Ref.~\cite{2015JHEP...09..110A}, where the mutual information after a global quench is studied.
We first comment on the case $d>L$, which is the main interest in Ref.~\cite{2015JHEP...09..110A}.
If we set $d>L$, then the holographic mutual information always vanishes as shown in the left of Fig.~\ref{fig:plot}.
This violates the quasi-particle picture described in Section \ref{univ_sec}. This discrepancy implies the strong scrambling of the correlations (e.g.~entanglement) between two intervals i.e.~entanglement cannot be carried by localized objects.

If we consider the reflected entropy in the same setup as in Ref.~\cite{2015JHEP...09..110A}, we always obtain 
\begin{equation}
I(A:B)=S_R(A:B)
\end{equation}
for both pure CFTs and integrable CFTs as shown in Fig.~\ref{fig:plot}.
This means that, in this configuration, we cannot extract new information by making use of the reflected entropy.

Although the behavior of reflected entropy is not particularly new in the case $d>L$, we can find new and interesting behavior in the case $d<L$ that is distinct from the mutual information (right of Fig.~\ref{fig:plot}).
First, we note that the mutual information behaves similarly between the pure CFTs and integrable CFTs.
The central difference is the time scale and magnitude of the turning point.
This implies that predictions from the quasi-particle picture break down at $t=\frac{L}{2}$; after that, part of the entanglement appears to vanish.
This {\it missing entanglement} can be also seen in the mutual information after a local quench \cite{2014PhRvD..89f6015A}. It is interesting to note that the explicit form of the missing entanglement entropy is given by
\begin{equation}
 2\frac{c}{3} \frac{2 \pi}{\b} \frac{d}{2}, \ \ \ \text{at } \fr{L+d}{2}<t<L-\fr{d}{2}.
\end{equation}
This means that the further separated the two subsystems are from each other, the greater the loss of the entanglement.

Against this backdrop, the reflected entropy can provide new insight into the correlations at intermediate times. 
In fact, the correlation {\it grows} after $t=\frac{L}{2}$, which can be observed by using the reflected entropy and not the mutual information.
We would like to comment that our result is consistent with the known inequality,
\begin{equation}
S_R(A:B) \geq I(A:B).
\end{equation}
It is natural to ask where the difference $S_R-I$ derives from\footnote{We should note that the difference between $S_R$ and $I$ in this scenario is parametrically large and should be taken very seriously. This is in contrast with previous differences between $S_R$ and $I$ that have been found in other dynamical settings that disappear in the scaling limit \cite{2019arXiv190607639K}.}. This brings us to the central puzzle that arises from our calculations. There are two possible interpretations that seem to be at odds with one another. We will lay out the evidence for both arguments. We hope this will motivate future work to fully understand the dynamical nature of information in chaotic systems.

\subsubsection{\texorpdfstring{$S_R-I$}{TEXT} is classical correlation}
In Refs.~\cite{2019PhRvL.122t1601B,2019arXiv190706646K,2019arXiv190906790K,2019arXiv190712555U}, it was argued that the reflected entropy is more sensitive to classical correlations than mutual information. Thus, we  should interpret $S_R-I$ as {\it classical correlations}. Classical correlations have also been argued to not appear in integrable systems, but only in non-integrable systems 
by examining mutual information and reflected entropy after local operator quenches
\cite{2019arXiv190706646K,2019arXiv190906790K}. 
The authors observed additional correlations detected by the reflected entropy that were independent of both the operator insertion and the cutoff, quantities that quantum correlations should certainly depend on. This only occurred for pure CFTs and not integrable theories.
In integrable CFTs, there are fewer unconstrained degrees of freedom. Therefore, the creation of local excitations may not lead to significant classical correlation. On the other hand, in irrational CFTs, there is an abundance of degrees of freedom and the classical correlation will more readily appear under excitation. Consequently, the mutual information will, in general, differ from the reflected entropy.

Another argument that the reflected entropy and entanglement wedge cross-section detect more classical correlations than the mutual information is the generic lower bound of the reflected entropy on the mutual information and the fact that mutual information should capture all quantum correlations. While the mutual information certainly detects some classical correlation, there is strong evidence that it captures \textit{all} quantum correlation because it bounds the connected correlator of local operators \eqref{connected_correlator}. One of the most convincing points is that the mutual information upper bounds axiomatic entanglement measures. In fact, it was shown in Ref.~\cite{2019arXiv190712555U} that the squashed entanglement is equivalent to half of the mutual information in holographic theories.  Presumably, any correlations not detected by these axiomatic measures should be classical\footnote{A caveat to this argument is that the physical meaning of these axiomatic measures is not well understood in quantum field theory. In finite dimensional quantum mechanics, the physical intuition is generally described in terms of counting Bell pairs which do not have a natural analog in the continuum. The same caveat applies to negativity, which is also not satisfactorily motivated in the continuum. }. 

One reason this is unexpected is that the monogamy of mutual information in holographic theories suggests that quantum correlations dominate all correlations in these theories \cite{2013PhRvD..87d6003H}. Thus, it would be surprising that classical correlations so heavily dominate during the ``plateau" region. This worry may be remedied by suggesting that mutual information is only sensitive to quantum correlations in holographic theories \cite{2019arXiv190712555U}. That is, the conclusion that holographic theories are dominated by quantum correlations is merely an artifact of using mutual information as a tool for total correlation and not more general total correlation measures.

It is important to note certain physical and practical implications of this conclusion. The authors of Ref.~\cite{2019arXiv190709581W} studied the operator entanglement of the reduced density matrix, a quantity closely related to the reflected entropy. Similar to our analysis, they found an extended plateau region for holographic theories. They argued that this acts as an ``entanglement barrier" in the sense that the two subsystems are so highly entangled that they become computationally intractable by using variational ansatzes such as matrix product states \cite{2017arXiv170208894L}. If it is correct to conclude that $S_R-I$ is classical information, then this ``barrier from chaos" is not as big of a barrier as originally thought
because classical correlations are not computational obstacles. Moreover, we find that the mutual information is even smaller than it was for integrable systems (missing entanglement). This implies that reduced density matrices of chaotic systems are the easiest to simulate i.e.~\textit{chaos shrinks the entanglement barrier}.

\subsubsection{\texorpdfstring{$S_R-I$}{TEXT} is quantum correlation}

While the above arguments strongly suggest this additional correlation to be classical, this interpretation is quite puzzling from another perspective. 
We know that in holographic conformal field theories, the logarithmic negativity is equivalent to half of the reflected entropy at R\'enyi index $1/2$ \cite{2019PhRvD..99j6014K,PhysRevLett.123.131603}
\begin{align}
    \mathcal{E} = \frac{S_R^{(1/2)}}{2}.
\end{align}
Even though computing R\'enyi entropies holographically requires the daunting task of solving Einstein's equations with codimension-two sources \cite{2016NatCo...712472D}, it is simple to bound the negativity by the (von Neumann) reflected entropy because R\'enyi entropies are non-increasing functions of R\'enyi index
\begin{align}
    \mathcal{E} \geq \frac{S_R}{2}.
\end{align}
Thus, we conclude that negativity must also show the surplus of correlations that are in the plateau region. It is quite peculiar that the negativity, which is only sensitive to quantum correlations, can be significantly larger than the mutual information, which is sensitive to both quantum and classical correlations. This would also contradict the previous analysis that the ``plateau" is a consequence of classical correlations. We suspect that this hints at the special features of the entanglement structure of holographic theories, analogous to constraints found in e.g. Ref.~\cite{2013PhRvD..87d6003H}. In particular, the negativity has previously always been found to be well behaved in quantum field theory \cite{2012PhRvL.109m0502C,2013JSMTE..02..008C,2014JSMTE..12..017C,2015PhRvB..92g5109W,2019arXiv190607639K}, so this potentially spurious behavior is very new.

The two possible resolutions to this tension require us to dramatically revise our understanding of the meanings of mutual information or negativity in quantum systems. We have the following two choices: (i) Mutual information does not count all quantum correlations shared between subsystems. (ii) Negativity drastically over counts quantum correlations and can become spurious in strongly coupled quantum field theory. Both scenarios are unsettling. While we currently find the second choice more appealing, we believe more work needs to be done to ease this tension. Other factors which can be potentially relevant to the missing entanglement are non-local spreading and possibly multipartite nature of quantum information\footnote{See Ref.~\cite{2019arXiv191107852A} for discussion regarding the implications of $S_R-I > 0$ for tripartite entanglement. Also see Ref.~\cite{2019CMaPh.tmp..293C} for relevant discussion on the multipartite nature of entanglement in holography.}.

%%%%%%%%%%%%%%%%%%%%%%%%%%%%%%%%%%%%%%%%%%%%%%%%%%%%%%%%%%%%%%%%%%%%%%%%%%%%%%%%%%%%%%%%%%%%%%
\subsection{More general setup}
%%%%%%%%%%%%%%%%%%%%%%%%%%%%%%%%%%%%%%%%%%%%%%%%%%%%%%%%%%%%%%%%%%%%%%%%%%%%%%%%%%%%%%%%%%%%%%

\begin{figure}[t]
\centering
  \includegraphics[width=6.0cm,clip]{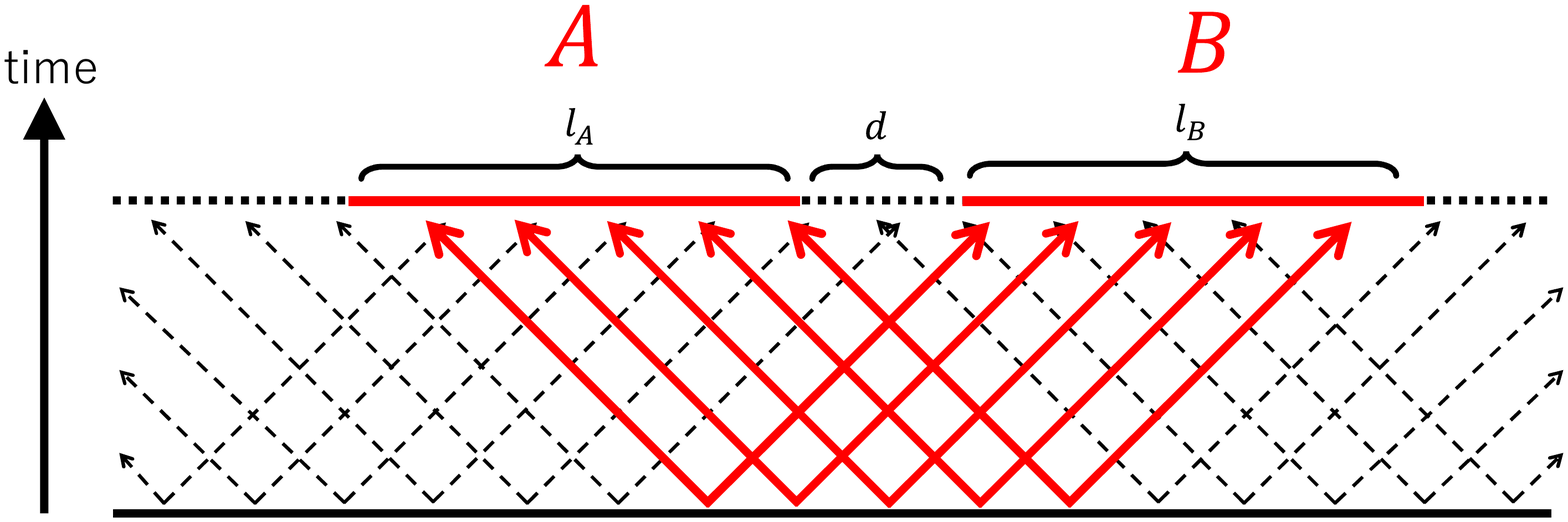}
  \includegraphics[width=8.0cm,clip]{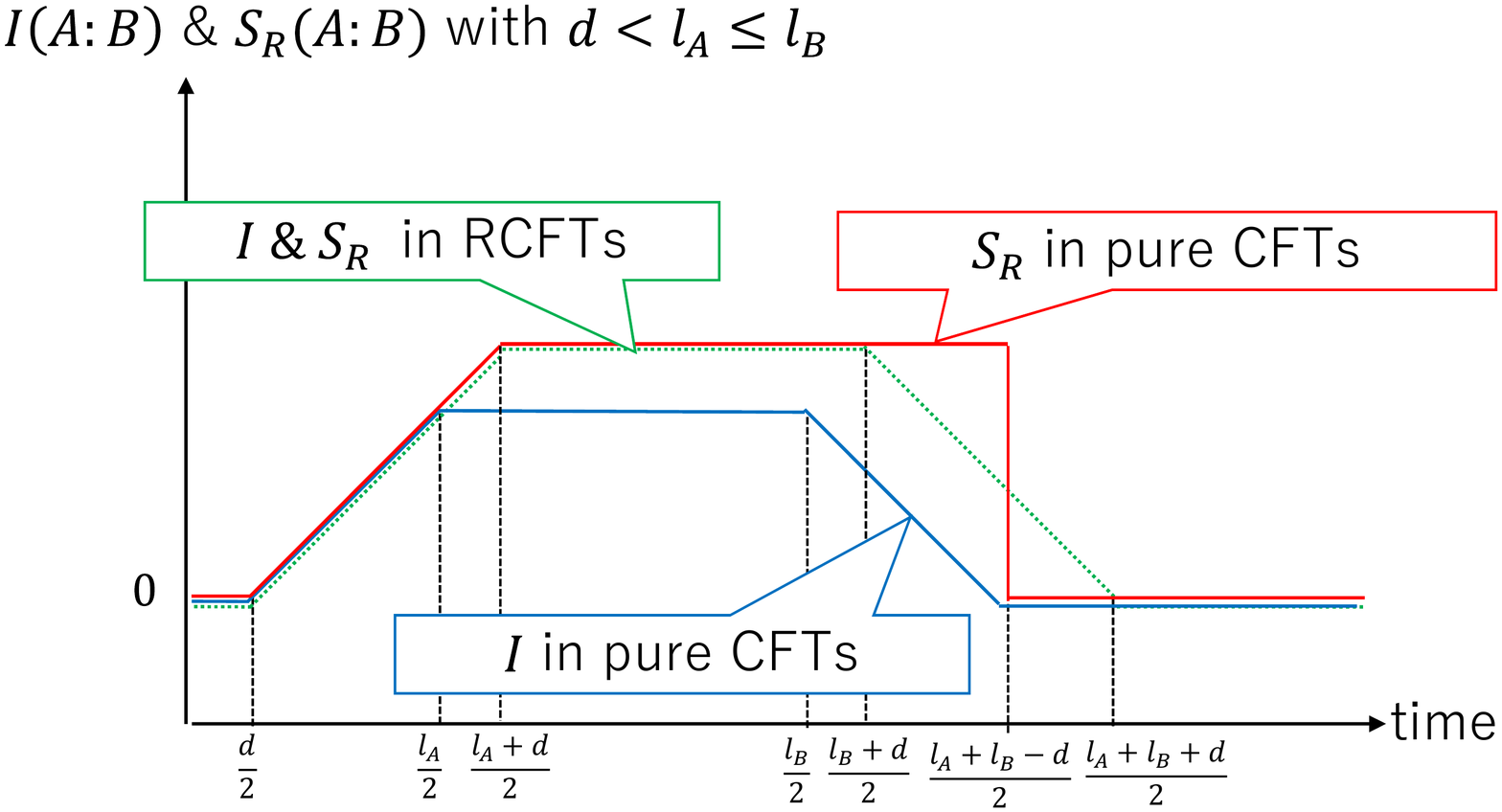}
 \caption{Left: The generalized setup with different subsystem sizes $l_A$ and $l_B$.
Right: The time evolution when $d<l_A \leq l_B$. One can find a new plateau that lasts for a time equal to $\fr{l_B-l_A}{2}$.
}
 \label{fig:setup2}
\end{figure}

In the above, we restricted ourselves to the case where the two subsystems have the same length, $L$, for simplicity, as in Fig.~\ref{fig:setup}.
One can generalize this to the case where two subsystems have difference lengths $l_A$ and $l_B$ (with $l_A \leq l_B$), which is illustrated in the left of Fig.~\ref{fig:setup2}.
Since the calculations of the mutual information and reflected entropy can be done in the same way as the above, we only give our results in the right of Fig.~\ref{fig:setup2}. A new feature in this case is that a plateau of duration $\fr{l_B-l_A}{2}$ appears for mutual information and reflected entropy for both rational and irrational CFTs. In other words, there is a {\it time delay} of duration $t =\fr{l_B-l_A}{2} $ after the saturation times (i.e.~, $t_s=\fr{l_A}{2}$ for $I$ in pure CFTs and $t_s=\fr{l_A+d}{2}$ for $S_R$ in pure CFTs and both $I$ and $S_R$ in RCFTs). This plateau may be understood by the quasi-particle picture in RCFTs, but as before, the plateau for pure CFTs is more mysterious. Later, we will be able to understand the pure CFT plateau using intuition from random unitary circuits.

%%%%%%%%%%%%%%%%%%%%%%%%%%%%%%%%%%%%%%%%%%%%%%%%%%%%%%%%%%%%%%%%%%%%%%%%%%%%%%%%%%%%%%%%%%%%%%
\subsection{Comments on odd entropy and negativity}
%%%%%%%%%%%%%%%%%%%%%%%%%%%%%%%%%%%%%%%%%%%%%%%%%%%%%%%%%%%%%%%%%%%%%%%%%%%%%%%%%%%%%%%%%%%%%%
The odd entropy for the time-dependent TFD state can also be evaluated by an eight-point correlation function of twist operators.
Therefore, we again need to look for a dominant conformal block and evaluate it in the light cone limit.
This procedure is similar to the previous sections, therefore, we do not show the detailed calculation in the present paper.
The result is identical to the reflected entropy (up to the normal factor of 2).
Because we do not have intuition for the physical interpretation of the difference between the odd entropy and the mutual information, we cannot currently make use of this fact to resolve the tensions from the previous section.
We hope that a further understanding of the odd entropy will help to answer questions for the quantum chaoticity, the missing entanglement, etc.~in the future.

Similarly, the logarithmic negativity for the TFD quench may be computed by a similar eight-point correlation function. Paralleling the derivation of holographic negativity in Ref.~\cite{PhysRevLett.123.131603}, the dominant conformal block for negativity will be equivalent to the dominant conformal block for the R\'enyi reflected entropy at $n=1/2$ even at finite central charge. While the single block dominance in Ref.~\cite{PhysRevLett.123.131603} arose from the large-c limit, here, single block dominance arises from the light cone limit, so $\beta\rightarrow 0$ is crucial for us to make the identification between negativity and reflected entropy.

%%%%%%%%%%%%%%%%%%%%%%%%%%%%%%%%%%%%%%%%%%%%%%%%%%%%%%%%%%%%%%%%%%%%%%%%%%%%%%%%%%%%%%%%%%%%%%
%%%%%%%%%%%%%%%%%%%%%%%%%%%%%%%%%%%%%%%%%%%%%%%%%%%%%%%%%%%%%%%%%%%%%%%%%%%%%%%%%%%%%%%%%%%%%%
\section{Holographic CFTs} \label{holo_sec}
%%%%%%%%%%%%%%%%%%%%%%%%%%%%%%%%%%%%%%%%%%%%%%%%%%%%%%%%%%%%%%%%%%%%%%%%%%%%%%%%%%%%%%%%%%%%%%
%%%%%%%%%%%%%%%%%%%%%%%%%%%%%%%%%%%%%%%%%%%%%%%%%%%%%%%%%%%%%%%%%%%%%%%%%%%%%%%%%%%%%%%%%%%%%%

Conformal field theories that admit weakly coupled gravity duals may be considered as the irrational theories considered in the previous section in the limit of large central charge. We found that in the light cone limit, the results were universal for irrational CFTs in that they only depend on an overall proportionality of the central charge and are agnostic to the specific operator content of the theories. In this section, we generalize these results beyond the light cone limit i.e.~to finite $\beta$ by performing calculations directly in the gravity theory. We find precise agreement in the $\beta \rightarrow 0$ limit. Given this consistency check, we extend our analysis of irrational CFTs to both local joining and inhomogeneous global quenches. We find large deviations from the quasi-particle results from Sec.~\ref{univ_sec} due to the strong scrambling behavior of holographic CFTs.

\begin{figure}
    \centering
    \includegraphics[width = \textwidth/2]{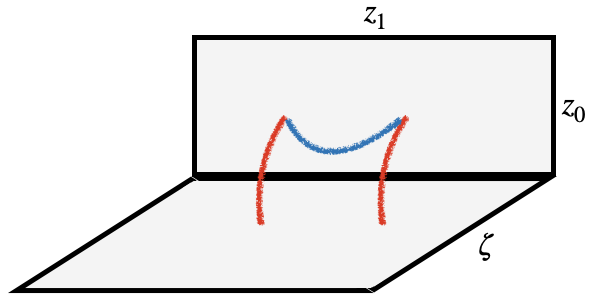}
    \caption{The extremal surface can either be in the connected regime (blue) or disconnected regime (red) where it terminates on the EoW brane.}
    \label{single_int}
\end{figure}

Fortunately, a simple way to compute any geometrized entanglement measure (e.g.~von Neumann entropy \cite{2006PhRvL..96r1602R,2006JHEP...08..045R,2007JHEP...07..062H,2018NatPh..14..573U,2018JHEP...01..098N,PhysRevLett.122.141601,2019arXiv190500577D,2019PhRvD..99j6014K,PhysRevLett.123.131603}) holographically when the correlation function involves only twist-fields on the upper half plane has been developed  \cite{2012JHEP...12..027R,2013arXiv1311.2562U,2016arXiv160407830M}. This is done by considering the proposed holographic dual to boundary conformal field theories \cite{2011PhRvL.107j1602T, 2011JHEP...11..043F}. The bulk dual to the upper half plane is simply half of empty $AdS$. There exists an end-of-world (EoW) brane at $t=0$ where Neumann boundary conditions for the metric must be imposed. The extremal surfaces may end on this EoW brane. We demonstrate the two possible configurations for the extremal surface homologous to a single interval in Fig.~\ref{single_int}. 

We start with the $AdS_3$ metric in Poincar\'e coordinates
\begin{align}
    ds^2 = \frac{d\zeta^2 + dz_+ dz_-}{\zeta^2}
\end{align}
where $z_{\pm} = z_1 \mp z_0$ are light cone coordinates on the asymptotic boundary. For boundary points $P_1 = (z_{+,1},z_{-,1})$ and $P_2 = (z_{+,2},z_{-,2})$, the two configurations have areas
\begin{align}
    \gamma_{con} =  \log \frac{{(z_{+,1}-z_{+,2})(z_{-,1}-z_{-,2})}}{\zeta_{min}^2}, \quad  \gamma_{dis} = \frac{1}{2}\log \frac{{z_{+,1}-z_{-,1}}}{\zeta_{min}^2}+ \frac{1}{2}\log\frac{{z_{+,2}-z_{-,2}}}{\zeta_{min}^2}
    \label{areas}
\end{align}
where $\zeta_{min}$ is the UV cutoff.

Of course, the CFT path integrals that we prepare are conformal maps from the upper half plane, so we must discuss how this mapping extends into the bulk.
We consider the (Lorentzian) conformal map which takes the original manifold to the upper half plane
\begin{align}
    z_{\pm} = f_{\pm}(x_{\pm}).
    % = e^{2\pi x_{\pm}/\beta}.
\end{align}
This can be shown to extend into the bulk as a large diffeomorphism \cite{2012JHEP...12..027R}
\begin{align}
    \zeta = 4z\frac{(f'_+ f'_-)^{3/2}}{4 f_+'f_-'+z^2f_+'' f_-''}, \quad 
    z_{\pm} = f_{\pm} -\frac{2z^2f_{\pm}'^2f_{\mp}''}{4 f_+'f_-'+z^2f_+'' f_-''}.
    \label{roberts_diff}
\end{align}

 In order to compute the entanglement entropy after the quench, we evaluate each term of \eqref{areas} using (\ref{roberts_diff}) at the boundary
\begin{align}
    z_{\pm} = f_{\pm}(x_{\pm}), \quad \zeta_{min} = a\sqrt{f_+'(x_{+})f_-'(x_{-})}
\end{align}
and take the minimum of the connected and disconnected configurations. We have introduced $a$ as the UV cutoff in the original theory. In general, the cutoff is transformed differently for the two different boundary points, so one needs to treat each separately. 

It is clear how we must proceed in order to evaluate the entanglement wedge cross section after quantum quenches. We repeat the above procedure, only changing the expressions of the areas of extremal surfaces (\ref{areas}). An additional complication is that we must consider more complicated configurations of intervals because $E_W$ is inherently a mixed state correlation measure. This leads to many more possible configurations of the entanglement wedge\footnote{We note that while the holographic correspondence between reflected entropy and the entanglement wedge cross-section has not been explicitly derived for dynamical spacetimes, it has passed several nontrivial tests \cite{2019arXiv190706646K,2019arXiv190906790K}. The results in this paper provide further consistency checks. We see no obstruction to a covariant derivation \'a la Ref.~\cite{2016JHEP...11..028D}.}.

Our basic building block for computing $E_W$ is the entanglement wedge cross section for two disjoint intervals in vacuum \cite{2018NatPh..14..573U}
\begin{align}
    E_W = \frac{c}{12} \log \left(\frac{1+\sqrt{x}}{1-\sqrt{x}} \right)+\frac{c}{12} \log \left(\frac{1+\sqrt{\bar{x}}}{1-\sqrt{\bar{x}}} \right), \quad x = \frac{(x_3-x_1)(x_4-x_2)}{(x_2-x_1)(x_4-x_3)}, \quad x^* = \bar{x}. 
    \label{EW_block}
\end{align}
For the various configurations, we will just need to determine the relevant boundary points and analytic continuations.

\subsection{Adjacent intervals}

\begin{figure}
    \centering
    \includegraphics[width = .24\textwidth]{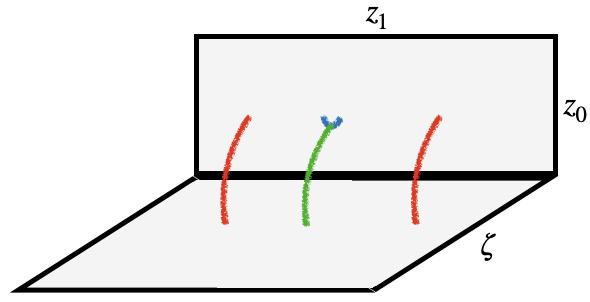}
    \includegraphics[width = .24\textwidth]{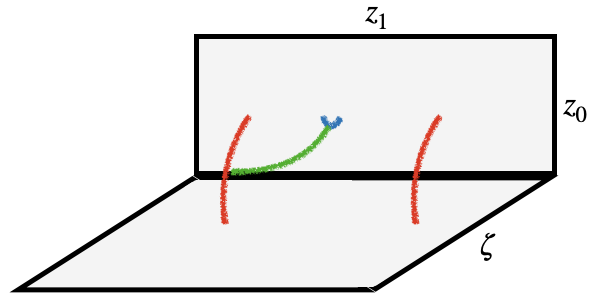}
    \includegraphics[width = .24\textwidth]{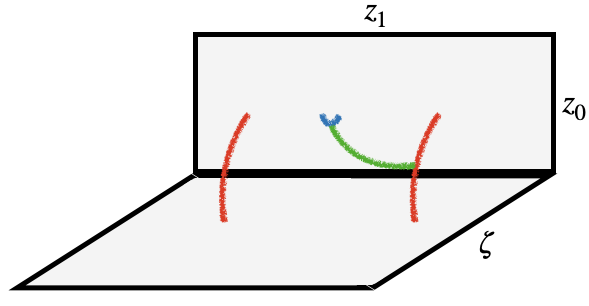}
    \includegraphics[width = .24\textwidth]{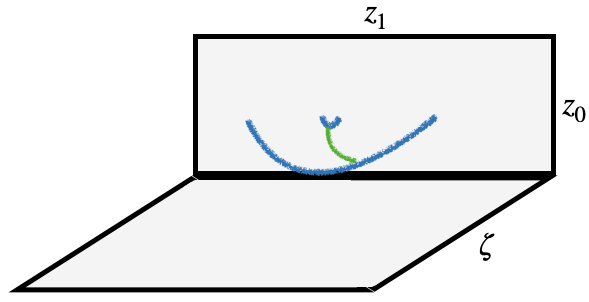}
    \caption{The four possible phases (I-IV left to right) of $E_W$ (green) for adjacent intervals. We exaggerate the distance between the intervals to emphasize that we need to treat the cutoff carefully.}
    \label{adj_configs}
\end{figure}

We begin with the simplest nontrivial case, adjacent intervals. 
There are four phases of $E_W$ that we must consider, enumerated in Fig.~\ref{adj_configs}. To determine which is dominant, we first perform the standard extremization to find the entanglement wedge. This consists of determining which geodesic in (\ref{areas}) is smaller. If we are in the disconnected phase, there are three configurations of $E_W$ (phases I, II, and III) that we must minimize over. If we are in the connected phase, $E_W$ must be in phase IV. Phase IV determines the late-time behavior.

\paragraph{Phase I} 

In Phase I, the extremal surface is in the disconnected regime and $E_W$ is connected to the EoW brane. Gravitationally, this channel is the simplest and similar to the disconnected holographic entanglement entropy discussed in Refs.~\cite{2013arXiv1311.2562U,2016arXiv160407830M}, though we need to be careful with the cutoff. 

The four boundary points are the position of the midpoint, ${x}_2$, another point an infinitesimal distance, $a$, away, $x'_2$, and their ``image points" reflected across the EoW brane. In this case,
\begin{align}
    x = \bar{x} = \frac{\left(f_+(\tilde{x}_+)+f_-(\tilde{x}_-) \right)\left(f_+((\tilde{x}+a)_+)+f_-((\tilde{x}+a)_-) \right)}{\left(f_+(\tilde{x}_+)+f_-((\tilde{x}+a)_-) \right)\left(f_+((\tilde{x}+a)_+)+f_-(\tilde{x}_-) \right)}.
\end{align}
Furthermore, we only take half of \eqref{EW_block} because $E_W$ connects to the EoW brane, not all the way to the image points. 

Let us evaluate this for the global quench
\begin{align}
    x = \frac{\cosh^2\left( \frac{2 \pi t}{\beta}\right)}{\cosh\left( \frac{2 \pi \left(t-\frac{d}{2}\right)}{\beta}\right)\cosh\left( \frac{2 \pi \left(t+\frac{d}{2}\right)}{\beta}\right)}
\end{align}
where the distance between the points, $d$, has thus far been left arbitrary.
In the adjacent intervals limit ($d\rightarrow a$), this becomes
\begin{align}
    x = 1 - \left(\frac{\pi a}{\beta \cosh \left( \frac{2\pi t}{\beta}\right)} \right)^2.
\end{align}
Furthermore, we can take the high temperature limit to find
\begin{align}
    E_W = \frac{c}{6}\left(\log \left( \frac{\beta}{\pi a}\right) + \frac{2 \pi t}{\beta} \right)
\end{align}
showing the early-time linear growth of $E_W$.

\paragraph{Phases II \& III} 

In phase II (III), the entanglement wedge cross section connects to the extremal surface. In this case, the relevant four boundary point are $x_1$ ($x_3$), $x_2$, $x_2'$, and $x_1$'s ($x_3$'s) image point.

After analytic continuation, the cross-ratios are
\begin{align}
    x = \frac{(f(x_{1+}) - f(x_{2+}))(f(x_{2+}') + f(x_{1-}))}{(f(x_{1+}) - f(x'_{2+}))(f(x_{2+}) + f(x_{1-}))}, \quad \bar{x} = \frac{(f(x_{1-}) - f(x_{2-}))(f(x_{2-}') + f(x_{1+}))}{(f(x_{1-}) - f(x'_{2-}))(f(x_{2-}) + f(x_{1+}))}.
    \label{cross_phaseII}
\end{align}
An analogous statement can be made for phase III.

Let us explicitly derive the phase II for the global quench. From \eqref{glob_quench_map} and \eqref{cross_phaseII}, we find
\begin{align}
    x = \frac{\sinh \left(\frac{\pi l_1}{\beta} \right)\cosh \left(\frac{2\pi \left(t + \frac{l_1+d}{2}\right)}{\beta} \right) }{\sinh \left(\frac{\pi (l_1+d)}{\beta} \right)\cosh \left(\frac{2\pi \left(t + \frac{l_1}{2}\right)}{\beta} \right)}, \quad  \bar{x} = \frac{\sinh \left(\frac{\pi l_1}{\beta} \right)\cosh \left(\frac{2\pi \left(t - \frac{l_1+d}{2}\right)}{\beta} \right) }{\sinh \left(\frac{\pi (l_1+d)}{\beta} \right)\cosh \left(\frac{2\pi \left(t - \frac{l_1}{2}\right)}{\beta} \right)}
\end{align}
In the adjacent intervals limit, $d\rightarrow a$, this becomes
\begin{align}
    x = 1 - \frac{a \pi}{\beta}\frac{\cosh \left(\frac{2\pi t}{\beta} \right) }{\sinh \left(\frac{\pi l_1}{\beta} \right)\cosh \left(\frac{2\pi \left(t + \frac{l_1}{2}\right)}{\beta} \right)}, \quad x = 1 - \frac{a \pi}{\beta}\frac{\cosh \left(\frac{2\pi t}{\beta} \right) }{\sinh \left(\frac{\pi l_1}{\beta} \right)\cosh \left(\frac{2\pi \left(t - \frac{l_1}{2}\right)}{\beta} \right)}.
\end{align}
Furthermore, if we wish to compare to the pure CFT results from the light cone limit, we must take $\beta\rightarrow 0 $, in which case
\begin{align}
    x = 1- \frac{2\pi a}{\beta} e^{-2\pi l_1/\beta},\quad \bar{x} =\begin{cases} 1- \frac{2\pi a}{\beta} e^{-4\pi (t-l_1/2)/\beta}, & t < l_1/2
    \\
    1- \frac{2\pi a}{\beta} , & t > l_1/2
    \end{cases}.
\end{align}
This phase is only dominant for $t> l_1/2$, so we find
\begin{align}
    E_W = \frac{c}{6}\left(\log \left(\frac{2\beta}{\pi a} \right)+ \frac{\pi l_1}{\beta}  \right).
\end{align}
Analogously, for phase III, we find
\begin{align}
    E_W = \frac{c}{6}\left(\log \left(\frac{2\beta}{\pi a} \right)+ \frac{\pi l_2}{\beta}  \right).
\end{align}

\paragraph{Phase IV} 
The final phase of $E_W$ for adjacent intervals does not involve the EoW brane. Rather, it is the standard $E_W$ for disjoint intervals. 

In Phase IV, the extremal surface is in the connected regime and $E_W$ connects orthogonally to the surface. We use a different cross-ratio such that $E_W$ is \cite{2018NatPh..14..573U}
\begin{align}
    E_W = \frac{c}{6}\log \left(1+2z+2\sqrt{z(z+1)} \right) , \quad z = \frac{\mathcal{L}\left[x, x_1\right]\mathcal{L}\left[x_2, x+a \right]}{\mathcal{L}\left[x+a, x\right]\mathcal{L}\left[x_2, x_1\right]},
    \label{cross_phaseIV}
\end{align}
where
\begin{align}
    \mathcal{L}[x_2,x_1] = \sqrt{(f(x_{+,2})-f(x_{+,1}))(f(x_{-,2})-f(x_{-,1}))}.
\end{align}
In the adjacent intervals limit that we are concerned with
\begin{align}
    E_W \simeq \frac{c}{6}\log \left( 4z \right).
\end{align}

For the global quench, we find
\begin{align}
    z = \frac{\sinh\left(\frac{\pi l_1}{\beta} \right)\sinh\left(\frac{\pi l_2}{\beta} \right)}{\sinh\left(\frac{\pi d}{\beta} \right)\sinh\left(\frac{\pi (l_1+l_2+d)}{\beta} \right)}
\end{align}
In the adjacent intervals limit, this becomes
\begin{align}
    z =\frac{\beta} {\pi a}\frac{\sinh\left(\frac{\pi l_1}{\beta} \right)\sinh\left(\frac{\pi l_2}{\beta} \right)}{\sinh\left(\frac{\pi (l_1+l_2)}{\beta} \right)}
\end{align}
Furthermore, in the high temperature limit, this is
\begin{align}
    z =\frac{\beta}{2\pi a} 
\end{align}
which leads to 
\begin{align}
    E_W = \frac{c}{6}\log\left(\frac{2\beta}{\pi a} \right)
\end{align}
Alternatively, we can take the $\beta\rightarrow0$ limit without the adjacent intervals limit. In this case
\begin{align}
    z = e^{-4\pi d/\beta}
\end{align}
Because $z$ is exponentially small small, the entanglement wedge is disconnected, so $E_W = 0$.

\subsection{Disjoint intervals}

\begin{figure}
    \begin{tabular}{cccc}
    & {Adjacent} & {Disjoint (close)}  & {Disjoint (far)}
\\
\rotatebox{90}{\hspace{0.1in} {Global (hom.)}} 
  &  \includegraphics[height = 2.75cm]{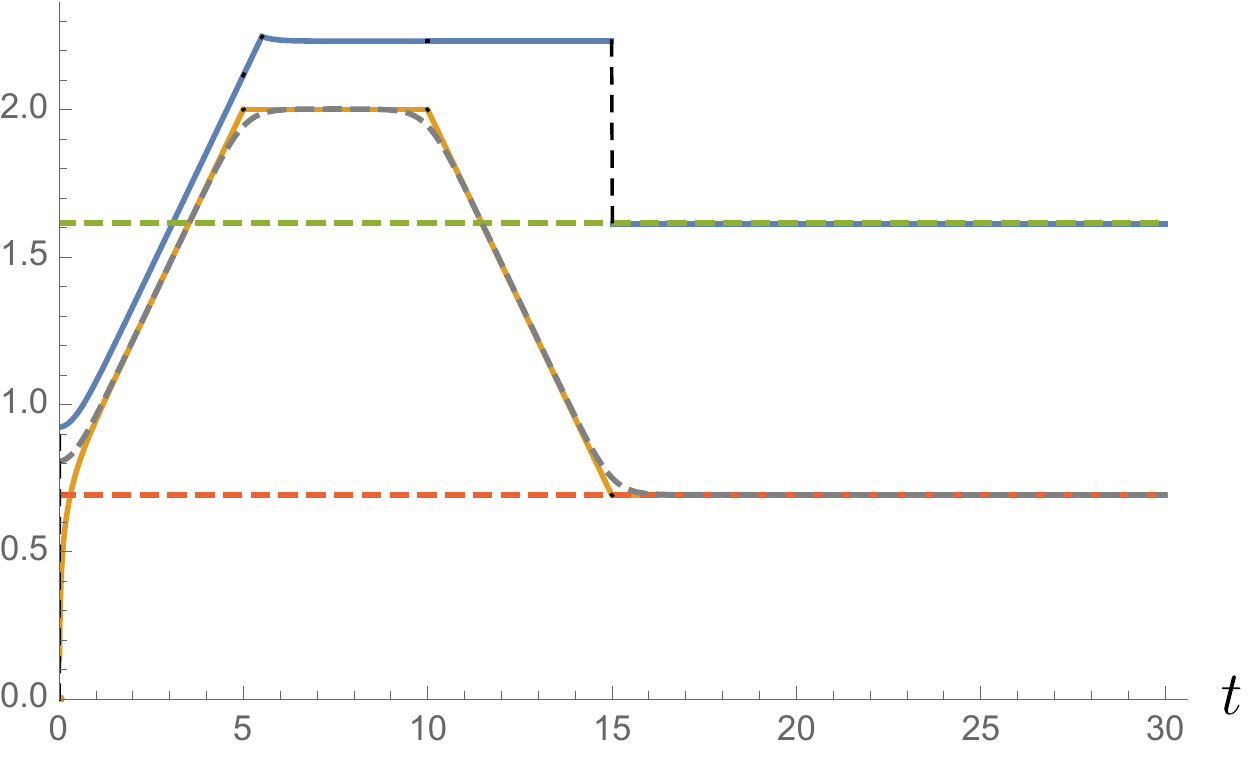} &
    \includegraphics[height = 2.75cm]{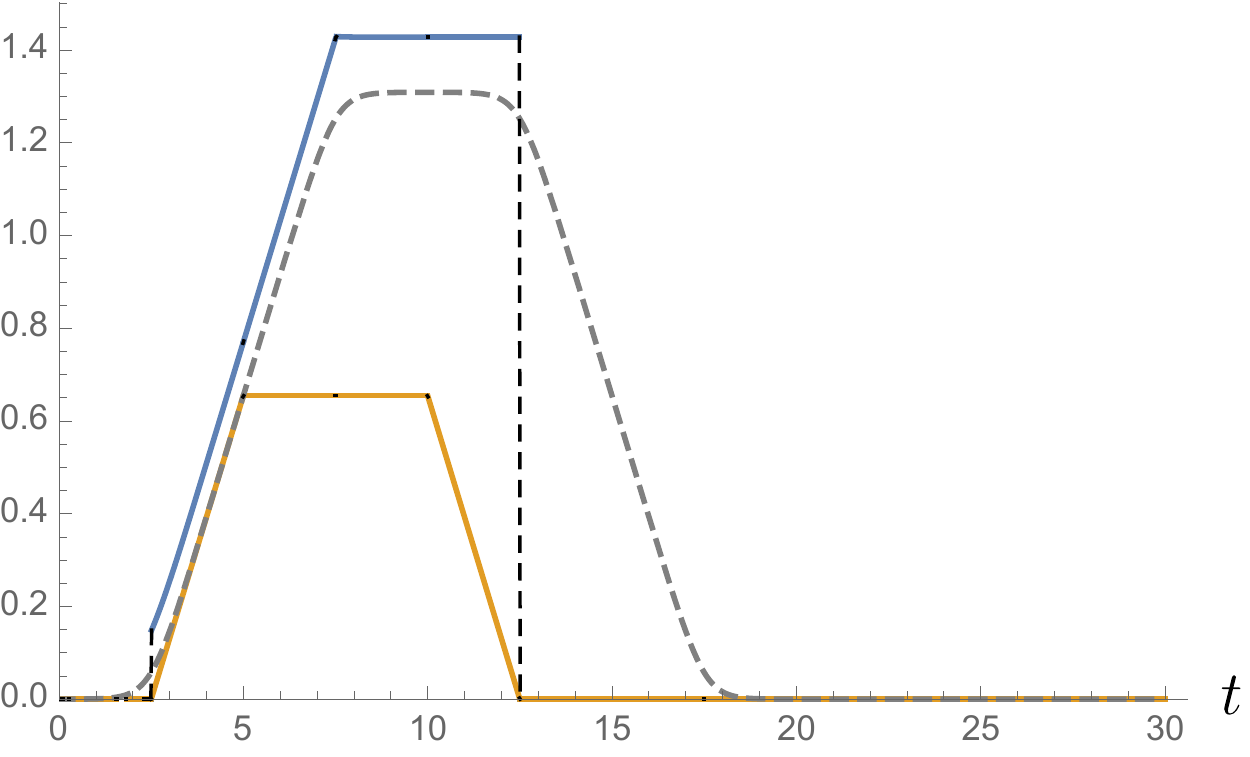} &
    \includegraphics[height =2.75cm]{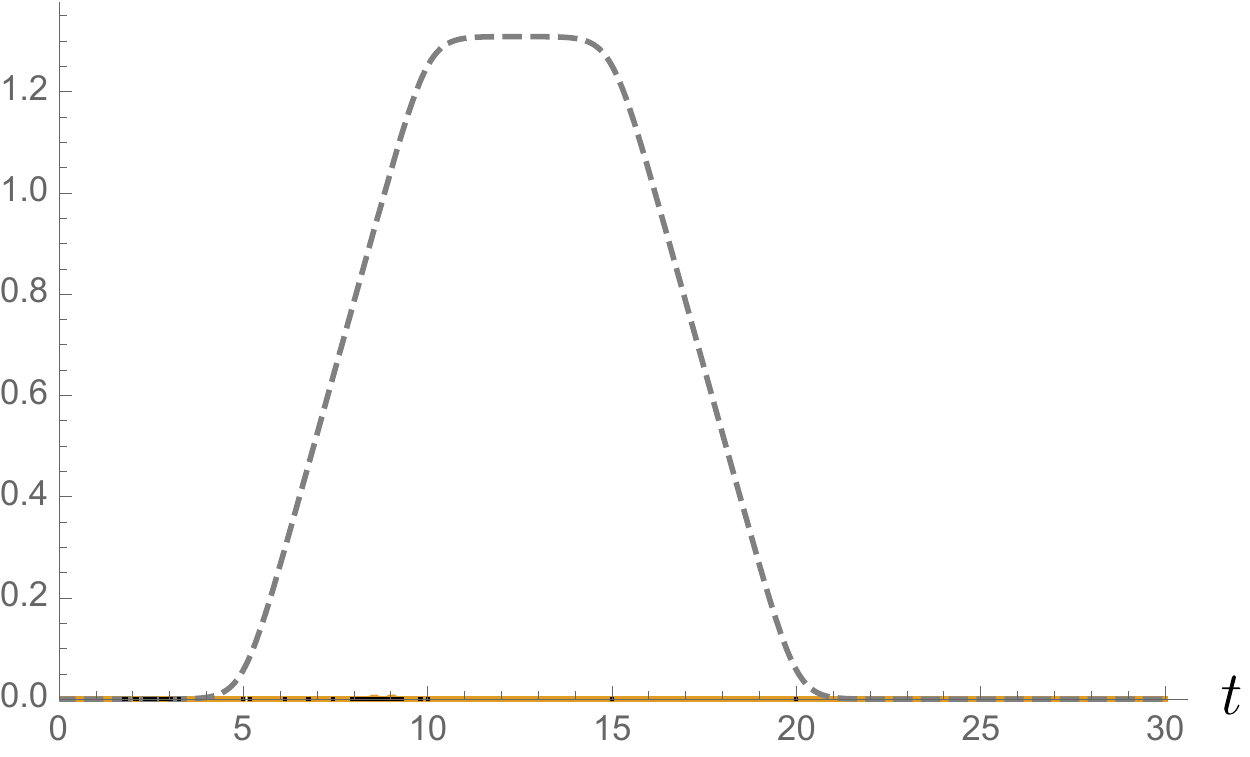} \\   
\rotatebox{90}{\hspace{-0.0in} {Global (inhom.)}}    &
 \includegraphics[height =2.75cm]{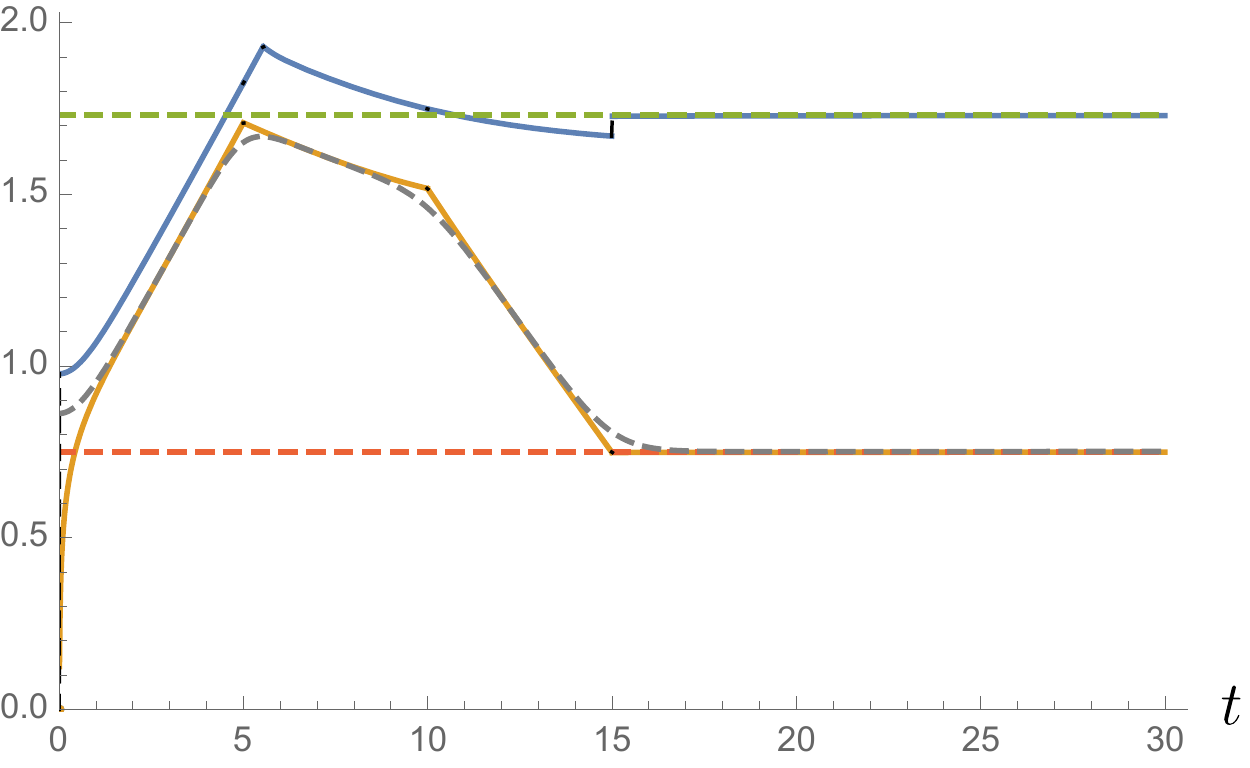} & 
 \includegraphics[height =2.75cm]{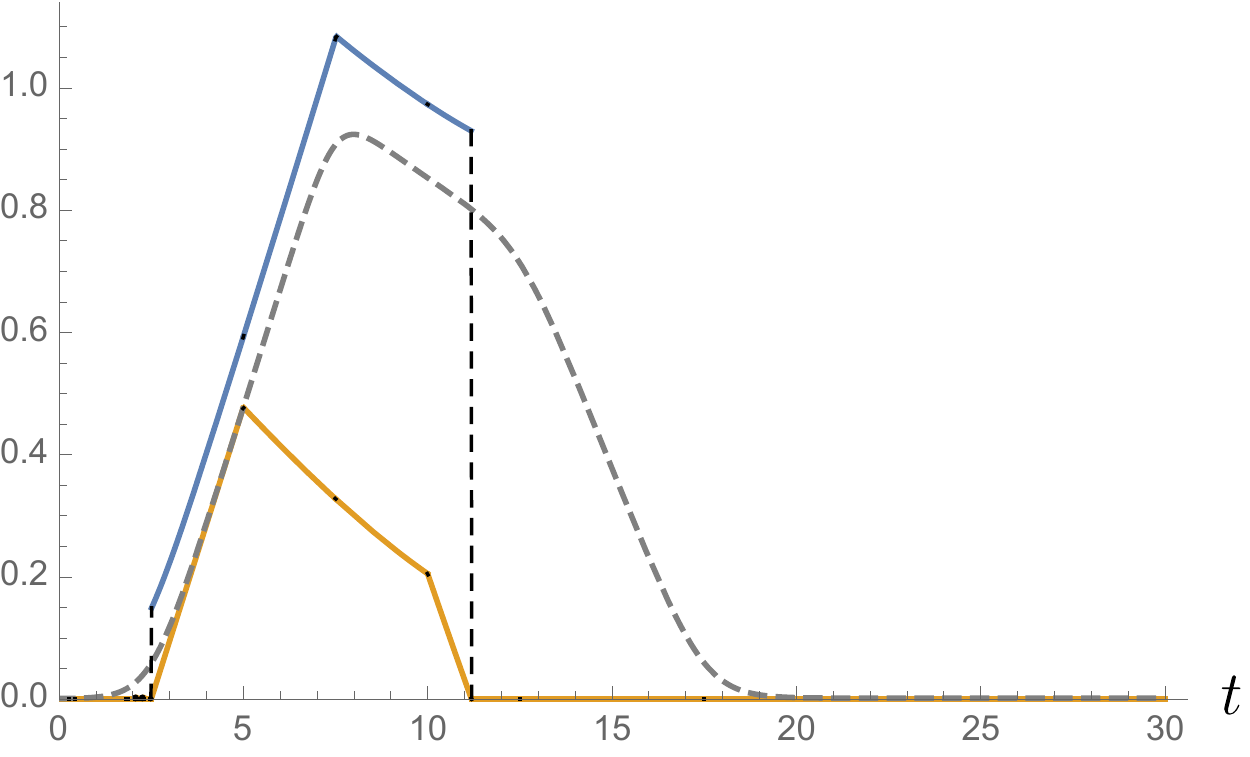} &
    \includegraphics[height =2.75cm]{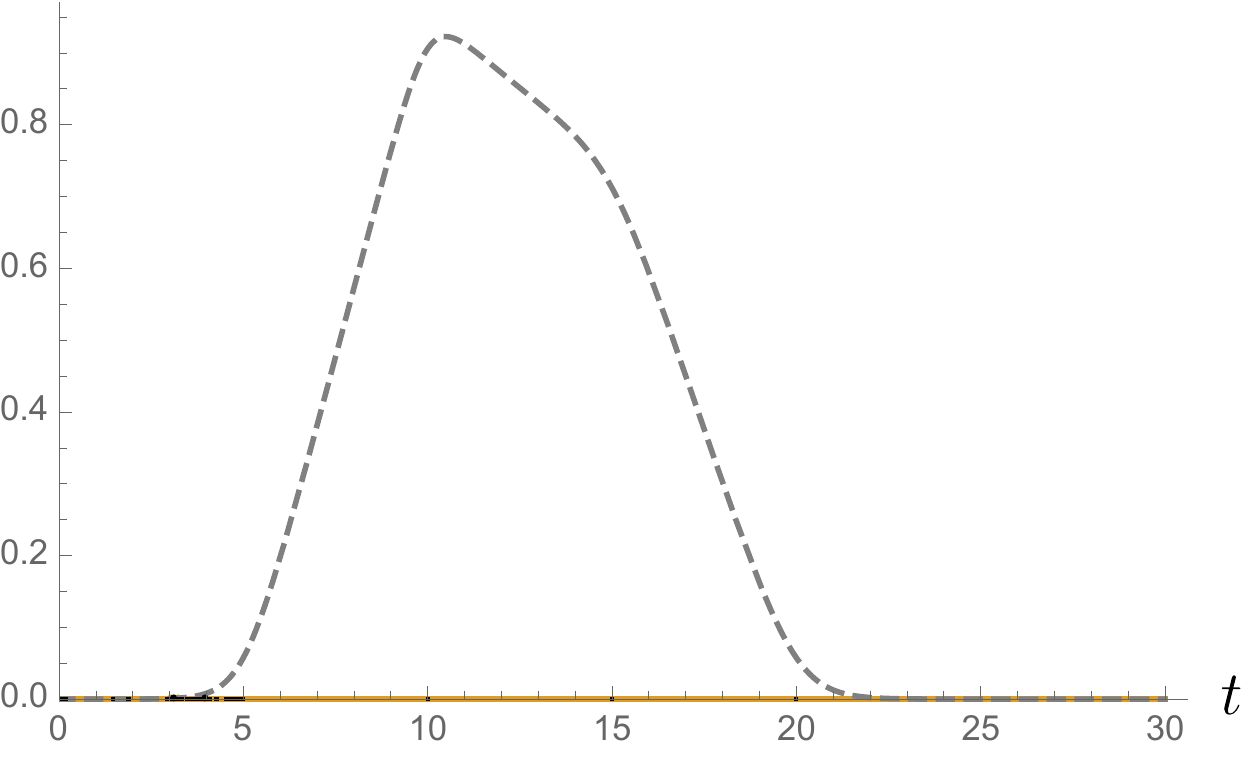} 
       \\ 
\rotatebox{90}{\hspace{0.35in} {Local}}    &
 \includegraphics[height =2.75cm]{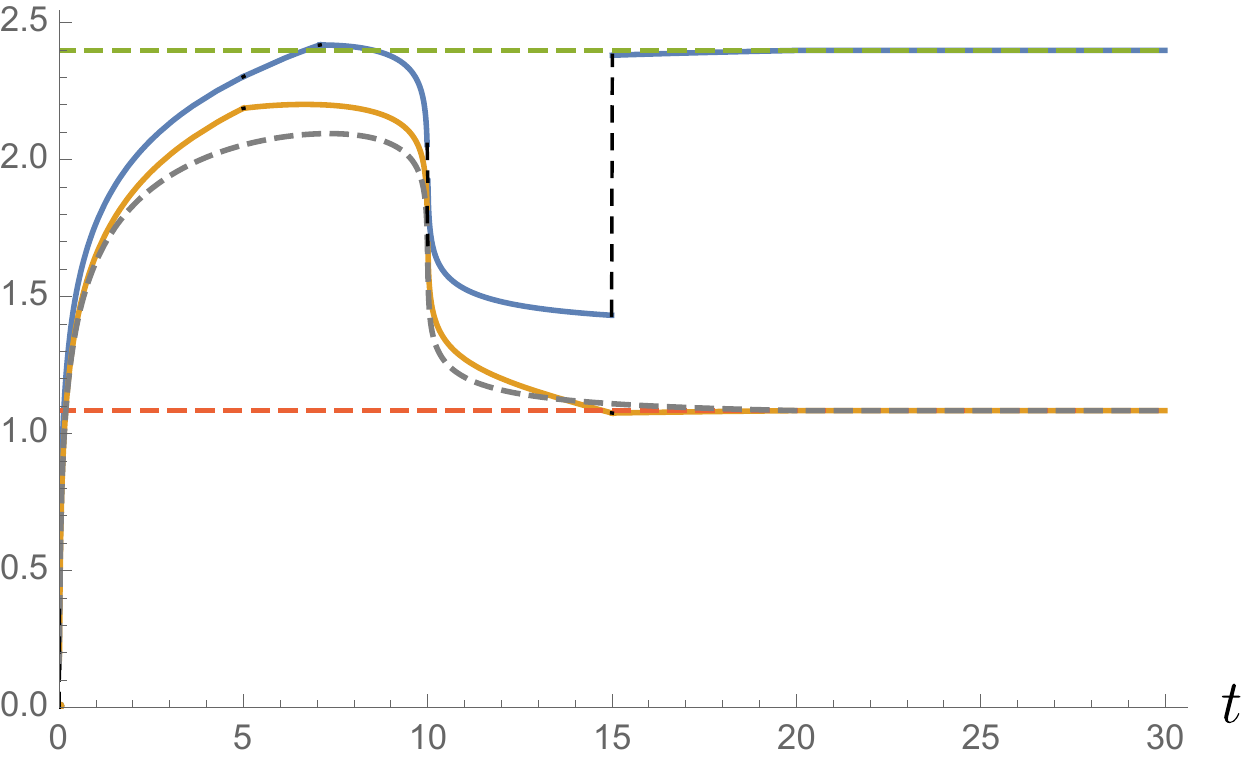} & 
 \includegraphics[height =2.75cm]{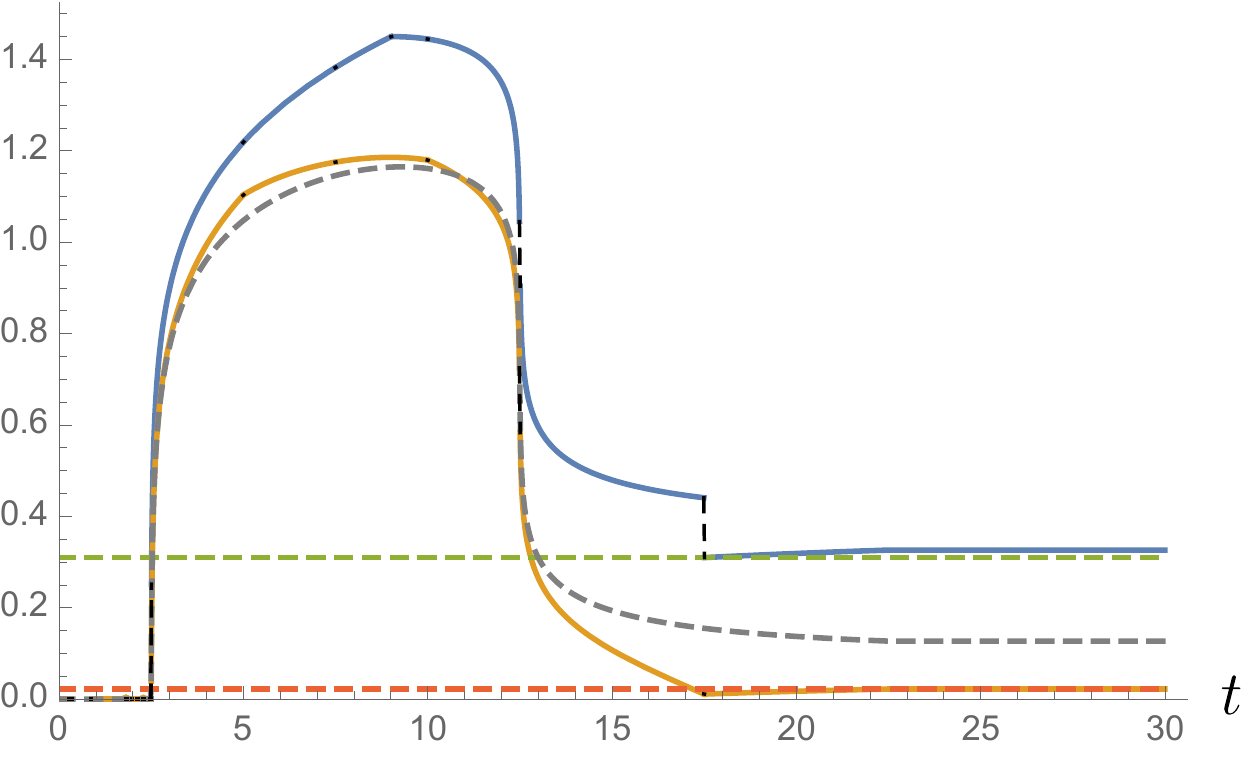} &
   \includegraphics[height =2.75cm]{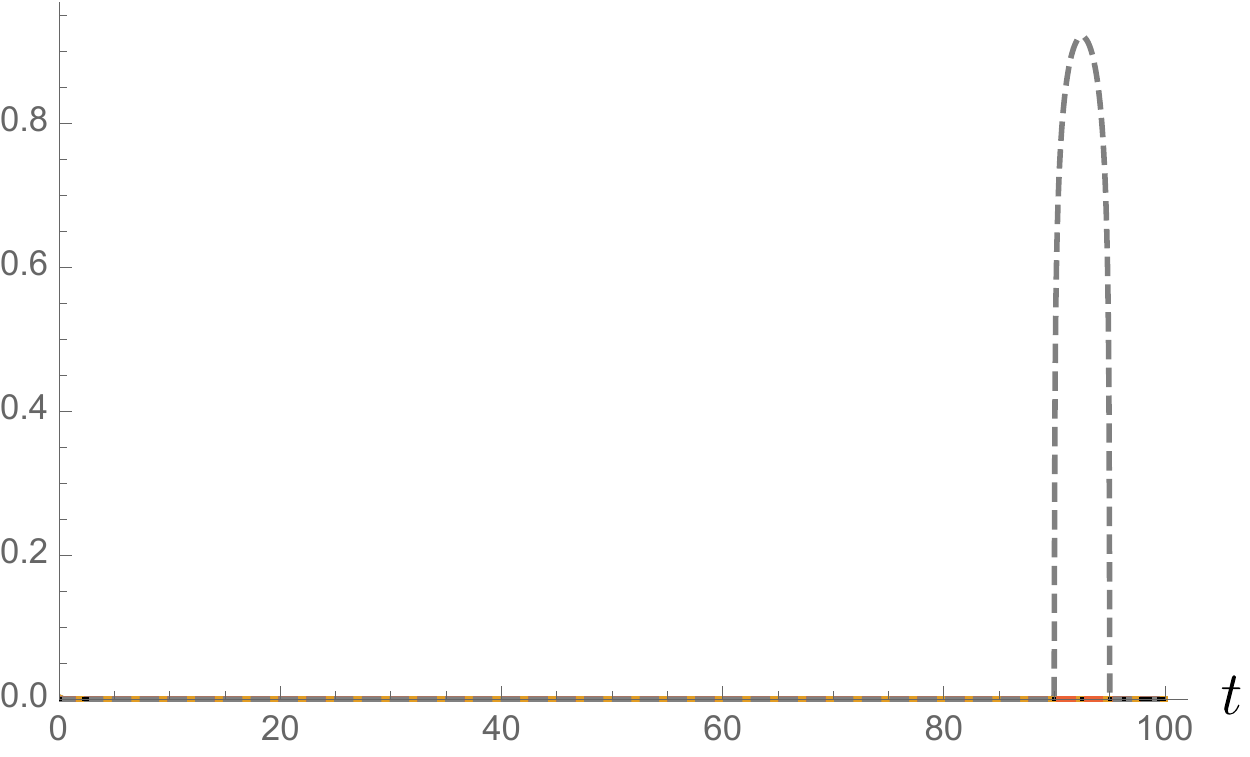} 
    \end{tabular}
    \caption{Half of mutual information (orange), $E_W$ (blue), universal result from (\ref{EW_quasi}) (grey dotted). For the global quenches, the red and green dotted lines are the thermal values. For the local quench, they are the ground state values. For the local quench, we need to take the intervals to be very far away and $\epsilon$ finite in order for $E_W$ to vanish at all times. This behavior was observed for mutual information in Ref.~\cite{2014PhRvD..89f6015A}.}
    \label{holo_quenches}
\end{figure}

We can generalize this study to disjoint intervals by considering all other configurations of $E_W$. It turns out that all of these give trivial $E_W$. Still, we must check if any of these configurations are dominant, in which case, $E_W$ will discontinuously jump to $0$. We have thus found $E_W$ after global quenches for disjoint intervals with completely generic parameters, generalizing the light cone limit results. We collect the results here in the high-temperature limit to be succinct. For adjacent intervals ($d \ll \beta$), we have
\begin{align}
    E_W = \begin{cases} 
     \frac{c}{6}\left(\log \left( \frac{\beta}{\pi a}\right) + \frac{2 \pi t}{\beta} \right) & t < \frac{\min[l_1, l_2]}{2}
     \\
    \frac{c}{6}\left(\log \left(\frac{2\beta}{\pi a} \right)+ \frac{\pi \min[l_1, l_2]}{\beta}
    \right), & \frac{\min[l_1, l_2]}{2} < t < \frac{l_1+l_2}{2}
    \\
    \frac{c}{6}\log\left(\frac{2\beta}{\pi a} \right), & t > \frac{l_1+l_2}{2}
    \end{cases}.
\end{align}
For disjoint intervals ($d \gg \beta$),
\begin{align}
    E_W = \begin{cases} 
    0 & t < \frac{d}{2} 
    \\
    \frac{\pi c}{3\beta}\left(t-\frac{d}{2} \right)+\frac{c}{6}\log 2 & \frac{d}{2} < t < \frac{l_1+d}{2}, \frac{l_2+d}{2}
    \\
    \frac{\pi c}{6\beta}\min[l_1,l_2]  +\frac{c}{6}\log 2 & \frac{\min[l_1,l_2]+d}{2}, < t < \frac{l_1+l_2-d}{2}
    \\
    0 & t > \frac{l_1+l_2-d}{2}
    \end{cases}.
\end{align}
This perfectly matches the CFT results \eqref{eq:SRresult} and \eqref{eq:SRresult_adjacent}.

Now that we have a reliable method for computing $E_W$ holographically, we can apply this formalism to inhomogenous global quenches and local joining quenches. The analytic expressions are straightforward to find but unenlightening. Instead, we plot representative cases in Fig.~\ref{holo_quenches}. We find clear violations of quasi-particle behavior. In particular, interesting intermediate behavior between quasi-particle picture and maximal scrambling occurs when we take the intervals to be disjoint but close. This is the ``missing entanglement" and ``mysterious correlations" discuessed earlier. We will investigate this using holographic mutual information in the following section.

\subsection{An intermediate phase of scrambling}

For the global quench, we analytically compute the lengths of the geodesics of all possible phases of the entanglement wedge for disjoint intervals. Luckily, all phases are composed of the two building blocks of disconnected and connected regimes seen in the computation for the single interval. The geodesic length for the connected regime is constant in time
\begin{align}
    \gamma_{con}^{ij} =  2\log \left[ \frac{\beta}{\pi \epsilon } \sinh \left(\frac{\pi}{\beta} |x_i - x_j| \right)\right]
\end{align}
which corresponds to the well-known finite temperature entanglement entropy of single interval once the $1/4 G_N$ factor is restored. In the limit of $l_1, l_2,d \gg \beta$, this reduces to
\begin{align}
    \gamma_{con}^{ij} =\frac{2\pi}{\beta} |x_i - x_j| + 2 \log \left[ \frac{\beta}{2\pi \epsilon }\right]
\end{align}
Thus, the late-time behavior should be captured by the connected regime if thermalization is to occur. On the other hand, the disconnected regime is time-dependent 
\begin{align}
    \gamma_{dis}^{ij} = 2\log \left[\frac{\beta}{\pi \epsilon} \sinh \left( \frac{2\pi t}{\beta}\right) \right].
\end{align}
The position-independence is consistent with the translational invariance of the problem at early times that are much smaller than the size of the subsystems. In the limit that $t \gg \beta$
\begin{align}
    \gamma_{dis}^{ij} = \frac{4\pi t}{\beta}  + 2 \log \left[ \frac{\beta}{2\pi \epsilon }\right].
\end{align}

Our job is then to minimize over all possible combinations of connected and disconnected geodesics
\begin{align}
    S_{A\cup B} &= \frac{1}{4G_N} \min [\gamma_{con}^{12}+\gamma_{con}^{34},\gamma_{con}^{13}+\gamma_{con}^{24},\gamma_{con}^{14}+\gamma_{con}^{23},\gamma_{con}^{12}+\gamma_{dis}^{34}   
    \nonumber
    \\
    &,\gamma_{con}^{13}+\gamma_{dis}^{24},\gamma_{con}^{14}+\gamma_{dis}^{23}
    ,\gamma_{dis}^{12}+\gamma_{con}^{34},\gamma_{dis}^{13}+\gamma_{con}^{24},\gamma_{dis}^{14}+\gamma_{con}^{23},\gamma_{dis}^{12}+\gamma_{dis}^{34}]
    \label{dis_options}
    \\
    &\simeq
    \frac{\pi c}{3\beta}\left(\min\left[ \min[2t,l_1] +\min[2t,l_2],\min[2t,l_1+d+l_2] +\min[2t,d]\right] \right)+ 4\log\left[\frac{\beta}{2\pi \epsilon} \right]
\end{align}
In the second line, we have taken all length scales to be much larger than $\beta$. Meanwhile, the entanglement entropy of the individual subsytems only has four options
\begin{align}
    S_{A }+S_B &= \frac{1}{4G_N} \min [\gamma_{con}^{12}+\gamma_{con}^{34},\gamma_{con}^{12}+\gamma_{dis}^{34}   
    ,\gamma_{dis}^{12}+\gamma_{con}^{34},\gamma_{dis}^{12}+\gamma_{dis}^{34}]
    \\
    &\simeq
    \frac{\pi c}{3\beta}\left(\min[2t,l_1] +\min[2t,l_2] \right)+ 4\log\left[\frac{\beta}{2\pi \epsilon} \right].
\end{align}
If any of these are dominant in (\ref{dis_options}), then the mutual information and the entanglement-wedge cross-sections are trivial. The analysis in Ref.~\cite{2015JHEP...09..110A} suggests that one of these four configurations is always dominant. This would mean that 
\begin{align}
    \min[2t,l_1+l_2+d] +\min[2t,d] < \min[2t,l_1] +\min[2t,l_2]
\end{align}
One can easily check that this is not always true. More precisely, there will be finite mutual information and a finite entanglement wedge cross-section when $d < \min[l_1, l_2]$. The mutual information is
\begin{align}
    I = \begin{cases}
    0 & 2t < d, 2t > l_1 +l_2-d
    \\
    \frac{\pi c}{3\beta}  (2t-d) &d < 2t< l_1,l_2
    \\
    \frac{\pi c}{3\beta} (l_1-d) & d,l_1<2t<l_2
    \\
    \frac{\pi c}{3\beta} (l_2-d) & d,l_2<2t<l_1
    \\
    \frac{\pi c}{3\beta} (l_1+l_2-d-2t) & d,l_2,l_1<2t<l_1+l_2-d
    \end{cases}
\end{align}
which describes the missing entanglement. We note similar intermediate behavior has been numerically observed in Vaidya spacetimes \cite{2015JHEP...09..114Z,2019JHEP...01..114Y}.

\section{Line-tension picture}
\label{randU_sec}

The dynamics of correlations in irrational CFTs (including holographic CFTs) is unintuitive and sharply contrasts the quasi-particle picture.
We now try to gain intuition for these results by appealing to the ``line-tension picture" \cite{2017PhRvX...7c1016N,2018arXiv180300089J,2018PhRvD..98j6025M,2018PhRvX...8b1013V,2019PhRvB..99q4205Z}, an effective description of entanglement \textit{production} that is conjectured to universally describe chaotic quantum systems analogous to the quasi-particle picture for integrable systems which describes entanglement \textit{spreading}. 
While proposed originally 
in the context of random unitary circuit models,
the line-tension picture 
has been shown to apply 
to holographic CFTs
and correctly reproduces 
the operator entanglement 
entropy and negativity of 
the time evolution operator
\cite{2019arXiv190607639K}.

In the line-tension picture,
the entanglement entropy may be phenomenologically described in terms of the ``energy" of an extremal codimension-one membrane homologous to the boundary region
\begin{align}
    S(x,t) = \int_{\mathcal{M}} \mathcal{T}(v)
    \label{linetension_eq}
\end{align}
where $\mathcal{T}(v)$ is the tension of the membrane. This line tension function in theory-dependent, though it has been argued to satisfy certain constraints. It has been analytically and numerically solved for in various systems \cite{2017PhRvX...7c1016N,2018arXiv180300089J,2019PhRvB..99q4205Z}. An analogous line-tension picture for logarithmic negativity was proposed in Ref.~\cite{2019arXiv190607639K} in which the same line tension function is used, but the membrane is now the extremal cross-section of the ``entanglement wedge" formed by the membrane in \eqref{linetension_eq}\footnote{We note that a similar construction for the operator entanglement of the reduced density matrix in Ref.~\cite{2019arXiv190709581W}. In fact, they were more precise about the origins of an $E_W$-like object in random unitary circuits by appealing to an effective statistical mechanics model where one considers the free energy of domain walls. An analogous procedure should apply here for negativity and reflected entropy. However, we note that the two protocols lead to slightly different results because the quantity in Ref.~\cite{2019arXiv190709581W} requires a minimization over the sum of entropy and $E_W$ because they study the operator state $\ket{\rho}$ rather than $\ket{\rho^{1/2}}$ needed for reflected entropy. 
Taking the replica limit to $\ket{\rho^{1/2}}$ is technically quite challenging, but we expect the contribution from the entropy drops out, leaving only $E_W$. In their notation, when $\alpha \rightarrow 1/2$, the second term of eqn.~(2.1) drops out. There are also physical arguments going into this expectation because we do not believe any physical system governed by a local Hamiltonian should be able to scramble information more effectively than a fully Haar random unitary circuit with infinite bond dimension. The derivation of this assumption is the topic of upcoming work.
}
\begin{align}
    \mathcal{E}(x,t) = \int_{E_W} \mathcal{T}(v).
    \label{neglinetension_eq}
\end{align}
The ``entanglement wedge" is the spacetime region bounded by the membrane $\mathcal{M}$ and the spacetime boundary. On its own, $\mathcal{M}$ is highly degenerate with many distinct configurations costing the same amount of energy. The rule for $E_W$ is then to take the global minimum over all possible $\mathcal{M}$'s that minimize \eqref{linetension_eq}.

The reflected entropy should also have a effective description, only with twice the line-tension in order to account for the canonical purification
\begin{align}
    S_R(x,t) = 2 \int_{E_W} \mathcal{T}(v).
    \label{linetension_eq_SR}
\end{align}

\begin{figure}
    \centering
    \includegraphics[height = 8cm]{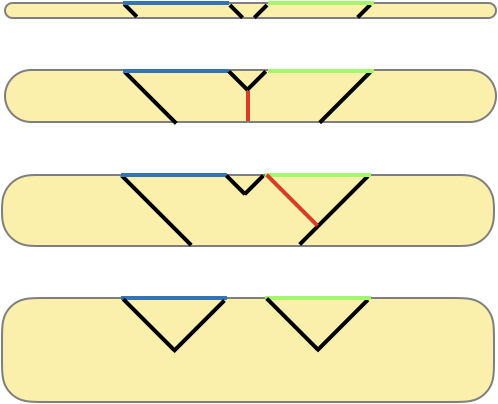}
    \includegraphics[height = 8cm]{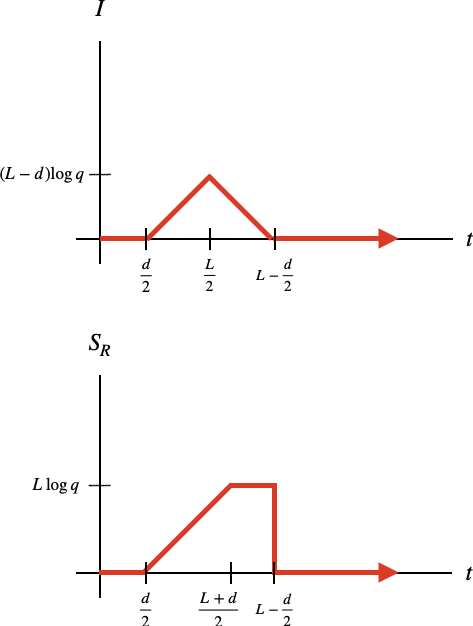}
    \caption{The four possible configurations of the minimal membranes are shown. The blue region is $A$, the green region is $B$, and time runs vertically. The black lines are membrane for the entanglement entropy while the red line is the cross-section. At very early times, the minimal membrane is disconnected again, so the cross-section is zero. Once the membrane becomes connected, the cross-section is nontrivial and attaches to the $t = 0$ time slice and its area therefore grows linearly in time. At intermediate times, the cross-section connects to the minimal membrane, hence remains constant. At late times, the minimal membrane is disconnected again, so the cross-section jumps to zero. We remind the reader that the membrane configurations are highly degenerate at large $q$ so we have only displayed one particular choice.}
    \label{membrane_dis_fig}
\end{figure}

We will now investigate this conjecture for brick-layered random unitary circuits in the limit of large bond dimension, $q$. In this limit, the line-tension may be explicitly derived to leading order by considering the minimal number of bonds cut in the circuit
\cite{2017PhRvX...7c1016N,2019arXiv190607639K}
\begin{align}
    \mathcal{T} = \begin{cases}
    \log q, & v < 1
    \\
    v \log q, & v > 1
    \end{cases}.
\end{align}
With this in hand, we can compute the negativity and reflected entropy. For simplicity, we start with adjacent intervals, each of length $L$. The possible configurations of the extremal membrane and its cross-section are shown in Fig.~\ref{membrane_dis_fig} and we find
\begin{align}
    S_R = \begin{cases}
    2t \log q, & t < L/2
    \\
    L \log q, & L/2 < t < L
    \\
    0, & t > L
    \end{cases}.
\end{align}
This precisely matches (up to additive constants from the cutoff) the irrational CFT result, including the plateau, once we make the identification of the bond dimension
\begin{align}
    q = e^{\pi c /3\beta}.
\end{align}
This value of $q$ precisely corresponds to the entropy density in the Cardy regime \cite{CARDY1986186}. The infinite bond dimension is clearly justified because $c>1$ and $\beta\rightarrow 0$ in our CFT analysis. We can similarly compute reflected entropy for disjoint intervals with the configurations tabulated in Fig.~\ref{membrane_dis_fig}. Here, we find
\begin{align}
    S_R = \begin{cases}
    0, & t < d/2
    \\
    2(t-d/2) \log q, & d/2<t <(L+d)/2
    \\
    L \log q, & (L+d)/2 < t < L-d/2
    \\
    0, & t > L-d/2
    \end{cases}.
\end{align}
Once again, this precisely matches the irrational CFT result.

So far, we have only applied the line-tension picture to global homogeneous quenches at infinite $q$. While interesting, this certainly does not capture all possible dynamical behavior of correlations in chaotic systems. In particular, finite $q$ corrections are important if we would like to compare to standard condensed matter systems. Furthermore, the line-tension picture should be systematically adaptable to the other quenches of interest. For inhomogeneous and local quenches, the line-tension will depend not only on the velocity but also the position in spacetime. This has been explicitly constructed for local operator entanglement\footnote{Local operator entanglement is the study of entanglement within the Hilbert space of operators. Study of these dynamics provides a state-independent diagnostic of chaos.} \cite{2018arXiv180300089J,loc_op_ent_draft}, but has not been addressed for the other quench protocols we have studied. We believe it is important to generalize this phenomenological picture to these generic settings such that they may explain all behavior in Fig.~\ref{holo_quenches}.

\begin{figure}
    \centering
    \includegraphics[width  = 12cm]{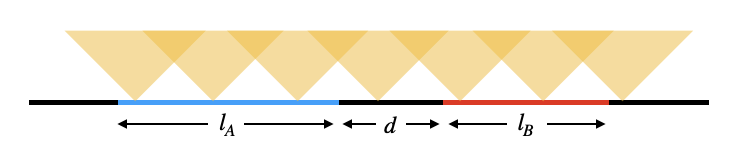}
    \includegraphics[height = 7cm]{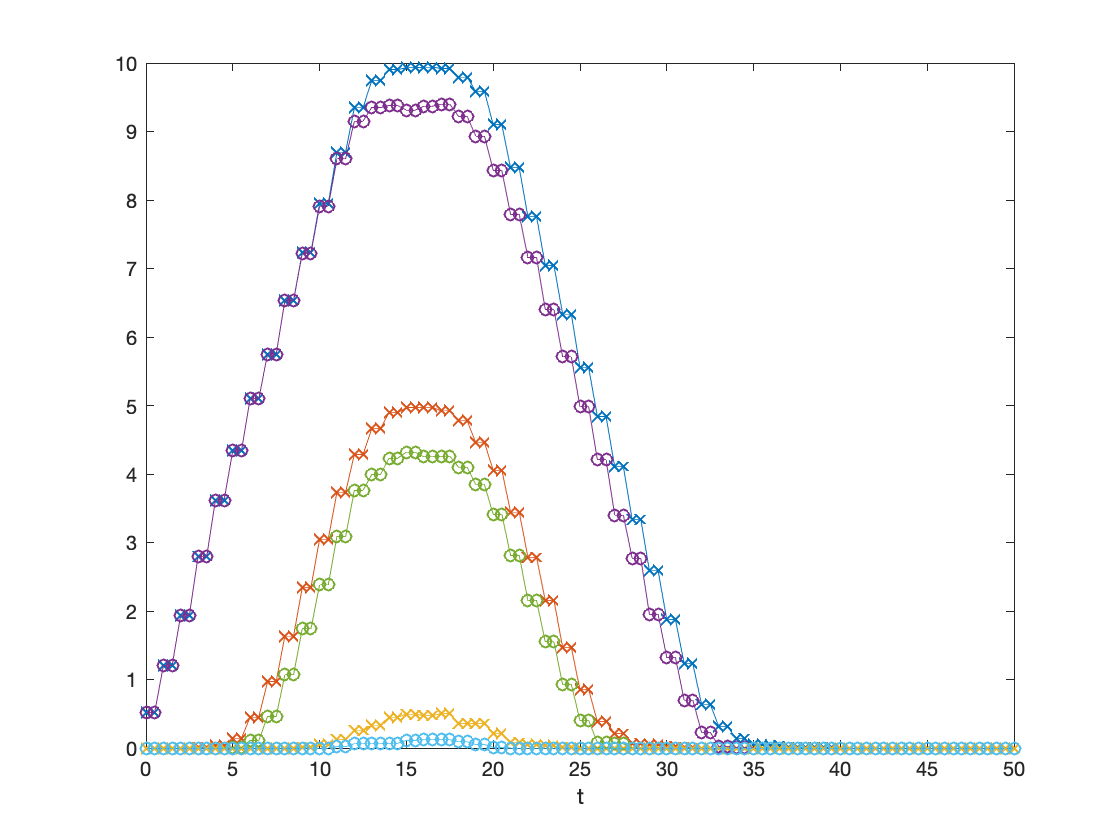} \quad
    \includegraphics[height = 7cm]{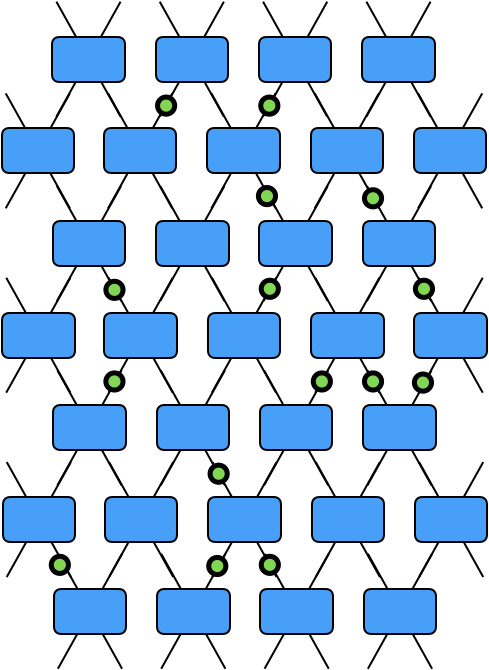}
    \caption{The mutual information ($\times$'s) and EPR pairs ($\circ$'s) between region $A$ (blue) and $B$ (red) after a global quantum quench. We start in the product state of all spin up and take $l_A = 30$, $l_B = 20$, $d = \{0,10,20\}$. On the right, we show a cartoon of the random Clifford circuit where CNOT gates (blue rectangles) are applied in a regular brick-layered fashion while phase and Hadamard gates (green circles) are applied randomly. We average over 100 disorder realizations.}
    \label{clifford_glob_quench}
\end{figure}

\subsection{Clifford circuits}
We would like to numerically test our predictions from the line-tension picture and holography using random unitary circuits with large bond dimension. Of course, this is computationally intractable due to the exponential number, $O(q^N)$, of parameters needed to be tracked. Instead, we are able to make progress by restricting to $q=2$. Furthermore, we only use random unitaries in the Clifford group consisting of CNOT, phase, and Hadamard gates. With this simplification, quantum states can be fully described by their stabilizers and the Gottesman-Knill theorem states that these circuits may be simulated in polynomial time \cite{1998quant.ph..7006G,2004PhRvA..70e2328A}. While random Clifford circuits have been shown to display certain aspects of scrambling and chaos such as KPZ behavior \cite{2017PhRvX...7c1016N} and linear scaling of tripartite operator mutual information \cite{2019arXiv190607639K}, they need to be treated with caution as they are merely unitary 3-designs \cite{2015arXiv151002769W, 2015arXiv151002619Z}. They map Pauli strings to Pauli strings and have pathological out-of-time-ordered correlators \cite{2019PhRvA..99f2334Z}. Even so, we find them to effectively model certain aspects of nonintegrable dynamics. We find their entanglement growth to strongly violate the quasi-particle picture. Furthermore, they display the missing entanglement seen in irrational CFTs and from the line-tension picture with maximal scrambling only occurring once the distance between the intervals is larger than the size of the intervals. This can be seen in Fig.~\ref{clifford_glob_quench} where we plot the mutual information and the number of EPR pairs shared between disjoint intervals. We are able to precisely distill the number of EPR pairs shared between arbitrary intervals in stabilizer states using the formalism developed in Ref.~\cite{2005NJPh....7..170A}. Due to the relative simplicity of these states, this is fully equivalent to any reasonable measure of quantum entanglement, such as logarithmic negativity. Because the mutual information is also sensitive to classical information, it is bounded below by the number of EPR pairs. We are able to detect the surplus of correlations from the mutual information by comparing the two quantities. For disjoint intervals, classical correlations can be present even when the number of EPR pairs vanishes. Notably, we do not find the plateau seen in irrational CFTs and the large-$q$ line-tension picture. This was also noted for smaller, fully Haar random, unitary circuits in Ref.~\cite{2019arXiv190709581W} and was  attributed to the small local Hilbert space size. Here, we attribute this to both small $q$ and the special stabilizer states.

\section{Discussion}
\label{discussion_sec}

In this work, we have provided a thorough analysis of mixed state correlation measures following quantum quenches. In doing so, we have provided intuition for the behavior of the newly introduced odd and reflected entropies. One main contribution is elucidating the theory dependence of each measure. For integrable theories, all of the measures behave identically, a signature of the ``all-bipartite entanglement structure" of the quasi-particle picture. In contrast, for irrational theories, we found negativity, odd entropy, and reflected entropy to become distinguished from the mutual information in that they have an extended period of mysterious correlation, the ``plateau," when the mutual information decreases. This observation is particularly unintuitive when considering negativity because negativity should only capture quantum correlations, while mutual information is also sensitive to classical correlations. It is thus surprising, though not a contradiction, that the negativity can be larger than the mutual information during this period. While the line-tension picture begins to explain this, we believe significantly more work is merited in order to resolve this tension. This is of significant practical importance due to the apparent phenomenon that chaotic systems destroy the entanglement barrier.

An interesting future direction is to determine how these quantities behave for theories that lie somewhere in between RCFTs and the irrational CFTs discussed in Section \ref{light cone_sec}, such as the compactified boson at irrational squared radius \cite{2017PhRvD..96d6020C,2017JPhA...50x4001C,2019arXiv191014575K}. Other interesting intermediate theories that are not conformal include chaotic spin chains and random unitary circuits at finite $q$. Perhaps by studying these, we can understand the emergence of the plateau. Furthermore, to fully understand the dynamics of information in 2D CFT, more generic non-equilibrium settings must be considered, such as those involving local operator insertions \cite{2014PhRvL.112k1602N}.
The interpolation from local to global quenches may provide new hints on the missing entanglement and mysterious correlations, which can be accomplished by considering, for example, multi-local excitation \cite{2019arXiv190508265C, 2019arXiv190803351K}.
We leave this to future work.

\acknowledgments
We thank Chris Akers, Tadashi Takayanagi, Kotaro Tamaoka, and Tianci Zhou for fruitful discussions and comments. We are grateful to Koenraad Audenaert and Martin Plenio for sharing their Matlab suite on stabiliser states. We thank the Yukawa Institute for Theoretical Physics at Kyoto University where this work was initiated during the workshop YITP-T-19-03 ``Quantum Information and String Theory 2019." YK is supported by the JSPS fellowship. SR is supported by a Simons Investigator Grant from the Simons Foundation.

\paragraph*{Note added} After the completion of this work, we were made aware of independent work that studied the time evolution of reflected entropy after a global quantum quench that will also appear on today's arXiv \cite{Moosa}.

% \bibliography{main}
% \bibliographystyle{JHEP.bst}
\providecommand{\href}[2]{#2}\begingroup\raggedright\endgroup

\end{document}